\documentclass[12pt]{article}
\usepackage{graphicx, amssymb, amsthm, amsbsy, amsmath}
\usepackage{xcolor}
\usepackage[normalem]{ulem}
\usepackage{url}
\usepackage[colorlinks=true,allcolors=blue]{hyperref}%
\usepackage{algorithm}
\usepackage{multirow}
\usepackage{algorithmic}
\usepackage{natbib}
\usepackage{epsfig}
\usepackage{subfigure}
\usepackage{cancel,ulem}

\usepackage[margin=1in]{geometry}

\DeclareGraphicsExtensions{.pdf,.eps.gz,.eps,.epsi.gz,.epsi,.ps,.ps.gz}
\long\def\symbolfootnote[#1]#2{\begingroup%
\def\thefootnote{\fnsymbol{footnote}}\footnote[#1]{#2}\endgroup}

\newtheorem{theorem}{Theorem}
\newtheorem{lemma}{Lemma}

\newtheorem{assumption}{{\bf Assumption}}

\def\A{{\bm A}}
\def\B{{\bm B}}

\def\E{{\bm E}}
\def\H{{\bm H}}
\def\I{{\bm I}}
\def\U{{\bm U}}
\def\V{{\bm V}}
\def\W{{\bm W}}
\def\X{{\bm X}}

\def\Z{{\bm Z}}

\def\b{{\bm b}}
\def\v{{\bm v}}

\def\u{{\bm u}}
\def\w{{\bm w}}
\def\x{{\bm x}}
\def\y{{\bm y}}
\def\z{{\bm z}}
\def\1{{\bm 1}}
\def\0{{\bm 0}}
\def\log{\hbox{log}}
\def\eop{\hfill $\Box$}

\newcommand{\bm}{\boldsymbol}
\newcommand{\bbeta}{{\bm\beta}}
\newcommand{\bepsilon}{{\bm\epsilon}}
\newcommand{\btheta}{{\bm\theta}}
\newcommand{\bgamma}{{\bm\gamma}}

\newcommand{\bOmega}{{\bm\Omega}}
\newcommand{\bGamma}{{\bm\Gamma}}
\newcommand{\bSigma}{{\bm\Sigma}}
\newcommand{\bPsi}{{\bm\Psi}}

\def\mP{\bm{\mathcal{P}}}
\def\mJ{\mathcal{J}}
\def\mS{\mathcal{S}}

\def\mG{\mathcal{G}}

\begin{document}
\title{High Dimensional Gaussian Graphical Regression Models with Covariates}
\vspace{0.2in}
\date{}
\author{
{Jingfei Zhang$^1$ and Yi Li$^2$}
\vspace{1.6mm}\\
\fontsize{11}{10}\selectfont\itshape
$^1$\,Department of Management Science, University of Miami, Coral Gables, FL 33146. \\
\fontsize{11}{10}\selectfont\itshape
$^2$\,Department of Biostatistics, University of Michigan, Ann Arbor, MI 48109. \\
}
\maketitle

\begin{abstract}
Though Gaussian graphical  models have been widely used in many scientific fields, relatively limited progress has been made to link graph structures to external covariates. We propose a Gaussian graphical regression model, which regresses both the mean and the precision matrix of a Gaussian graphical model on covariates. In the context of co-expression quantitative trait locus (QTL) studies, our {method can determine how genetic variants and clinical conditions modulate the subject-level network structures, and recover both the population-level and subject-level gene networks.}  Our framework encourages sparsity of covariate effects on both the mean and the precision matrix. In particular for the precision matrix, we stipulate simultaneous sparsity, i.e., group sparsity and element-wise sparsity, on effective covariates and their effects on network edges, respectively. We establish variable selection consistency first under the case with known mean parameters and then a more challenging case with unknown means depending on external covariates, and {establish in both cases the $\ell_2$ convergence rates and the selection consistency of the estimated precision parameters.}
The utility and efficacy of our proposed method is demonstrated through simulation studies and an application to a co-expression QTL study with brain cancer patients. 
\end{abstract}

\noindent{Keywords: subject-specific Gaussian graphical model; {Gaussian graphical model with covariates}; non-asymptotic convergence rate; sparse group lasso; co-expression QTL.}

\newpage
\baselineskip=21.5pt

\section{Introduction}
\label{sec:introduction}

%Background 
Gaussian graphical models, which shed light on the dependence structure among a set of response variables, have been applied to studies of, for example, gene regulatory networks from gene expression data \citep{fan2009network,li2012sparse,chen2016asymptotically}, brain connectivity networks from functional magnetic resonance imaging (fMRI) data \citep{li2018nonparametric,zhang2019mixed}, and firm-level financial networks from stock market data \citep{kolar2010sparse}.
Most existing models consider a homogeneous population obeying a common graphical model \citep{meinshausen2006high,yuan2007model,friedman2008sparse,peng2009partial} or several stratified graphical models \citep{guo2011joint,danaher2014joint}.

{In some applications,} graph structures may depend on individuals' characteristics, leading to the notion of subject-specific graphical models. 
In gene expression networks, external covariates, such as genetic variants, clinical  and environmental factors, may affect both the expression levels of individual genes and the co-expression relationships among genes. In biology, genetic variants that alter  co-expression relationships are referred to as co-expression quantitative trait loci (QTLs), and identifying them is of keen scientific interest \citep{wang2012snpxge2, wang2013statistical,van2018single,van2018integrative}. Other factors such as cellular states and environmental conditions may also alter gene regulatory networks \citep{luscombe2004genomic}. With these relevant external covariates, a fundamental interest, therefore, is to ascertain how they modulate the subject-level network structures, and recover both the population-level and subject-level gene networks. Characterizing such {gene regulatory networks} is key in developing gene therapies that target specific gene or pathway disruptions \citep{van2018integrative}. 

Though the literature on graphical models has been steadily growing \citep[for example,][]{meinshausen2006high,yuan2007model,friedman2008sparse,peng2009partial,fan2009network,xie2020identifying}, relatively few  frameworks permit subject-specific graphical model estimation {with theoretical justifications}. 
Several works \citep{rothman2010sparse,yin2011sparse,li2012sparse,lee2012simultaneous,cai2012covariate,JMLR:v17:16-004,chen2016asymptotically} considered covariate-dependent Gaussian graphical models, wherein the mean of the nodes depends on covariates, while the network structure is constant across all of the subjects.
\cite{guo2011joint} and  \cite{danaher2014joint} jointly estimated several group-specific Gaussian graphical models, where the graph structure is allowed to vary with discrete covariates; \cite{liu2010graph} proposed a graph-valued regression, which partitions the covariate space into several subspaces and fits separate Gaussian graphical models for each subspace using graphical lasso. 
{As noted by \cite{cheng2014sparse}, it may be} difficult to interpret the relationship between the covariates and the graphical models, as even the adjacent covariate subspaces may differ much. 
\cite{kolar2010sparse} considered a nonparametric approach for conditional covariance estimation {with  continuous covariates}. \cite{cheng2014sparse} considered a conditional Ising model for binary data where the log-odds is modeled as a linear function of external covariates. 
\cite{ni2019bayesian} considered a conditional DAG model that allows the graph structure to vary with a finite number of discrete or continuous  covariates, {and assumed a known hierarchical ordering of the nodes. Such pre-knowledge may not always be available in practical settings.}

%Our proposal
We propose a Gaussian graphical regression model that allows the network structure to vary with  external covariates (discrete and continuous) {of high dimensions}. Specifically, both the mean and the precision matrix are modeled as functions of covariates, enabling  estimation of subject-specific graphical models; see Figure \ref{fig:illu}. To facilitate estimation, we show that our proposed model can be formulated as a sequence of linear regression models that include the interactions between  response variables (e.g., gene expressions) and external covariates (e.g., genetic variants); Section \ref{subsec:igmm}.
Our model accommodates the setting where both {response variables and external covariates are high dimensional}, which is frequently encountered in genetic studies, {and includes the existing conditional mean Gaussian graphical model  \citep[e.g.,][]{yin2011sparse} as a special case}.
To estimate coefficients in the covariate-dependent precision matrix, we impose a sparse group lasso penalty that encourages effective covariates to be sparse and their effects on edges to be sparse as well.

The simultaneously sparse structure leads to a parsimonious model with estimability and interpretability, and also brings considerable theoretical challenges that are to be tackled as follows. 
We first consider a simpler setting where the mean coefficients are known; this allows us to focus on estimating the precision matrix coefficients that are simultaneously sparse. 
Recent techniques developed for the sparse group lasso under the usual linear regression setting \citep{cai2019sparse} may not be directly applicable, as the design matrix in our setting includes high-dimensional interaction terms and {non sub-Gaussian rows.}
We then investigate a more challenging setting with unknown  mean coefficients. In this case,  estimating the precision matrix is more delicate with errors arising from the estimation of mean coefficients.
For both cases, we derive the non-asymptotic rates of convergence in $\ell_2$ norm and establish selection consistency,  ensuring that we correctly select edges in both the population- and subject-level networks with probability going to 1.

%Our contribution
Our work contributes to both methodology and theory. 
As to \textit{methodology}, we propose a flexible subject-specific graphical model that depends on a large number of external covariates. We employ a combined sparsity structure that encourages effective covariates and the effect of effective covariates on the network to be simultaneously sparse. %Such a sparsity structure assumption 
With respect to \textit{theory}, we carry out a thorough investigation of the simultaneously sparse estimator, by deriving tight non-asymptotic estimation error bounds and establishing variable selection consistency. Our work addresses the theoretical challenges arising from regressing both the means and the precision matrices on external covariates. Moreover, as the simultaneously sparse regularizer is non-decomposable, the existing techniques using decomposable regularizers and null space properties \citep{negahban2012unified} %\textcolor{red}
{are not applicable}; see Section \ref{sec:theory}. Thus, our techniques {may} advance high-dimensional regression with simultaneously sparse structures. Finally, though motivated by a biological application, our method provides a general regression framework of associating networks with external covariates and is  broadly applicable to other scientific fields that involve networks.

%Organization of the paper
The rest of the article is organized as follows. Section \ref{sec:model} introduces the Gaussian graphical regression model and Section \ref{sec:estimation} discusses model estimation with known mean coefficients. Section \ref{sec:theory} investigates theoretical properties of the estimator from Section \ref{sec:estimation}. Section \ref{sec:extension} presents a two-step estimation procedure and the related theoretical properties with unknown  mean coefficients. 
Section \ref{sec:simulation} reports the simulation results, and Section \ref{sec:realdata} conducts a co-expression QTL analysis using a brain cancer genomics data set. Section \ref{sec:discussion} concludes the paper with a brief discussion. 
%All of the technical proofs and lemmas are relegated to the Supplementary Materials.

\section{Graphical Regression Models}
\label{sec:model}
\subsection{Notation and Preamble}
We start with some notation. Given a vector $\x = (x_1, \ldots, x_d) \in\mathbb{R}^d$, we use {$\Vert\x\Vert_0$, $\Vert\x\Vert_1$, $\Vert\x\Vert_2$ and $\Vert\x\Vert_{\infty}$ to denote the $\ell_0$, $\ell_1$, $\ell_2$ and $\ell_{\infty}$ norms, respectively,}
{and use $\langle\x_1,\x_2\rangle$ to denote the inner product of $\x_1,\x_2\in\mathbb{R}^d$.}
We write $[d]=\{1,2,\ldots,d\}$. Given an index set $\mS\in[d]$, we use $\x_{\mS}\in\mathbb{R}^{|\mS|}$ to denote the sub-vector of $\x$ corresponding to index $\mS$. For a matrix $\X\in\mathbb{R}^{d_1\times d_2}$, we let $\Vert\X\Vert$ and $\Vert\X\Vert_{\max}=\max_{ij}X_{ij}$ denote the spectral norm and element-wise max norm, respectively. Given $\mS \in[d_2]$, we use $\X_{\mS}\in\mathbb{R}^{d_1\times|\mS|}$ to denote the sub-matrix with columns indexed in $\mS$. We use $\lambda_{\min}(\cdot)$ and $\lambda_{\max}(\cdot)$ to denote the smallest and largest eigenvalues of a matrix, respectively. For two positive sequences $a_n$ and $b_n$, write $a_n\precsim b_n$ {or $a_n=\mathcal{O}(b_n)$} if there exist $c>0$ and $N>0$ such that $a_n<cb_n$ for all $n>N$, {and $a_n=o(b_n)$ if $a_n/b_n\rightarrow 0$ as $n\rightarrow\infty$}; write $a_n\asymp b_n$ if $a_n\precsim b_n$ and $b_n\precsim a_n$.

Suppose $\X=(X_1,\ldots,X_p)\sim\mathcal{N}_p(\bm{0},\bSigma)$. 
Denote the precision matrix $\bSigma^{-1}$ by $(\sigma^{ij})_{p\times p}$. 
Under a Gaussian distribution, $\sigma^{ij}\neq 0$ is equivalent to $X_i$ and $X_j$ being conditionally dependent given all other $X$ variables \citep{lauritzen1996graphical}.
Let $\X_{-j}=\{X_k:k\in[p],\,k\neq j\}$. \cite{meinshausen2006high} and \cite{peng2009partial} related  $(\sigma^{ij})_{p\times p}$ to the coefficients in this linear regression model: \begin{equation}\label{eqn:model1}
X_j=\sum_{k\neq j}^p\beta_{jk}X_k+\epsilon_j,\quad j\in[p],
\end{equation}
where $\epsilon_j$ is independent of $\X_{-j}$ if and only if $\beta_{jk}=-\sigma^{jk}/\sigma^{jj}$; for such defined $\beta_{jk}$, it holds that $\text{Var}(\epsilon_j)=1/\sigma^{jj}$.
Consequently, estimating the conditional dependence structure (i.e., finding nonzero $\sigma^{jk}$'s) can be viewed as a model selection problem (i.e., finding nonzero $\beta_{jk}$'s) under the regression setting in \eqref{eqn:model1}.

Let $\U=(U_1,\ldots,U_q)^\top$ be a $q$-dimensional vector of covariates. One may consider a covariate-dependent Gaussian graphical model \citep{rothman2010sparse,yin2011sparse,li2012sparse,lee2012simultaneous,cai2012covariate,chen2016asymptotically}: 
\begin{equation}\label{eqn:cgmm}
\X\vert\U=\u  \,\sim \,  \mathcal{N}_p(\bm\mu(\u),\bSigma), 
\end{equation}
where $\bm\mu(\u)=\bGamma\u$ and $\bGamma\in\mathbb{R}^{p\times q}$. 
In expression QTL studies, the $j$th row of $\bGamma$ specifies how the $q$ genetic regulators affect the expression level of the $j$th gene. 
Denote $\bGamma=(\bgamma_1,\ldots,\bgamma_p)^\top$.
Similar to \eqref{eqn:model1}, we have that
\begin{equation}\label{eqn:model2}
X_j=\u^\top\bgamma_j+\sum_{k\neq j}^p\beta_{jk}(X_k-\u^\top\bgamma_k)+\epsilon_j,\quad j\in[p],
\end{equation}
where $\epsilon_j$ is independent with $\X_{-j}$ if and only if $\beta_{jk}=-\sigma^{jk}/\sigma^{jj}$. With such defined $\beta_{jk}$,  $\text{Var}(\epsilon_j)=1/\sigma^{jj}$.

\subsection{High Dimensional Gaussian Graphical Regression with Covariates}
\label{subsec:igmm}
\begin{figure}[!t]
	\centering
	\includegraphics[trim=0 0.75cm 0 0, scale=0.725]{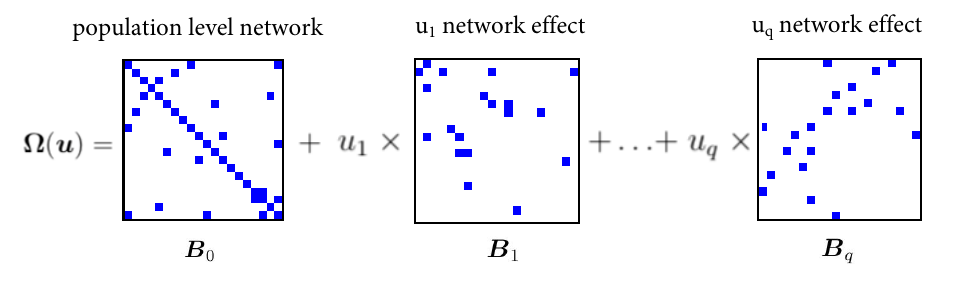}
	\caption{An illustration of the subject-specific Gaussian graphical model.}
	\label{fig:illu}
\end{figure}

With a $p$-dimensional response vector $\X=(X_1,\ldots,X_p)$ and a $q$-dimensional covariate vector $\U=(U_1,\ldots,U_q)^\top$, we assume that 
\begin{equation}\label{eqn:igmm}
\X\vert \U=\u  \, \sim \,  \mathcal{N}_p(\bm\mu(\u),\bSigma(\u)),
\end{equation}
where $\bm\mu(\u)=\bGamma\u$ and $\bSigma(\u)$ are the conditional mean vector and covariance matrix, respectively, and $\bOmega(\u)=\bSigma^{-1}(\u)$ is the precision matrix  linked to $\u$ via
\begin{equation*}
%\bOmega(\u)_{jk}=
%\begin{cases}
%\sigma^{jj}     & \quad j=k,\\
%-\omega_{jk}(\u) & \quad j\neq k,\\
% \end{cases}
\bOmega(\u)=\B_0+\sum_{h=1}^q\B_hu_h. 
\end{equation*}
Here, $\B_0,\B_1,\ldots,\B_q$ are symmetric $p\times p$ coefficient matrices,
where $\B_0$ characterizes the population level regulatory network, and $\B_h$ encodes the effect of $u_h$ on the regulatory network.
Specifically, for the $(j,k)$th entry, we have $\bOmega(\u)_{jk}=[\B_{0}]_{jk}+\sum_{h=1}^q[\B_{h}]_{jk} \times u_h$, where $[\B_{h}]_{jk}$ denotes the $(j,k)$th entry of $\B_{h}$. {We assume $\bOmega(\u)_{jj}=\sigma^{jj}$ for any $j$ and comment on it underneath \eqref{reg}.} With this assumption, the partial correlation between $Z_j$ and $Z_k$ can be expressed as $\rho_{jk}(\u)=-\frac{\bOmega(\u)_{jk}}{\sqrt{\sigma^{jj}\sigma^{kk}}}$.
See sufficient conditions on $\B_h$'s and $\u$ in Section \ref{sec:discussion} for a positive definite $\bOmega(\u)$. By specifying  $\bOmega(\u)_{jk}$'s to linearly depend on $\u$, the proposed model allows both the sparsity patterns and the %\textcolor{red}{\sout{nonzero values (i.e., }}
strengths of dependence in $\bOmega(\u)$ to vary with {external} covariates; see Figure \ref{fig:illu}. 
Model \eqref{eqn:igmm} is identifiable as long as the number of effective covariates (i.e., nonzero $\B_h$'s) is less than $n$ \citep{wu2020survey}.

%Model \eqref{eqn:igmm} is general: it includes  the covariate-adjusted Gaussian graphical model \eqref{eqn:cgmm} as a special case when {$\bSigma(\u)=\bSigma$}, that is, when all subjects share a common  network characterized by {$\bSigma^{-1}$}; if additionally $\bGamma=\0$, then $\X$ is independent of $\U$ and \eqref{eqn:igmm} encompasses the regular Gaussian graphical model where $\X\sim \mathcal{N}_p(\0,\bSigma)$ as a special case. 

As in \eqref{eqn:model1} and \eqref{eqn:model2}, model (\ref{eqn:igmm}) entails estimation of $\bGamma$ and $\bOmega(\u)$ via the following regression models, termed \textit{Gaussian graphical regression}:
\begin{equation}\label{reg}
X_j=\u^\top\bgamma_j+\sum_{k\neq j}^p\beta_{jk0}(X_k-\u^\top\bgamma_k)+\sum_{k\neq j}^p\sum_{h=1}^q\beta_{jkh}\underbrace{u_h\times(X_k-\u^\top\bgamma_k)}_{\text{{interaction term}}} +\epsilon_j,
\end{equation}
where $\beta_{jkh}=-[\B_{h}]_{jk}/\sigma^{jj}$and $\text{Var}(\epsilon_j)=1/\sigma^{jj}$, for all $j,k$ and $h$.
%{$\beta_{jkh}={\beta'_{jkh}}/\sigma^{jj}$}
Model \eqref{reg} provides a regression framework for estimating the mean and precision parameters in  \eqref{eqn:igmm}, by adding to \eqref{eqn:model1} or \eqref{eqn:model2} the interactions between $\X_{-j}$ and $\u$. {Correspondingly,} the partial correlation between $X_j$ and $X_k$, conditional on all other $X$ variables, {is modeled as} a function of $\u$, forming the basis of Gaussian graphical regression. 
The diagonal elements of $\bOmega(\u)$ (i.e., $\sigma^{jj}$'s) are connected to the residual variances in \eqref{reg}, that is, $\text{Var}(\epsilon_j)=1/\sigma^{jj}$. From this perspective, assuming $\sigma^{jj}$ to be free of $\u$ may be viewed as assuming the residual variance of $Z_j$, after removing effects of $\u$, $\Z_{-j}$ and the interactions between $\u$ and $\Z_{-j}$, to not dependent on $\u$, which is plausible in the context of regression. {However, as \eqref{eqn:model2} is a regression-type representation of the precision matrix, caution must be exercised when {comparing the residual terms in \eqref{eqn:model2} to the error terms in a standard regression problem;}
%treating its residual terms the same way as in a standard regression problem; 
see more discussions in Section \ref{sec:discussion}.}
Obviously, model (\ref{reg}) includes models (\ref{eqn:model1}) and (\ref{eqn:model2}) as special cases with $\beta_{jkh}=0$ for all $j, k$ and $h$.

Given $\U=\u$, write $\Z=\X-\bGamma\u =
(Z_1, \ldots, Z_p)$, and re-express \eqref{reg}  as
\begin{equation}\label{reg2}
Z_j=\sum_{k\neq j}^p\beta_{jk0}Z_k+\sum_{k\neq j}^p\sum_{h=1}^q\beta_{jkh}u_hZ_k +\epsilon_j.
\end{equation} 
Denote $\bbeta_j=(\b_{j0},\b_{j1},\ldots,\b_{jq})^\top\in\mathbb{R}^{(p-1)(q+1)}$, where %$\b_{j0}=(\beta_{j10},\ldots,\beta_{jp0})\in\mathbb{R}^{p-1}$ encodes connections of node $j$ {with other nodes} at the population level and
$\b_{jh}=(\beta_{j1h},\ldots,\beta_{jph})\in\mathbb{R}^{p-1}$ for all $h$;
%encodes the effects of {covariate $u_h$ on connections of node $j$ with other nodes}; 
see a more organizational and functional view of $\bbeta_j$ below:
\begin{equation}\label{eqn:illustrate}
\bbeta_j=(\underbrace{\beta_{j10},\ldots,\beta_{jp0},}^{^{\text{\footnotesize group 0}}}_{\substack{\textbf{\footnotesize $\b_{j0}:$ population level\quad} \\ \textbf{\footnotesize edges of node $j$}}}
\underbrace{\beta_{j11},\ldots,\beta_{jp1},}^{^{\text{\footnotesize group 1}}}_{\substack{\textbf{\footnotesize $\b_{j1}:$ $u_1$'s effect on} \\ \textbf{\footnotesize edges of node $j$}}}\ldots,
\underbrace{\beta_{j1q},\ldots,\beta_{jpq}}^{^{\text{\footnotesize group q}}}_{\substack{\textbf{\footnotesize $\b_{jq}:$ $u_q$'s effect on} \\ \textbf{\footnotesize edges of node $j$}}})^\top.
\end{equation}

\begin{figure}[!t]
	\centering
	\includegraphics[scale=0.75]{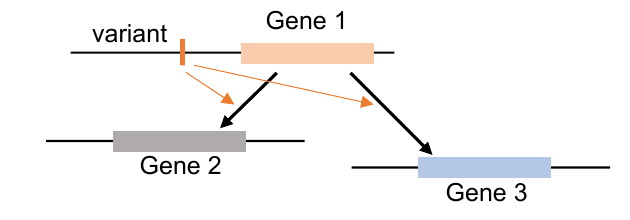}
	\caption{An illustration of gene co-expressions: the genetic variant is a trans-eQTL modulating co-expressions of pairs (1,2) and (1,3), where Gene 1 is an upstream gene and Genes 2 and 3 are downstream genes.}
	\label{illu2}
\end{figure}

When both $p$ and $q$ are large, to ensure the estimability of $\bbeta_j$, we impose on it simultaneous group sparsity and element-wise sparsity.  With groups illustrated in \eqref{eqn:illustrate},
  we assume $\bbeta_j$ is \textit{group sparse},   stipulating that effective covariates are sparse,  i.e., only a few covariates may impact edges and those impactful covariates are termed effective covariates.
We further assume $\bbeta_j$ is \textit{element-wise sparse}. 
%with covariate effects on gene co-expressions being sparse. 
That is, effective covariates may influence only a few edges.
These simultaneous sparsity assumptions are  well supported by  genetic studies \citep{van2018single}.
We exclude $\b_{j0}$ from the group sparsity constraint (but not the element-wise sparsity constraint), as it determines the population level regulatory network.
%The simultaneous sparsity assumptions bring substantial challenges to our theoretical development, as the corresponding regularizer is non-decomposable (see more discussions in Section \ref{sec:theory}). 
%The simultaneous sparsity assumptions bring substantial challenges to our theoretical development, as the corresponding regularizer is non-decomposable, and classic techniques that use decomposable regularizers and null space properties may no longer be suitable \citep{cai2019sparse}. By sharpening bound of the stochastic term in the estimation, we show in Section \ref{sec:theory} that the proposed simultaneously sparse estimator {has an improved $\ell_2$ convergence rate than that with a separate sparse structure alone (i.e., group sparse or element-wise sparse) when the true underlying coefficients are simultaneously sparse. \citep{cai2019sparse} }
With covariate $\u$ and sparsity on $\bbeta_j$'s, it is possible that $(X_j,X_k)$ and $(X_k,X_s)$ are conditionally dependent, while $(X_j,X_s)$ are conditionally independent. This type of structures is biological plausible. Consider as an example our motivating data application in co-expression QTL identification. It is possible for genetic variants located near {a  gene  (say, Gene $1$)},
called the trans-acting expression quantitative trait loci (trans-eQTLs) \citep{fehrmann2011trans},
to alter how intensively Gene $1$ may regulate (e.g. activate, inhibit) {two downstream Genes $2$ and $3$}, {while not altering the coexpression between Genes $2$ and $3$. 
In fact, the two downstream Genes $2$ and $3$ can be independent  conditional on the rest of the gene network, regardless of what the upstream trans-eQTLs might be \citep{kolberg2020co, gong2018pancanqtl, brynedal2017large}; see Figure \ref{illu2} for an illustration.} 

Model \eqref{reg2} can be viewed as an interaction model. Our later development does not abide by the common hierarchical principle for the inclusion of interactions, that is, an interaction is allowed only if the  main effects are present \citep{hao2018model,she2018group}. 
This is because gene co-expressions may occur only for certain genetic variations \citep{wang2013statistical,van2018single}, in which case, $\beta_{jkh}$ (i.e., effect of $u_h$ on edge $(j,k)$) can be nonzero while $\beta_{jk0}$ is zero (i.e., population level edge $(j,k)$). Section \ref{sec:discussion} discusses modifications of our proposal if hierarchy is to be enforced.

To ease the exposition of key ideas, we first assume a known $\bGamma$ in the ensuing development, and focus on the estimation of %\textcolor{red}{\sout{$\{\beta_{jkh}\}_{j,k\in[p],h\in\{0\}\cup[q]}$} 
$\bbeta_j$'s. In Section \ref{sec:extension}, we drop this assumption, develop an estimation procedure and derive theory when $\bGamma$ is unknown.

\section{Estimation}
\label{sec:estimation}
With $n$ independent observations, denoted by $\mathcal{D}=\{(\u^{(i)},\x^{(i)}), i\in[n]\}\in\mathbb{R}^p\times\mathbb{R}^q$, and $\z^{(i)}=\x^{(i)}-\bGamma\u^{(i)}$.
Also denote the samples of the $j$th $\z$ variable by  $\z_j=(z_j^{(1)},\ldots,z_j^{(n)})^\top$ for $j\in[p]$ and the samples of the $h$th  $\u$ covariate by $\u_h=(u_h^{(1)},\ldots,u_h^{(n)})^\top$ for $h\in[q]$. The Gaussian graphical regression model on the $j$th response variable can be written as
\begin{equation}\label{eqn:greg}
\z_j=\sum_{k\neq j}^p\beta_{jk0}\z_k+\sum_{k\neq j}^p\sum_{h=1}^q\beta_{jkh}\u_h\odot\z_k+\bm\epsilon_j,
\end{equation}
where  {$\bm\epsilon_j\sim \mathcal{N}_p(\0,1/\sigma^{jj}\I)$} and  $\odot$ denotes the element-wise product of two equal-length vectors. We partition 
{the vector of $\bbeta_j$ into $q+1$ blocks indexed by $(0),(1),\ldots,(q)\subset\{1,\ldots,(p-1)(q+1)\}$, such} that $(\bbeta_j)_{(0)}=\b_{j0}$ and $(\bbeta_j)_{(h)}=\b_{jh}$, $h\in[q]$.

Denote the squared error loss function by
\begin{equation*}
\ell_j(\bbeta_j|\mathcal{D})=\frac{1}{2n}\Vert\z_j-\W_{-j}\bbeta_j\Vert_2^2,
\end{equation*}
where $\W_{-j}=[\z_1,\z_1\odot\u_1,\ldots,\z_1\odot\u_q,\ldots,\z_{j-1}\odot\u_q,\z_{j+1},\z_{j+1}\odot\u_1\ldots,\z_p\odot\u_q]$ is an $n\times(p-1)(q+1)$ matrix. 
To estimate $\bbeta_j$, we consider %a sparse group lasso penalized loss function: 
\begin{equation}\label{eqn:obj}
\ell_j(\bbeta_j|\mathcal{D})+\lambda\Vert\bbeta_j\Vert_1+\lambda_{g}\Vert\bbeta_{j,-0}\Vert_{1,2},
\end{equation}
where $\Vert\bbeta_{j,-0}\Vert_{1,2}=\sum_{h=1}^q\Vert(\bbeta_j)_{(h)}\Vert_2$ and $\lambda,\lambda_g\ge0$ are tuning parameters. The convex regularizing terms, $\Vert\bbeta_j\Vert_1$ and $\Vert\bbeta_{j,-0}\Vert_{1,2}$, encourage element- and group-wise sparsity, respectively, though the group sparse penalty is not applied to $(\bbeta_j)_{(0)}$. 
The combined sparsity penalty in \eqref{eqn:obj} is termed the \textit{sparse group lasso} penalty \citep{simon2013sparse,li2015multivariate}.

As \eqref{eqn:obj} is convex, it can be optimized by using the existing gradient descent algorithms for sparse group lasso \citep{simon2013sparse,vincent2014sparse}, even when both $p$ and $q$ are large. 
Since the optimizers do not guarantee the symmetry of $\bOmega(\u)$, we propose a post-processing step, {similar to \cite{meinshausen2006high} and \cite{cheng2014sparse}}. 
{Denote by $\hat\beta^0_{jkh}=-\hat\sigma^{jj}{\hat\beta_{jkh}}$, where {$\hat\beta_{jkh}$} is estimated from \eqref{eqn:obj} and $\hat\sigma^{jj}$ from \eqref{eqn:variance} {for all $j,k$ and $h$.} 
%We later show that both $\hat\beta^0_{jkh}$ and $\hat\beta^0_{kjh}$ are consistent for $[\B_{h}]_{jk}$, and the difference between the network structure estimated using $\hat\beta^0_{jkh}$ and $\hat\beta^0_{kjh}$ vanishes as $n$ increases (see Theorem \ref{thm2}). 
With finite samples, we consider the following approach to enforce symmetry:
\begin{equation}\label{eq:sepmin}
[\B_{h}]_{jk}=[\B_{h}]_{kj}=\hat\beta^0_{jkh}1_{\left\{|\hat\beta^0_{jkh}|<|\hat\beta^0_{kjh}|\right\}}+\hat\beta^0_{kjh}1_{\left\{|\hat\beta^0_{jkh}|>|\hat\beta^0_{kjh}|\right\}}.
\end{equation}
Symmetrization  can also be achieved via
\begin{equation}\label{eq:sepmax}
[\B_{h}]_{jk}=[\B_{h}]_{kj}=\hat\beta^0_{jkh}1_{\left\{|\hat\beta^0_{jkh}|\ge|\hat\beta^0_{kjh}|\right\}}+\hat\beta^0_{kjh}1_{\left\{|\hat\beta^0_{jkh}|\le|\hat\beta^0_{kjh}|\right\}},
\end{equation}
but it is less conservative as $[\hat\B_{h}]_{jk}$ is nonzero if either $\hat\beta^0_{jkh}$ or $\hat\beta^0_{kjh}$ is nonzero, compared to \eqref{eq:sepmin} wherein $[\hat\B_{h}]_{jk}$ is nonzero if both $\hat\beta^0_{jkh}$ and $\hat\beta^0_{kjh}$ are nonzero. 
Though both are asymptotically equivalent (see Theorem \ref{thm2}), 
\eqref{eq:sepmin} has a better finite sample performance \citep{meinshausen2006high}, especially when $p$ is large relative to $n$.}

Two parameters $\lambda$ and $\lambda_g$ in \eqref{eqn:obj} require tuning; in our procedure, they are jointly selected via $L$-fold cross validation. As in \citet{simon2013sparse} and \citet{cai2019sparse}, we rewrite $\lambda=\alpha\lambda_0$ and $\lambda_g=(1-\alpha)\lambda_0$, where $\alpha $ reflects the weight of the lasso penalty relative to the group lasso penalty and $\lambda_0$ reflects the total amount of regularization. We assess a set of values for $\alpha\in[0,1]$, with $\alpha=0$ and $1$ corresponding to lasso and group lasso, respectively; for each $\alpha$, a sequence of $\lambda_0$ values are considered to obtain the whole regularization path  \citep{vincent2014sparse}. Finally, we choose the combination of $(\alpha, \lambda_0)$ that minimizes the cross validation error.  In our implementations, we consider $\alpha\in\{0,0.1,0.2,\ldots,0.9,1\}$ and $L=5$, and note that the result is fairly robust to the choices of $\alpha$ (see Section \ref{sec:simulation}).
%With finite samples, we select $\lambda$ and $\lambda_g$ in \eqref{eqn:obj} via BIC, similar to what {has been considered in the graphical model literature }\citep{yuan2007model,foygel2010extended}. 
%Specifically, among a set of working parameter values, we choose the combination of $(\lambda, \lambda_g)$ that minimizes
%\begin{equation}\label{eqn:ebic}
%\text{BIC}(\lambda,\lambda_g) = 
%$$n\log\{\ell_j(\hat\bbeta_j|\mathcal{D})\}+s_j\times\log\,n,$$
%\end{equation}
%where 
%$\ell_j(\hat\bbeta_j|\mathcal{D})$ is the loss function in \eqref{eqn:obj}
%$\hat\bbeta_j$ is the estimate of $\bbeta_j$ under the working tuning parameters, and $s_j$ is the number of nonzero elements in $\hat\bbeta_j$. 
 
%As network estimation is often performed as part of an exploratory analysis, model selection should also be guided by  interpretability, stability, and false discovery rate control \citep{meinshausen2010stability,danaher2014joint}.

\section{Theoretical Properties}
\label{sec:theory}
In this section, we derive the non-asymptotic $\ell_2$ convergence rate of the sparse group lasso estimator from \eqref{eqn:obj} and establish variable selection consistency. Our theoretical investigation is challenged by several unique aspects of the model. First, as the design matrix $\W_{-j}\in\mathbb{R}^{n\times(p-1)(q+1)}$ includes high-dimensional interaction terms between $\z^{(i)}$ and $\u^{(i)}$, and the variance of $\z^{(i)}$ is a function of $\u^{(i)}$, characterizing the joint distribution of each row in $\W_{-j}$ is difficult and 
%the complex joint distribution of each row
requires a delicate treatment.  
Second, as the combined penalty term $\lambda\Vert\bbeta_j\Vert_1+\lambda_{g}\Vert\bbeta_{j,-0}\Vert_{1,2}$ is not decomposable, the classic techniques for decomposable regularizers and null space properties \citep{negahban2012unified} {are not applicable}. Standard treatments of the stochastic term \citep{bickel2009simultaneous,lounici2011oracle,negahban2012unified} such as $\langle\bepsilon,\W_{-j}\Delta\rangle\le\Vert\W_{-j}^\top\bepsilon\Vert_{\infty}\Vert\Delta\Vert_1$, where $\Delta\in\mathbb{R}^{(p-1)(q+1)}$, can only yield an {$\ell_2$} convergence rate  comparable to that from the lasso or the group lasso.  
Utilizing the statistical properties and the computational optimality of the sparse group lasso estimator in \eqref{eqn:obj}, we derive two interrelated bounds on the stochastic term. The first bound characterizes the cardinality measure of the covariate space, while the second one utilizes the Karush–Kuhn–Tucker condition and properties of the combined regularizer.
Combining these bounds, we  give a sharp upper bound on the stochastic term, and show our proposed estimator possesses an improved $\ell_2$ error bounds compared to the lasso and the group lasso when the true coefficients are simultaneously sparse; see Section S3.1.

Denote the true parameters by $\bbeta_j$ for all $j$, though in some contexts we use
them to denote the corresponding arguments in functions. 
Let $\mS_j$ be the element-wise support set and $\mG_{j}$ be the group-wise support set of $\bbeta_j$, i.e, { $\mS_j=\{l:(\bbeta_j)_l\neq 0, l\in[(p-1)(q+1)]\}$ and $\mG_j=\{h:(\bbeta_j)_{(h)}\neq \0, h\in[q]\}$.} Moreover, let $s_j=|\mS_j|$, $s_{j,g}=|\mG_{j}|$, {and assume $s_j\ge 1$}; it follows that $s_{j,g}\le s_j$, $j\in[p]$.
When there is no ambiguity, we write $\W$ without noting its dependence on $j$.
%for notational ease. 
Denote by $\sigma^2_{\epsilon_j}=1/\sigma^{jj}$. 
We state a few regularity conditions and recall $\bSigma(\u^{(i)})=\text{Cov}(\z^{(i)})$, $i\in[n]$. 

\begin{assumption}
\label{ass1}
Suppose $\u^{(i)}$ are i.i.d. mean zero random vectors with a covariance matrix satisfying $\lambda_{\min}(\text{Cov}(\u^{(i)}))\ge 1/\phi_0$ for some constant $\phi_0>0$. Moreover, there exists a constant $M>0$ such that $|u^{(i)}_h|\le M$ for all $i$ and $h$.
%there exists a constant $L>0$ such that $P(\vert\z^\top\u^{(i)}\vert>t)\le2e^{-t^2/L}$ for $t>0$, $\Vert\z\Vert_2=1$
\end{assumption}
\begin{assumption}
\label{ass2}
Suppose $\phi_1\le\lambda_{min}(\text{Cov}(\z^{(i)}))\le\lambda_{max}(\text{Cov}(\z^{(i)}))\le\phi_2$ for some constants $\phi_1,\phi_2>0$.
%Let $\bSigma_{\W}=\mathbb{E}(\W^\top\W/n)$. There exist positive constants $\phi_1$, $\phi_2$ such that $\lambda_{min}(\bSigma_{\W})\ge1/\phi_1>0$ and $\lambda_{min}(\bOmega(\u^{(i)}))\ge1/\phi_2>0$, $i\in[n]$. 
\end{assumption}

\noindent
%Assumption \ref{ass1} stipulates that, with  $\{\u^{(i)}\}_{i\in[n]}$ being fixed, $\{\z^{(i)}\}_{i\in[n]}$ is the only source of randomness in \eqref{eqn:greg}. The boundedness condition on $\Vert\u^{(i)}\Vert_2$ controls the moments of $\W_{i.}^\top\a$ for $\Vert\a\Vert_2=1$ and $\a\in\mathbb{R}^{(p-1)(q+1)}$, where $\W_{i.}$ is the $i$th row of $\W$. It also implies that the $\vert u_h^{(i)}\vert\le M_1$ or equivalently $\Vert\u^{(i)}\Vert_{\infty}\le M_1$.The boundedness condition on the inner product controls the moments of $\W_{i.}^\top\a$ for $\Vert\a\Vert_2=1$ and $\a\in\mathbb{R}^{(p-1)(q+1)}$, where $\W_{i.}$ is the $i$th row of $\W$, and is analogous to a bounded moment condition on the design matrix \citep{vershynin2010introduction,vershynin2012close}.Particularly by taking $\v=\bm e_h$, a directional vector with the $h$th entry being 1 and 0 elsewhere, $|\langle\u^{(i)},\v\rangle|<M_1$ implies that the $\vert u_h^{(i)}\vert\le M_1$ or equivalently $\Vert\u^{(i)}\Vert_{\infty}\le M_1$.
Assumption \ref{ass1} stipulates that the covariates are element-wise bounded, which is needed in characterizing the joint distribution of each row in $\W$. This condition is not restrictive as genetic variants are often coded to be $\{0,1\}$ or $\{0,1,2\}$ \citep{chen2016asymptotically}. 
Assumptions \ref{ass1} and \ref{ass2} impose bounded eigenvalues on $\text{Cov}(\u^{(i)})$ and $\text{Cov}(\z^{(i)})$ as commonly done in the high-dimensional regression literature \citep{chen2016asymptotically,hao2018model,cai2019sparse}.

\begin{assumption}
\label{ass2.2}
The dimensions $p,q$ and sparsity $s_j$ satisfy $\log\,p+\log\,q=\mathcal{O}(n^\delta)$ and $s_j=o(n^\delta)$ for $\delta\in[0,1/6]$.
\end{assumption}

\noindent
Assumption \ref{ass2.2} is a sparsity condition, allowing both $\log\,p$ and $\log\,q$ to grow at a polynomial order of $n$. Moreover, the number of nonzero entries $s_j$ can also grow with $n$. This condition and $\delta\in[0,1/6]$ are useful when establishing a {restricted eigenvalue} condition \citep{bickel2009simultaneous} for $\W^\top\W/n$ and when bounding the stochastic term $\langle\bepsilon,\W\Delta\rangle$.

Let $s_{\lambda}$ denote the number of nonzero entries in a candidate model such that $s_j<s_{\lambda}\le n$. Given an $s_{\lambda}$ satisfying the  conditions in Theorem \ref{thm1}, we choose $\lambda_{\max}$ and $\lambda_{\min}$ to be the upper and lower limits of $\lambda_0$ for each $\alpha$,  respectively corresponding to {an empty model with no variables selected and a sparse model with $s_{\lambda}$ variables selected.}

\begin{theorem}\label{thm1}
Suppose that Assumptions \ref{ass1}-\ref{ass2.2} hold, $s_{\lambda}(\log\,p+\log\,q)=\mathcal{O}(\sqrt{n})$ and $n\ge A_1\{s_j\log(ep)+s_{j,g}\log(eq/s_{j,g})\}$ for some constant $A_1>0$.
Then  $\hat\bbeta_j$, $j\in[p]$, in \eqref{eqn:obj} with 
\begin{equation}\label{eqn:lambda}
\lambda=C\sigma_{\epsilon_j}\sqrt{\log(ep)/n+s_{j,g}\log(eq/s_{j,g})/(ns_j)},\quad \lambda_{g}=\sqrt{s_j/s_{j,g}}\lambda, 
\end{equation}
satisfies, with probability at least $1-C_1\exp[-C_2\{s_j\log(ep)+s_{j,g}\log(eq/s_{j,g})\}]$, 
\begin{equation}\label{eqn:bound1}
\|\hat\bbeta_j-\bbeta_j\|_2^2\precsim \frac{\sigma_{\epsilon_j}^2}{n}\left\{s_j\log(ep)+s_{j,g}\log(eq/s_{j,g})\right\}+\frac{\sigma_{\epsilon_j}^2}{n},
\end{equation}
where $C$, $C_1$, and $C_2$ are positive constants.
\end{theorem}

Theorem \ref{thm1} shows that our proposed estimator enjoys an improved $\ell_2$ error bound over both the lasso and the group lasso under simultaneous sparsity. Specifically, given that the dimension of $\bbeta_j$ is $(p-1)(q+1)$ and $s_{j,g}\le s_j$, applying the regular lasso regularizer $\lambda\Vert\bbeta_j\Vert_1$ alone would yield an error bound of $(s_j/n)\log(pq)$ \citep{negahban2012unified}, which is slower than that in \eqref{eqn:bound1} when $\log p/\log q=o(1)$ and $s_{j,g}/s_j=o(1)$, {corresponding to group sparsity}. 
Moreover, when $p>n+1$, estimating with the group lasso regularizer $\lambda_{g}\Vert\bbeta_{j,-0}\Vert_{1,2}$ alone, which excludes $(\bbeta_j)_{(0)}$, is not feasible, because the dimension of the latter (i.e., $p-1$) exceeds $n $. If we utilize a group lasso regularizer $\lambda_{g}\Vert\bbeta_{j}\Vert_{1,2}$ that includes $(\bbeta_j)_{(0)}$,
the estimator would have an $\ell_2$ error bound of $(s_{j,g}/n)\log\,q+(s_{j,g}/n)p$ \citep{lounici2011oracle}, which is slower than that in \eqref{eqn:bound1} when $log q/p=o(1)$ and $s_j/ s_{j,g}=o(p/\log p)$, corresponding to within-group sparsity. 
While the optimality of these error bounds warrants further investigation, the combined regularizer $\lambda\Vert\bbeta_j\Vert_1+\lambda_{g}\Vert\bbeta_{j,-0}\Vert_{1,2}$ may improve upon both the regular lasso and group lasso regularizers, when the true underlying coefficients are both element-wise and group sparse. In Theorem \ref{thm1}, the condition $s_{\lambda}(\log\,p+\log\,q)=\mathcal{O}(\sqrt{n})$ upper bounds the size of candidate models, which in turn helps to bound $\langle\bepsilon,\W\Delta\rangle$. {The parameter $s_{\lambda}$ can be set to $c\sqrt{n}/\max\{\log\,p,\log\,q\})$ for some $c>0$; by Assumption \ref{ass2.2}, it follows that $s_j=o(s_{\lambda})$.}

%The rate in Theorem \ref{thm1} is comparable with that in \citet{cai2019sparse}, which studied the convergence rate of the sparse group lasso estimator under a regular linear regression setting. As we mentioned, their technique is not {directly} applicable to our setting which includes high-dimensional interaction terms and rows that are not jointly sub-Gaussian.
Some group lasso literature \citep{yuan2006model,lounici2011oracle} noted that the grouped $\ell_1$ penalty should compensate for the group size. It might be the case that $\lambda_g$ {is} adjusted by $\sqrt{p-1}$, as each group in $\bbeta_j$ is of size $p-1$. Indeed, with $(\bbeta_j)_{(0)}=\0$ and no element-wise sparsity within the nonzero groups
%, that is, $\Vert(\bbeta_j)_{(h)}\Vert_0=p-1$ for $h\in\mG_j$,  
$\sqrt{s_j/s_{j,g}}$ becomes $\sqrt{p-1}$ in \eqref{eqn:lambda}.
Interestingly, our theoretical investigation reveals that, for the combined regularizer $\lambda\Vert\bbeta_j\Vert_1+\lambda_{g}\Vert\bbeta_{j,-0}\Vert_{1,2}$, $\lambda_{g}=\sqrt{s_j/s_{j,g}}\lambda$  suffices to suppress the noise term; see (S10).

%\citet{cai2019sparse}  studied the convergence rate of the sparse group lasso estimator in a regular linear regression setting. With a different proof strategy, their theoretical analysis yields a convergence rate of $\sigma^2\left\{s_j\log(es_{j,g}p)+s_{j,g}\log(eq/s_{j,g})\right\}/n$, which is slower than \eqref{eqn:bound1}, especially when $s_{j,g}$ is large. Again with regular linear regression, \cite{cai2019sparse} showed that the minimax lower bound for the estimation error is $\sigma^2\left\{s_j\log(es_{j,g}p/s_j)+s_{j,g}\log(eq/s_{j,g})\right\}/n$, which is matched by the upper \\ bound in \eqref{eqn:bound1} if $s_j\asymp s_{j,g}$.

We next show that our proposed sparse group lasso estimator achieves  variable selection consistency under a mutual coherence condition. {Let $\bSigma_{\W}=\mathbb{E}(\W^\top\W/n)$.}
\begin{assumption}[Mutual coherence]
\label{ass3} 
Denote by $\eta_j=1+\sqrt{s_j/s_{j,g}}$, $j\in[p]$. We assume that for some positive constant $c_0>6\phi_1/\phi_0$, the covariance matrix $\bSigma_{\W}$ satisfies that 
$$
\max_{k\neq l}\vert\bSigma_{\W}(k,l)\vert\le\frac{1}{c_0(1+8\eta_j)s_j},
$$
where $\bSigma_{\W}(k,l)$ denotes the $(k,l)$th element of $\bSigma_{\W}$.
\end{assumption}
\noindent
Assumption \ref{ass3} specifies that the correlation between columns in $\W$ cannot be excessive.
%, which has been considered in establishing the $\ell_{\infty}$ norm convergence of the lasso and the group lasso \citep{lounici2011oracle}. For example, this condition is satisfied if the columns of $\U$ are orthogonal. Similar correlation conditions include the weak mutual coherence condition \citep{bunea2007sparsity}, neighborhood stability condition \citep{meinshausen2006high} and the irrepresentability condition \citep{zhao2006model}; see \cite{van2009conditions} for a discussion of these relationships.

Specifically, by the law of total probability, we write
\begin{eqnarray*}
\max_{k\neq l}\vert\bSigma_{\W}(k,l)|
&=&\max_{\substack {l_1,l_2,l_3,l_4\\
(l_1,l_2)\neq(l_3,l_4)}}\mathbb{E}\left\{\mathbb{E}\left(z_{l_1}^{(1)}z_{l_2}^{(1)}u_{l_3}^{(1)}u_{l_4}^{(1)}|\u^{(1)}\right)\right\}\\
&\le&\max_{l_1\neq l_2}[\text{Cov}(\z^{(1)})]_{l_1,l_2}\times\max_{l_3\neq l_4}[\text{Cov}(\u^{(1)})]_{l_3,l_4}.
\end{eqnarray*}
Hence, Assumption \ref{ass3} holds when the correlations among $\{Z_j\}_{j\in[p]}$ and among $\{U_h\}_{h\in[q]}$ are not too large.
A trivial sufficient condition is $\text{Cov}(\u^{(1)})=\I$. Furthermore, if $s_j=\mathcal{O}(1)$, Assumption \ref{ass3} is satisfied when $\max_{l_1\neq l_2}[\text{Cov}(\z^{(1)})]_{l_1,l_2}\times\max_{l_3\neq l_4}[\text{Cov}(\u^{(1)})]_{l_3,l_4}$ is less than some positive constant.
Similar correlation conditions include the neighborhood stability condition \citep{meinshausen2006high} and the irrepresentability condition \citep{zhao2006model}; see \cite{van2009conditions} for a discussion of these relationships.

\begin{theorem}\label{thm2}
{Suppose Assumptions \ref{ass1}-\ref{ass3} hold. If $\log\,p\asymp\log\,q$} and $n\ge A_1\{s_j\log(ep)+s_{j,g}\log(eq/s_{j,g})\}$ for some constant $A_1>0$, then for $j\in[p]$, the estimator $\hat\bbeta_j$ in \eqref{eqn:obj} with $\lambda$ and $\lambda_g$ as in \eqref{eqn:lambda} satisfies
\begin{equation}\label{eqn:infybound}
\|\hat\bbeta_j-\bbeta_j\|_{\infty}\le\left\{3\phi_1\eta_j+\frac{18\phi^2_1(1+4\eta_j)^2\eta_j}{{\phi_0}(c_0{\phi_0}-2\phi_1)(1+8\eta_j)}\right\}\lambda,
\end{equation}
with probability at least $1-C'_1\exp(-C'_2\log\,p)$, where $C'_1$, $C'_2$ are some positive constants.
Define $\hat{\mathcal{S}}_j=\left\{k:\,|(\hat\bbeta_{j})_k|>\left\{3\phi_1\eta_j+\frac{18\phi^2_1(1+4\eta_j)^2\eta_j}{{\phi_0}(c_0{\phi_0}-2\phi_1)(1+8\eta_j)}\right\}\lambda\right\}$.
In addition, if the minimum signal strength satisfies 
\begin{equation}\label{eqn:betamin}
\min_{l\in\mathcal{S}}|(\bbeta_{j})_l|>2\left\{3\phi_1\eta_j+\frac{18\phi^2_1(1+4\eta_j)^2\eta_j}{{\phi_0}(c_0{\phi_0}-2\phi_1)(1+8\eta_j)}\right\}\lambda,
\end{equation}
we have that $\mathbb{P}(\hat \mS_j=\mS_j)\ge 1-C'_1\exp(-C'_2\log\,p)$, $j\in [p]$.
%\textcolor{red}{changed to $\ge$.}
\end{theorem}

For the recovery of true signals in high-dimensional regression, minimum signal strength conditions such as \eqref{eqn:betamin} are necessary \citep{zhang2009some}. 
%Following Remark \ref{re:thm1_1},  \eqref{eqn:betamin} implies that, when $s_j\asymp s_{j,g}$, the sparse group lasso estimator can accommodate weaker signals compared to the regular lasso \citep{wainwright2009sharp} and the group lasso \citep{lounici2011oracle},  a desirable property of the sparse group lasso estimator.
The condition of $\log\,p\asymp\log\,q$ allows $p$ and $q$ to grow at a polynomial rate relative to each other,  ensuring a tighter bound on $\Vert\W^\top\bepsilon_j\Vert_{\infty}$; see  \cite{chen2016asymptotically}.  
Moreover, the selection consistency result in Theorem \ref{thm2} holds for both estimates {in \eqref{eq:sepmax} and \eqref{eq:sepmin}}, as \eqref{eqn:infybound} characterizes the relationship between the fitted values and the true parameters. 

With $\hat\bbeta_j$, a natural estimate of the variance  $\sigma^2_{\epsilon_j}=1/\sigma^{jj}$ would be
\begin{equation}\label{eqn:variance}
\hat\sigma^2_{\epsilon_j}=\frac{1}{n-\hat s_j}\Vert\z_j-\W\hat\bbeta_j\Vert_2^2=\frac{1}{n-\hat s_j}\z_j^\top\left(\bm I_{n\times n}-\mathcal{P}_{\hat\mS_j}\right)\z_j,
\end{equation}
where $\mathcal{P}_{\hat\mS_j}$ is the projection matrix onto the column space of $\W_{\hat\mS_j}$.
The estimator in \eqref{eqn:variance} can alternatively be written as $\hat\sigma^2_{\epsilon_j}=\frac{1}{n-\hat s_j}(1-\gamma_n^2)\bepsilon^\top\bepsilon$, where $\gamma_n^2=\bepsilon^\top\mathcal{P}_{\hat\mS_j}\bepsilon/\bepsilon^\top\bepsilon$ represents the fraction of bias in $\hat\sigma^2_{\epsilon_j}$.
Under conditions in Theorem 2 and using a result in (S10), we get $
\gamma^2_n\asymp{\sigma_{\epsilon_j}^2}\left\{s_j\log(ep)+s_{j,g}\log(eq/s_{j,g})\right\}/n$. Therefore, $\hat\sigma^2_{\epsilon_j}$ is  consistent, provided that $\sigma_{\epsilon_j}^2\left\{s_j\log(ep)+s_{j,g}\log(eq/s_{j,g})\right\}/n\rightarrow 0$.

\section{Estimation with Unknown \texorpdfstring{$\bGamma$}{TEXT}: a Two-Step Procedure}
\label{sec:extension}
We present a two-step estimation procedure when $\bGamma$ is unknown, followed by its theoretical properties. Assuming a sparse $\bGamma$, Step 1 estimates $\bGamma$ using an $\ell_1$-penalized regression; in Step 2, {we approximate each $\z^{(i)}$ with $\hat\z^{(i)}=\x^{(i)}-\hat\bGamma\u^{(i)}$, where $\hat\bGamma$ is estimated from the first step, and estimate $\bbeta_j$ based on $\hat \z^{(1)},\ldots,\hat \z^{(n)}$ by using the procedure described in Section \ref{sec:estimation}.}
The two-step procedure is computationally feasible, particularly when both $p$ and $q$ are large, and has been considered 
for covariate-adjusted Gaussian graphical models \citep{cai2012covariate,yin2013adjusting,chen2016asymptotically}. 

\medskip
\noindent
\textbf{Step 1.} Denote {the covariate matrix} by $\H=[\u^{(1)},\ldots,\u^{(n)}]^\top$ and the sample of the $j$th variable by $\x_j=(x_j^{(1)},\ldots,x_j^{(n)})^\top$.
We first estimate %\textcolor{red}{the mean coefficient} 
$\bGamma$ with
\begin{equation}\label{eqn:obj2}
\hat\bgamma_j=\arg\min_{\bgamma\in\mathbb{R}^{q}}\frac{1}{2n}\Vert\x_j-\H\bgamma\Vert_2^2+\lambda_1\Vert\bgamma\Vert_1,
\end{equation}
and denote the estimates by  $\hat\bGamma=(\hat\bgamma_1,\ldots,\hat\bgamma_p)^\top$.

%=(\hat z_1^{(i)},\ldots,\hat z_p^{(i)})^\top
\medskip
\noindent
\textbf{Step 2.} With $\hat\bGamma$ obtained from Step 1, {we calculate $\hat\z^{(i)}=\x^{(i)}-\hat\bGamma\u^{(i)}$
%$ for all $i\in[n]$ 
and estimate} $\bbeta_j$ via
\begin{equation}\label{eqn:obj3}
\hat\bbeta_j=\arg\min_{\bbeta_j\in\mathbb{R}^{(p-1)(q+1)}}\frac{1}{2n}\Vert\hat\z_j-\hat\W_{-j}\bbeta_j\Vert_2^2+\lambda\Vert\bbeta_j\Vert_1+\lambda_g\Vert\bbeta_{j,-0}\Vert_{1,2},
\end{equation}
where {$\hat\z_j=(\hat z_j^{(1)},\ldots,\hat z_j^{(n)})^\top$ and} $\hat\W_{-j}=[\hat\z_1\odot\u_1,\ldots,\hat\z_1\odot\u_q,\ldots,\hat\z_{j-1}\odot\u_q,\hat\z_{j+1}\odot\u_1\ldots,\hat\z_p\odot\u_q]$, {with $\hat\z_j$ and $\hat\W_{-j}$ respectively approximating $\z_j$ and  $\W_{-j}$.} When there is no ambiguity, we  write $\hat\W$ without emphasizing its dependence on $j$. 
Both \eqref{eqn:obj2} and \eqref{eqn:obj3} are  convex, and can be optimized efficiently \citep{simon2013sparse,vincent2014sparse}.

Step 1 poses a regular lasso penalty on $\bGamma$, as commonly done in the covariate-adjusted Gaussian graphical model literature \citep{rothman2010sparse,cai2012covariate,yin2013adjusting,chen2016asymptotically}. 
%Because each regression in Step 1 is only of dimension $q$, compared to $(p-1)(q+1)$ in Step 2,  a lasso penalty may suffice to reduce the sample complexity; see \citet{cai2019sparse} on sample complexity comparisons between (group) lasso and sparse group lasso. 
{Note that each regression in Step 1 is of dimension $q$.} When $q$ is large, it may be necessary to consider a sparse group penalty (as in Step 2) that encourages $\bGamma$ to be both element-wise and group sparse.

The two-step procedure involves three  parameters $\lambda_1$, $\lambda$ and $\lambda_g$ that need to be tuned.  We  tune $\lambda_1$ in Step 1 via $L$-fold cross validation, and then tune $\lambda$ and $\lambda_g$ jointly in Step 2 using the same procedure as in Section \ref{sec:estimation}. Sequential tuning is common \citep[e.g.][]{danaher2014joint}, and gives a good numerical performance in our experiments.

The theoretical development {for the  two-step procedure} is challenging. Step 1 involves a regularized regression $\x_j=\H\bgamma_j+\z_j$ with heteroskedastic errors %\textcolor{red}{\sout{$z^{(i)}_j\sim\mathcal{N}(\0,\bSigma(\u^{(i)})_{jj})$, $i\in[n]$} 
as the variance of $z^{(i)}_j$ is a function of $\u^{(i)}$.
%, which ar more difficult than the usual i.i.d errors.
In Step 2, both the response vector $\hat\z_j$ and the design matrix $\hat\W$ inherit approximation errors from
$\hat\bGamma-\bGamma$, which further complicates the analysis of the sparse group lasso estimator. We show that $\hat\bbeta_j$ in \eqref{eqn:obj3} can achieve the same convergence rate as that in Theorem \ref{thm1} (i.e., the noiseless case), and thus enjoys the oracle property (as if $\bGamma$ were known).
%We show that, using the same conditions as in Theorem \ref{thm1}, the convergence rate of $\hat\bbeta_j$ in \eqref{eqn:obj3} is slowed by a term of $\sqrt{t\log\,q}$ with { $t=\max_j\Vert\bgamma_j\Vert_0$}, which reflects the estimation error from the first step. We further show that if correlations between columns in the design matrix are small, $\hat\bbeta_j$ can achieve the same convergence rate as that in Theorem \ref{thm1} (i.e., the noiseless case), and thus enjoys the oracle property.  

\begin{assumption}
\label{ass4} 
There exists a constant $M_2>0$ such that $\Vert\bbeta_j\Vert_1\le \sigma_{\epsilon_j}M_2$ for all $j$. Moreover, there exists a constant $\phi'_0>0$ such that $\lambda_{\max}(\text{Cov}(\u^{(i)}))\le \phi'_0$.
\end{assumption}

\noindent
The boundedness of $\Vert\bbeta_j\Vert_1$ controls the approximation errors in $\hat\z_j$ when analyzing the second step of the estimation procedure.
%, where we consider $\hat\z_j=\hat\W\bbeta_j+\bepsilon_j+(\hat\z_j-\z_j)+(\W-\hat\W)\bbeta_j$. 
Similar conditions have been considered in other two-step procedures \citep{cai2012covariate,chen2016asymptotically}.
This assumption can be relaxed to allow $M_2$ to diverge, in which case $M_2$ appears in the convergence rates in Theorem \ref{thm3} and \ref{thm4}.

%\begin{assumption}[Restricted Eigenvalue]
%\label{ass6}
%Let $\mathcal{C}_j=\{\v\in\mathbb{R}^{q}:\Vert\v_{\mT^c_j}\Vert_1\le2\Vert\v_{\mT_j}\Vert_1\}$, {where $\mT_j$ is the index set of nonzero entries in $\bgamma_j$}. There exists $\kappa_j>0$ such that,  for all $j$,
%$$\frac{\Vert\H\v\Vert^2_2}{n}\ge\kappa_j\Vert\v\Vert_2^2,\quad \forall\v\in\mathcal{C}_j.$$
%\end{assumption}
%Assumption \ref{ass6} is needed for quantifying  the error when estimating {$\bGamma$}. If the gram matrix $\H^\top\H/n$ satisfies this restricted eigenvalue  condition, the $\ell_1$ regularized estimation leads to the desired prediction and estimation error rates \citep{bickel2009simultaneous,negahban2012unified}.

\begin{theorem}\label{thm3}
Suppose that {conditions in Theorem \ref{thm1} and Assumption \ref{ass4}} are satisfied, $t=o(n^{1/3})$ and {$n\ge A_2\{s_j\log(ep)+s_{j,g}\log(eq/s_{j,g})\}$ for some constant $A_2>0$.}
Let $\lambda_1=14\phi_2\sqrt{\tau_1\log\,q/n}$ for any $\tau_1>0$. The minimizer $\hat\bgamma_j$ in \eqref{eqn:obj2} satisfies
\begin{equation}\label{eqn:bound2}
\|\hat\bgamma_j-\bgamma_j\|^2_2\precsim \frac{t\log\,q}{n},\quad\frac{1}{n}\|\hat\z_j-\z_j\|^2_2\precsim  \frac{t\log\,q}{n},
\end{equation}
with probability at least $1-3\exp(-\tau_1\log \,q)$, $j\in[p]$.
The minimizer $\hat\bbeta_j$ in \eqref{eqn:obj3} with $\lambda$ and $\lambda_g$ as in \eqref{eqn:lambda}
%\begin{equation}\label{eqn:lambda2}
%\lambda=C\sigma_{\epsilon_j}\sqrt{\{\log(ep)/n+s_{j,g}\log(eq/s_{j,g})/(ns_j)\}},\quad %lambda_{g}=\sqrt{s_j/s_{j,g}}\lambda, 
%\end{equation}
satisfies with probability at least $1-C_3\exp[C_4\{\log\,p-(\tau_1-1)\log\,q\}]$,
\begin{equation}\label{eqn:bound3}
\|\hat\bbeta_j-\bbeta_j\|_2^2\precsim \frac{\sigma_{\epsilon_j}^2}{n}\left\{s_j\log(ep)+s_{j,g}\log(eq/s_{j,g})\right\}+\frac{\sigma_{\epsilon_j}^2}{n},
\end{equation}
for some positive constants $C_3$, $C_4$.
\end{theorem}
%Moreover, if Assumption \ref{ass3} holds and {$n\ge A_3\{s_j\log(ep)+s_{j,g}\log(eq/s_{j,g})\}$ for some constant $A_3>0$}, then $\hat\bbeta_j$ in \eqref{eqn:obj3} with $\lambda_1$ and $\lambda_{g}$ in \eqref{eqn:lambda} satisfies 
%\begin{equation}\label{eqn:bound4}
%\|\hat\bbeta_j-\bbeta_j\|_2^2\precsim \frac{\sigma_{\epsilon_j}^2}{n}\left\{s_j\log(ep)+s_{j,g}\log(eq/s_{j,g})\right\}+\frac{\sigma_{\epsilon_j}^2}{n},
%\end{equation}
%with probability at least $1-C'_3\exp[C'_4\{\log\,p-(\tau_1-1)\log\,q\}]$, for some positive constants $C'_3$ and $C'_4$. 

When $\bGamma$ is unknown, as opposed to the oracle regression equation $\z_j=\W\bbeta_j+\bepsilon_j$, we only have access to the noisy equation $\hat\z_j=\hat\W\bbeta_j+\E_j$, where $\E_j=\bepsilon_j+(\hat\z_j-\z_j)+(\W-\hat\W)\bbeta_j$. 
The condition $t=o(n^{1/3})$ is needed to control the errors from the estimation in the first step, which in turn controls the error $(\hat\z_j-\z_j)+(\W-\hat\W)\bbeta_j$. It is seen that the rate in \eqref{eqn:bound3} is the same as that derived in the oracle case (as if $\bGamma$ were known) in \eqref{eqn:bound1}.
%Compared to the tuning parameters in \eqref{eqn:lambda} under the noiseless case (with $\bGamma$ known), the tuning parameters in \eqref{eqn:lambda2} are multiplied by an additional term $\sqrt{t\cancel{\log\,q}}$. This additional term in $\lambda$ and $\lambda_g$ is needed to suppress the noise terms, including both $\langle\bepsilon_j,\hat\W\Delta\rangle$ and $(\hat\z_j-\z_j)+(\W-\hat\W)\bbeta_j$, where $\Delta=\hat\bbeta_j-\bbeta_j$. Consequently, the convergence rate in \eqref{eqn:bound3} is also multiplied by a term of $t\cancel{\log q}$. Under Assumption \ref{ass3}, the rate in \eqref{eqn:bound4} is the same as the oracle rate in \eqref{eqn:bound1}. This is achievable as the stochastic term $\langle\bepsilon_j,\hat\W\Delta\rangle$ can be better bounded when the influence of the ``noise" covariates (i.e., $\hat\W_{\mS^c}$) on the effective covariates (i.e., $\hat\W_{\mS}$) is controlled under Assumption \ref{ass3}. Conditions similar to Assumption \ref{ass3} were also considered in \cite{cai2012covariate} and \cite{yin2013adjusting} to establish the oracle inequality of the covariate-adjusted precision matrix estimator obtained from the two-step estimation procedure.

\begin{theorem}\label{thm4}
Suppose that Assumptions \ref{ass1}-\ref{ass4} {hold},  $\lambda_1=14\phi_2\sqrt{\tau_1\log\,q/n}$ for any $\tau_1>0$,  {$n\ge A_3\{s_j\log(ep)+s_{j,g}\log(eq/s_{j,g})\}$ for some constant $A_3>0$, $\log\,p\asymp\log\,q$} and  $t=o(n^{1/3})$. For $j\in[p]$, the sparse group lasso estimator $\hat\bbeta_j$ in \eqref{eqn:obj3} with $\lambda$ and $\lambda_g$ as in \eqref{eqn:lambda} satisfies,
\begin{equation}\label{eqn:infybound2}
\|\hat\bbeta_j-\bbeta_j\|_{\infty}\le\frac{9}{2}\left\{\phi_1\eta_j+\frac{12\phi^2_1(1+3\eta_j)^2}{{\phi_0}(c_0{\phi_0}-6\phi_1)(1+8\eta_j)}\right\}\lambda,
\end{equation}
with probability at least $1-C_5\exp[C_6\{\log\,p-(\tau_1-1)\log\,q\}]$, for some positive constants $C_5$, $C_6$.
Define $\hat{\mathcal{S}}_j=\left\{k:\,|\hat\bbeta_{j,k}|>\frac{9}{2}\left\{\phi_1\eta_j+\frac{12\phi^2_1(1+3\eta_j)^2}{{\phi_0}(c_0{\phi_0}-6\phi_1)(1+8\eta_j)}\right\}\lambda\right\}$.
If, in addition, the minimum signal strength satisfies 
\begin{equation}\label{eqn:betamin2}
\min_{k\in\mathcal{S}}|\bbeta_{j,k}|>9\left\{\phi_1\eta_j+\frac{12\phi^2_1(1+3\eta_j)^2}{{\phi_0}(c_0{\phi_0}-6\phi_1)(1+8\eta_j)}\right\}\lambda,
\end{equation}
then $\mathbb{P}(\hat \mS_j=\mS_j)\ge 1-C_5\exp[C_6\{\log\,p-(\tau_1-1)\log\,q\}]$, $j\in [p]$.
\end{theorem}

%The additional sparsity condition $t=o(\sqrt{n/\log\,q})$ is needed to establish the $\ell_{\infty}$ bound for $\hat\bbeta_j$ in the presence of the error in $\hat\bGamma-\bGamma$. 
Compared to the minimal signal strength condition \eqref{eqn:betamin} in the noiseless case, the condition in \eqref{eqn:betamin2} is slightly stronger. Similar to the case where $\bGamma$ is known, (S36) leads to that 
$\hat\sigma^2_{\epsilon_j}=\frac{1}{n-\hat s_j}\Vert\hat\z_j-\hat\W\hat\bbeta_j\Vert_2^2$ is  consistent, if $\sigma_{\epsilon_j}^2\left\{s_j\log(ep)+s_{j,g}\log(eq/s_{j,g})\right\}/n\rightarrow 0$.

\section{Simulations}
\label{sec:simulation}
We investigate the finite sample performance of our proposed method by comparing it with some competing solutions. Specifically, we evaluate three competing methods. We first consider our proposed Gaussian graphical model regression method defined in \eqref{eqn:obj3}, referred to as \texttt{RegGMM} hereafter. 
%The estimator with the separate-max and separate-min post-processing steps is referred to as \texttt{RegGMM} and \texttt{RegGMM}$_{\min}$, respectively. 
We also consider a lasso estimator 
\begin{equation}\label{lasso}
\arg\min_{\bbeta_j\in\mathbb{R}^{(p-1)(q+1)}}\frac{1}{2n}\Vert\hat\z_j-\hat\W_{-j}\bbeta_j\Vert_2^2+\lambda\Vert\bbeta_j\Vert_1,
\end{equation}
and a group lasso estimator 
\begin{equation}\label{grlasso}
\arg\min_{\bbeta_j\in\mathbb{R}^{(p-1)(q+1)}}\frac{1}{2n}\Vert\hat\z_j-\hat\W_{-j}\bbeta_j\Vert_2^2+\lambda_g(\Vert(\bbeta_j)_{(0)}\Vert_1+\sqrt{p-1}\Vert\bbeta_{j,-0}\Vert_{1,2}),
\end{equation}
where the total number of groups is $(p-1)+q$. 

We simulate $n$ samples $\{(\u^{(i)},\x^{(i)}), i\in[n]\}$ from  \eqref{eqn:igmm}, where each sample has $\x^{(i)}\in\mathbb{R}^p$ (e.g., genes) and external covariate $\u^{(i)}\in\mathbb{R}^q$ (e.g., SNPs),  including  discrete and continuous covariates. 
Discrete covariates are generated from $\{0,1\}$ with equal probabilities, and continuous covariates are generated from $\text{Uniform}[0,1]$. For $\bGamma\in\mathbb{R}^{p\times q}$, we randomly set $s_{\bGamma}$ of its entries to 0.25, and the rest to zero.
%Let $s_{\bGamma}$ denote this proportion of nonzero entries in $\bGamma$. 
\begin{figure}[!t]
	\centering
	\includegraphics[trim=7.5mm 0 0 0, scale=0.385]{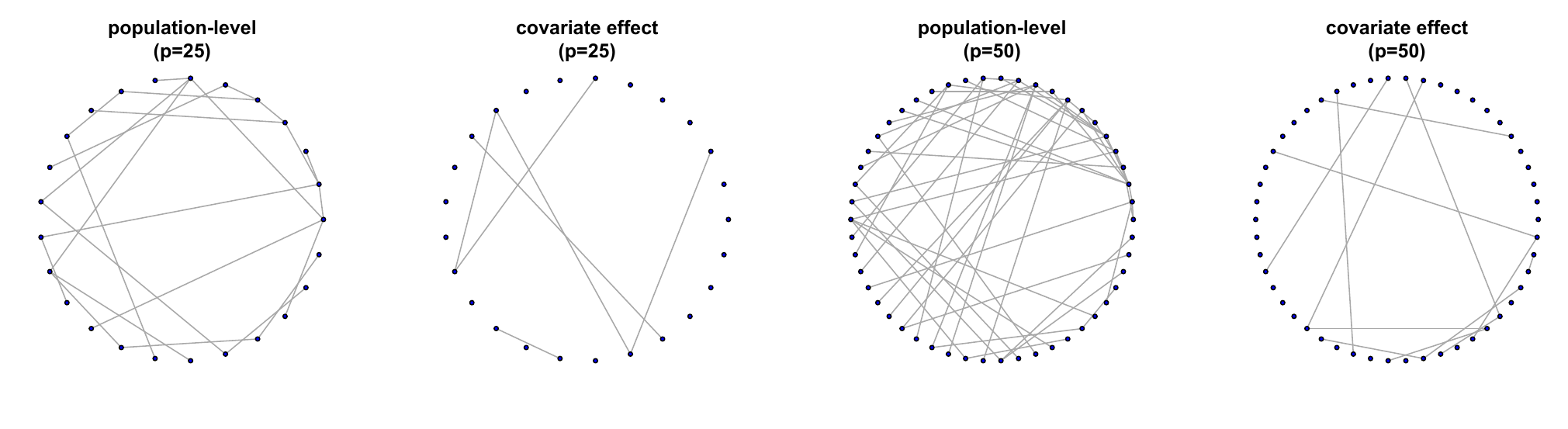}
	\caption{ {Graphs  corresponding to the population-level effects and covariate effects with $p=25$ or $50$. When illustrating covariate effects, we randomly pick only one of  the $q_e$ effective covariates.}}
	\label{fig:simnet}
\end{figure}

The population level network is assumed to follow a scale-free network model, with the degrees of nodes generated from a power-law distribution \citep{clauset2009power} with parameter 2.5. We randomly select $q_e$ out of $q$ covariates to have nonzero effects, and the graphs for these $q_e$ covariates follow an Erdos-Renyi model with edge probability $v_e$;
see the graph structures in Figure \ref{fig:simnet}. We set $\sigma^{jj}=1$ for $j\in[p]$.
The initial nonzero coefficients 
$\beta_{jkh}$
%for all $j,k$ and $h$, 
are generated from $\text{Uniform}([-0.5,-0.35]\cup[0.35,0.5])$. 
For each $j$, we  rescale $\{\beta_{jkh}\}_{k\neq j\in[p], h\in\{0\}\cup[q]}$ by dividing each entry by $\sum_{k\neq j\in[p], h\in\{0\}\cup[q]}|\beta_{jkh}|$. 
{After rescaling, for each $j,k$ and $h$, we use the average of $\beta_{jkh}$ and $\beta_{kjh}$ to fill the entries at $jkh$ and  $kjh$.} 
This process results in symmetry with diagonal dominance and, thus,  ensures the positive definiteness of the precision matrices. We set $s_{\bGamma}=125$, $q_e=5$, $v_e=0.01$, and consider $n=200,400$, $p=25,50$ and $q=50,100$, {with 1,224 to 4,949 parameters to estimate.}

For each simulation configuration, we generate {200} independent data sets, within each of which we randomly set half of the $q$ covariates to be discrete and the rest  continuous. 
Given $\u^{(i)}$, we are able to determine $\bOmega(\u^{(i)})$ and $\bSigma(\u^{(i)})$; the $i$th sample $\x^{(i)}$ is generated from $\mathcal{N}(\bGamma\u^{(i)},\bSigma(\u^{(i)}))$, $i\in[n]$.
When comparing  the estimates of $\bbeta_j$'s obtained by the competing methods, 
%In \eqref{grlasso}, each element in $(\bbeta_j)_{(0)}$ is treated as one group, as we do not wish to regularize population-level connections of node $j$ (i.e., elements in $(\bbeta_j)_{(0)}$) as one group. 
%For both the lasso and group lasso estimators, we report results from the separate-max post-processing procedure, as it results in smaller estimation errors. 
{we  report the results after post-processing as in \eqref{eq:sepmin}.}
For a fair comparison, tuning parameters in all of the methods are selected via 5-fold cross validation.

To evaluate the estimation accuracy, we report the estimation errors $\| \bGamma-\hat\bGamma \|_F$ (the Frobenius norm) and $\sum_{j=1}^p\|{\hat\bbeta_j-\bbeta_j} \|_2$, {where $\hat\bbeta_j$'s, with a slight overuse of notation, denote the estimates of $\bbeta_j$'s obtained by various methods}. Also reported is the average estimation error of the precision matrix defined as $\sum_{i=1}^n\Vert\hat{\bOmega}_i-\bOmega_i\Vert_{F,\text{off}}^2/n$, where $\bOmega_i=\B_0+\sum_{h=1}^q\B_h\u^{(i)}_{h}$ and $\hat\bOmega_i$ is estimated from a given method. For the selection accuracy, we report the true positive rate (TPR) and false positive rate (FPR). Results for estimating the mean coefficient $\bGamma$ are also good, and are given in Section S1.2 in the interest of space.
%and  the F$_1$ score, calculated as $2\text{TP}/(2\text{TP}+\text{FP}+\text{FN})$, where \text{TP} is the true positive count, \text{FP} is the false positive count, and \text{FN} is the false negative count. The highest possible value of F$_1$  is 1, indicating perfect selection.

\begin{table}[!t]
\setlength{\tabcolsep}{2.5pt}
\centering
{\renewcommand{\arraystretch}{0.9}
\begin{tabular}{|c|c|c|cccc|} \hline
$n$ &  $p$, $q$  &  Method   & TPR$_{\bbeta}$ & FPR$_{\bbeta}$ & Error of $\bbeta$ & Error of $\bOmega$ \\\hline
\multirow{12}{*}{200} & \multirow{3}{*}{\begin{tabular}[c]{@{}l@{}} $p=25$\\ $q=50$\end{tabular}} 
& \texttt{RegGMM}   &  0.817 (0.004)       & {\bf 0.003} (0.000)   & {\bf 1.378} (0.006)      & {\bf 2.011} (0.018)           \\
&& lasso            & {\bf 0.820} (0.005)  & {\bf 0.003} (0.000)   & 1.541 (0.007)            & 2.500 (0.022) \\
&& group lasso      &  0.756 (0.004)       & 0.030 (0.002)         & 2.101 (0.005)            & 5.130 (0.033) \\\cline{2-7}

& \multirow{3}{*}{\begin{tabular}[c]{@{}l@{}}$p=25$\\ $q=100$\end{tabular}} 
& \texttt{RegGMM}   & {\bf 0.777} (0.005)   & {\bf 0.002} (0.000)   & {\bf 1.417} (0.006)       &  {\bf 2.147} (0.016)           \\
&& lasso            & 0.753 (0.005)         & {\bf 0.002} (0.000)   & 1.622 (0.005)             & 2.791 (0.018) \\
&& group lasso      &  0.721 (0.004)        & 0.013 (0.000)         & 2.103 (0.006)             & 5.023 (0.039) \\\cline{2-7}                  
& \multirow{3}{*}{\begin{tabular}[c]{@{}l@{}} $p=50$\\ $q=50$\end{tabular}} 
& \texttt{RegGMM}   & {\bf 0.624} (0.004)   & 0.003 (0.000)         & {\bf 2.228} (0.005)      & {\bf 5.036} (0.024)           \\
&& lasso            & 0.546 (0.005)         & {\bf 0.002} (0.000)   & 2.396 (0.006)            & 5.827 (0.029) \\
&& group lasso      &  0.579 (0.002)        & 0.030 (0.000)         & 4.219 (0.008)            & 26.652 (0.163) \\\cline{2-7}

& \multirow{3}{*}{\begin{tabular}[c]{@{}l@{}}  $p=50$\\ $q=100$\end{tabular}} 
& \texttt{RegGMM}   & {\bf 0.597} (0.003)   & {\bf 0.001} (0.000)   & {\bf 2.292} (0.005)      & {\bf 5.332} (0.020)      \\
&& lasso            & 0.473 (0.004)         & {\bf 0.001} (0.000)   & 2.514 (0.005)            & 6.412 (0.023) \\
&& group lasso      & 0.550 (0.002)         & 0.013 (0.000)         & 4.220 (0.008)            & 26.331 (0.210) \\\hline

\multirow{12}{*}{400} & \multirow{3}{*}{\begin{tabular}[c]{@{}l@{}} $p=25$\\ $q=50$\end{tabular}} 
& \texttt{RegGMM}   & {\bf 0.983} (0.001)  & {\bf 0.003} (0.000)     & {\bf 0.907} (0.003)      & {\bf 0.893} (0.006)  \\
&& lasso            & {\bf 0.983} (0.001)  & {\bf 0.003} (0.000)     & 1.016 (0.003)            & 1.118 (0.006) \\
&& group lasso      & 0.928 (0.002)        & 0.033 (0.000)           & 1.555 (0.002)            & 2.556 (0.010) \\\cline{2-7}

& \multirow{3}{*}{\begin{tabular}[c]{@{}l@{}}  $p=25$\\ $q=100$\end{tabular}} 
& \texttt{RegGMM}   & 0.959 (0.002)        & {\bf 0.001} (0.000)   & {\bf 0.997} (0.003)     & {\bf 1.069} (0.007)\\
&& lasso            & {\bf 0.960} (0.002)  & 0.002 (0.000)         & 1.113 (0.003)           & 1.329 (0.008) \\
&& group lasso      &  0.900 (0.003)       & 0.016 (0.000)         & 1.616 (0.003)           & 2.754 (0.011) \\\cline{2-7}                  
& \multirow{3}{*}{\begin{tabular}[c]{@{}l@{}}$p=50$\\ $q=50$\end{tabular}} 
& \texttt{RegGMM}    & {\bf 0.900} (0.002) & {\bf 0.002} (0.000)    & {\bf 1.632} (0.003)      & {\bf 2.741} (0.011) \\
&& lasso             & 0.892 (0.002)       & {\bf 0.002} (0.000)    & 1.736 (0.003)            & 3.096 (0.012) \\
&& group lasso       & 0.769 (0.002)       & 0.042 (0.000)          & 3.107 (0.004)            & 10.755 (0.033) \\\cline{2-7}

& \multirow{3}{*}{\begin{tabular}[c]{@{}l@{}}  $p=50$\\ $q=100$\end{tabular}} 
& \texttt{RegGMM}   & {\bf 0.894} (0.002)  & {\bf 0.001} (0.000)    & {\bf 1.690} (0.003)      & {\bf 2.935} (0.011)  \\
&& lasso            & 0.876 (0.002)        & {\bf 0.001} (0.000)    & 1.826 (0.003)            & 3.419 (0.011) \\
&& group lasso      & 0.714 (0.002)        & 0.018 (0.000)          & 3.148 (0.004)            & 10.984 (0.033) \\\hline
\end{tabular}}
\caption{Estimation accuracy of $\bbeta_j$'s in simulations with varying sample size $n$, network size $p$ and covariate dimension $q$. The three methods  are \texttt{RegGMM}, the lasso estimator in \eqref{lasso} and the group lasso estimator in \eqref{grlasso}. Marked in boldface are those achieving the best evaluation criteria in each setting.}\label{tab2}
\end{table}

Table \ref{tab2} reports the average criteria for estimating $\bbeta_j$'s, with standard errors in the parentheses, over 200 data replications. It shows that the  proposed \texttt{RegGMM} outperforms the competing methods in both estimation accuracy and selection accuracy for different sample sizes $n$, network sizes $p$ and covariate dimensions $q$. 
This is consistent with our theoretical findings. Moreover, the estimation errors of \texttt{RegGMM} decrease as $n$ increases, or as $p$ and $q$ decrease,  confirming the theoretical results in Theorem \ref{thm3}.
%In contrast, under  the scenarios examined, the group lasso estimator cannot identify any effective covariate, possibly because the nonzero entries in the nonzero groups in $\bbeta_{j,-0}$ are too sparse. As such, the group lasso estimator yields low false positive rates as well as low true positive rates. When $p$ and $q$ are both large (i.e., $p,q\ge50$), \texttt{RegGMM} and the group lasso have comparable false positive rates, while \texttt{RegGMM} gives better true positive rates. 
{For additional analyses, we evaluate some higher dimensional cases by increasing $p$, $q$  to 300 or 400, and present the results in Section S1.3.} In Section S1.4, we compare several benchmark solutions, including the standard Gaussian graphical model estimated using the neighborhood selection method \citep{meinshausen2006high} and the graphical lasso estimation method \citep{friedman2008sparse}, the conditional mean Gaussian graphical model \citep{cai2012covariate} and the stratified Gaussian graphical model  \citep{danaher2014joint}.

\begin{figure}[!t]
	\centering
	\includegraphics[trim=1cm 1cm 1cm 0, scale=0.375]{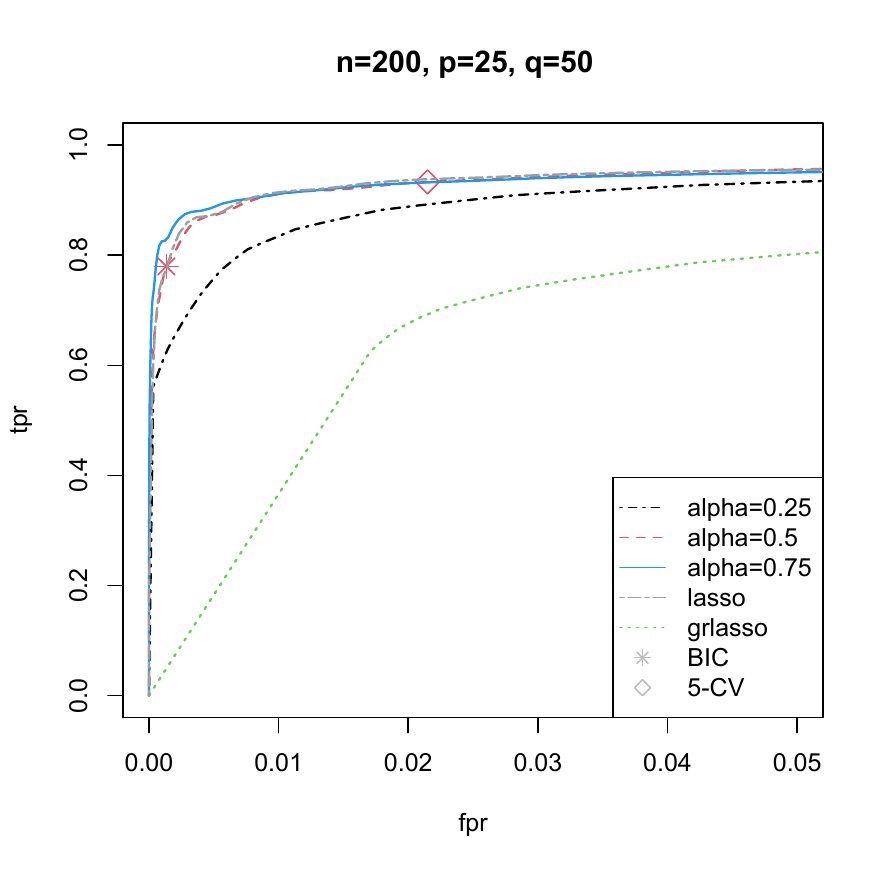}
	\includegraphics[trim=0cm 1cm 1cm 0, scale=0.375]{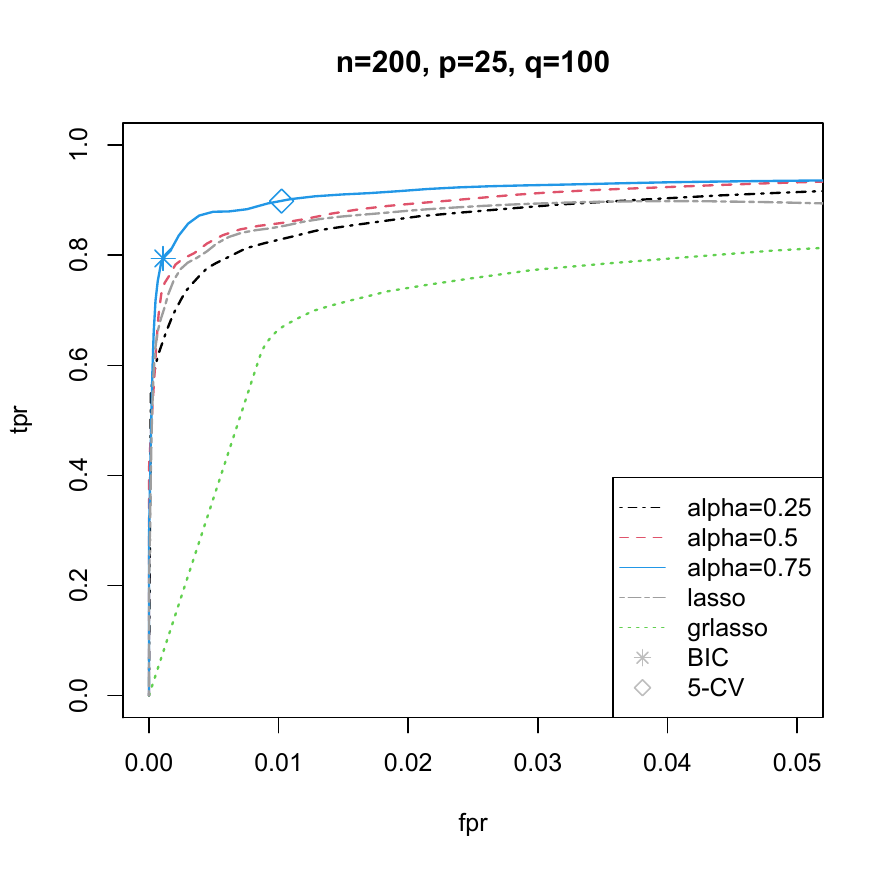}
	\includegraphics[trim=0cm 1cm 1cm 0, scale=0.375]{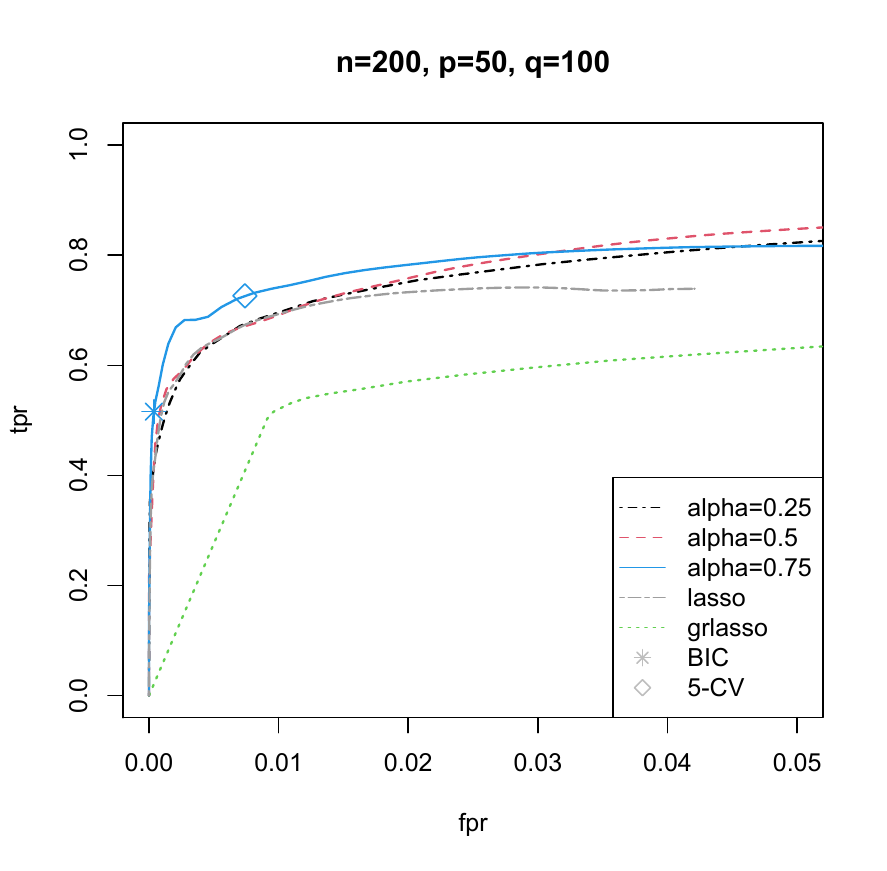}
	\caption{The ROC curves for \texttt{RegGMM} under $\alpha=0.25,0.5,0.75$, lasso and the group lasso (grlasso). For \texttt{RegGMM}, the values selected by BIC and five-fold cross validation are marked on the curves.}
	\label{fig:pre_roc}
\end{figure}

Next, we present several ROC curves,  plotting the true positive rate against the false positive rate across a fine grid of tuning parameters. In each curve, the true positive and false positive rates are averaged over $p$ regressions and over 200 data replicates. Specifically, to compare various methods in  the accuracy of selecting coefficients for the precision matrices, Figure \ref{fig:pre_roc} shows the ROC curves for \texttt{RegGMM} with $\alpha=0.25,0.5,0.75$, lasso and the group lasso (grlasso). We also compare  BIC and cross validation for selecting the optimal tuning parameter. The penalty term in BIC is $\log n\times\hat s_j$, with $\hat s_j$ being the number of nonzero elements in $\hat\bbeta_j$. It appears that cross validation strikes a reasonable balance between the true and false positive rates, especially when $p,q$ are large; \texttt{RegGMM} performs better than the lasso and group lasso estimator; and selection in the precision coefficient estimation is not overly sensitive to $\alpha$, which characterizes the weight of the lasso penalty relative to the group lasso penalty.

Finally, we investigate the computation cost that may occur during the tuning process. Table \ref{tab:time} shows the five-fold cross validation computation time for one node (or gene) and all $p$ nodes for a given $\alpha$. The simulations were run on an iMac with a 3.6 GHz Intel Core i9 processor. As the number of parameters is $\mathcal{O}(p^2q)$, the total computing cost is expected to be roughly quadratic in $p$ and linear in $q$, as seen in Table \ref{tab:time}. Our method enables a parallel implementation over the $p$ node-wise regressions and the working values of $\alpha$, in which case the computing time on each core is, for example, 16 seconds when $p=50$ and $q=100$.

\begin{table}[!t]
\centering
\renewcommand{\arraystretch}{0.9}\begin{tabular}{|c|c|c|c|}\hline    
$n$  &  $(p, q)$  
&{\renewcommand{\arraystretch}{0.75}\begin{tabular}[c]{@{}c@{}}computation time (s)\\ one node\end{tabular}}
&{\renewcommand{\arraystretch}{0.75}\begin{tabular}[c]{@{}c@{}}computation time (s)\\ all nodes\end{tabular}}\\\hline
\multirow{4}{*}{200} 
& (25,50)   & 2.819 (0.058)  & 70.475  (1.450)          \\\cline{2-4}
& (25,100)  & 5.372 (0.116)  & 134.300 (2.900)           \\\cline{2-4}
& (50,50)   & 8.343 (0.118)  & 140.950 (5.901)          \\\cline{2-4}
& (50,100)  & 15.550 (0.231) & 777.500 (11.552)          \\\hline
\end{tabular}
\caption{Five-fold cross validation computation time for one node (or gene) and all $p$ nodes for a given $\alpha$.}\label{tab:time}
\end{table}

\section{Co-expression QTL Analysis}
\label{sec:realdata}

Our application focuses on glioblastoma multiforme (GBM), the most aggressive and fatal subtype of brain cancer \citep{bleeker2012recent}, as featured in  the REMBRANDT trial (GSE108476) with a subcohort of $n=178$ GBM patients. 
Since existing therapies remain largely ineffective \citep{bleeker2012recent}, it is imperative to explore more effective  treatment, such as new gene therapies \citep{kwiatkowska2013strategies}. {Understanding the molecular underpinning of the disease is the key.
In the study, all} of  these 178  patients
had undergone  microarray and single nucleotide polymorphism (SNP) chip profiling, with both gene expression and SNP data available for analysis. 
Specifically, the extracted RNA from each tumor sample was processed using microarrays with 23,521 genes assayed on each array. 
Genomic DNA  from each sample was hybridized to SNP chips, which covers 116,204 SNP loci with a mean intermarker distance of 23.6kb. 
The raw data were pre-processed and normalized using standard pipelines; see \citet{gusev2018rembrandt} for more details. 

We study a set of $p=73$ genes  (response variables) that belong to the human glioma pathway in the Kyoto Encyclopedia of Genes and Genomes (KEGG) database \citep{kanehisa2000kegg}; 
see Figure S1.
The covariates  include local SNPs (i.e., SNPs that fall within 2kb upstream and 0.5kb downstream of the gene) residing near these 73 genes,  resulting in a total of 118 SNPs.
SNPs  are coded  with ``0" indicating homozygous in the major allele 
and ``1"  otherwise. 
For each patient, age and gender are included in analysis. Consequently, there are $q=120$ covariates, bringing a total of $73\times36\times121 = 317,988$ model parameters (including  intercepts).
Our main objective is to recover both the population-level and subject-level gene networks, and to {examine if and how age, gender and SNPs modulate the subject-level networks}. 

We have evaluated several benchmark methods in Section S1.4 of the Supplementary Materials; however, these methods are not designed to  and  cannot detect eQTL variants. Therefore, we have elected to apply the proposed two-step procedure in Section \ref{sec:extension} to this data set. 
It is common in penalized regressions to standardize predictors to ensure they be on the same scale \citep{tibshirani1997lasso}. For example, the covariates in the model are standardized to have mean zero and variance one. The scheme does not alter interpretations of the model; see discussions in Section S1.5.
Tuning parameters in both steps of the estimation procedure are selected via 5-fold cross validation, {and post-processing, as in \eqref{eq:sepmin}, generates the final estimates}. Out of the 120 covariates considered, 9 SNPs are estimated to have nonzero effects on the network. 

\begin{figure}[!t]
	\centering
	\includegraphics[trim=0 5mm 0 0, scale=0.3]{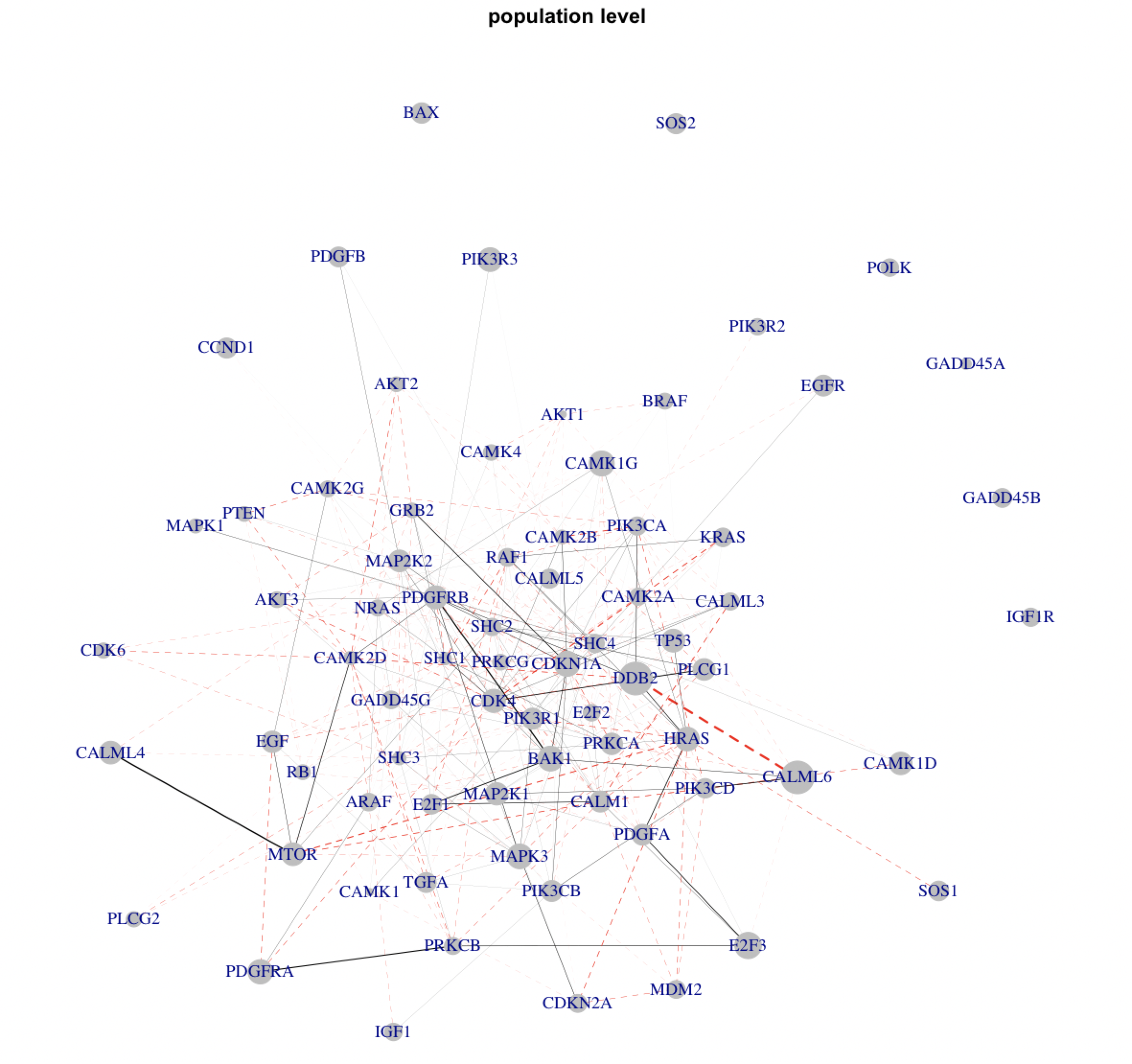}
	\caption{The graph corresponding to the population-level effect. The node sizes are proportional to mean expression levels and the edge weights are proportional to $\hat\B_0$. Edges with positive (negative) effects on partial correlations are shown in red dashed (black solid) lines.}
	\label{fig:pop}
\end{figure}

We first examine the population level network. 
Most of the well-connected genes in Figure \ref{fig:pop} are known to be associated with cancer. For example, PIK3CA is a protein coding gene and is one of the most highly mutated oncogenes identified in human cancers \citep{samuels2004oncogenic}; mutations in the PIK3CA gene are found in many types of cancer, including cancer of the brain, breast, ovary, lung, colon and stomach \citep{samuels2004oncogenic}. The PIK3CA gene is a part of the PI3K/AKT/MTOR signaling pathway, which is one of the core pathways in human GBM and other types of cancer \citep{cancer2008comprehensive}. TP53 is also a highly connected gene in the estimated network. This gene encodes a tumor suppressor protein containing transcriptional activation, and is the most frequently mutated gene in human cancer; the P53 signaling pathway is also one of the core pathways in human GBM and other types of cancer \citep{cancer2008comprehensive}. 
In Figure \ref{fig:pop}, we can identify several core pathways in human GBM including the PI3K/ AKT/MTOR, Ras-Raf-MEK-ERK, calcium and p53 signaling pathways; see Table \ref{tab:gene} for genes included in each pathway. These findings are in agreement to the existing literature on GBM genes and pathways \citep{cancer2008comprehensive,brennan2013somatic,maklad2019calcium}.

\begin{table}[!t]
\setlength{\tabcolsep}{3pt}
{\small
\begin{tabular}{lll}
\hline name                                        &         genes                                                                                      &   references                        \\\hline
\begin{tabular}[c]{@{}l@{}}PI3K/ AKT/MTOR \\ signaling pathway\end{tabular} & \begin{tabular}[c]{@{}l@{}}PIK3CA, PIK3CB, PIK3CD,\\ PIK3R3, PTEN,  AKT1, AKT2, AKT3,\\ MTOR, IGF1, PRKCA\end{tabular}                    & \citet{cancer2008comprehensive} \\\hline
\begin{tabular}[c]{@{}l@{}}Ras-Raf-MEK-ERK\\ signaling pathway\end{tabular} & \begin{tabular}[c]{@{}l@{}}EGF, EGFR, GRB2,\\ SOS1, SOS2, IGF1 \\ SHC1, SHC2, SHC3, SCH4 \\ MAPK1, MAPK3, MAP2K1, MAP2K2\\ HRAS, KRAS, NRAS,\\ RAF1, ARAF, BRAF, PRKCA\end{tabular}  & \citet{brennan2013somatic} \\\hline
\begin{tabular}[c]{@{}l@{}}calcium (Ca$^{+2}$) \\ signaling pathway\end{tabular}        & \begin{tabular}[c]{@{}l@{}}CALM1,CALML3, CALML4, CALML5,\\ CALML6, CAMK1,CAMK4, CAMK1D,\\ CAMK1G,CAMK2A, CAMK2B, \\ CAMK2D,CAMK2G, PRKCA\end{tabular} & \citet{maklad2019calcium}\\\hline
\begin{tabular}[c]{@{}l@{}}p53 \\ signaling pathway\end{tabular}  & \begin{tabular}[c]{@{}l@{}}TP53, MDM2, DDB2, PTEN, IGF1\\ CDK4, CDK6, CDKN1A, CDKN2A\end{tabular} & \citet{cancer2008comprehensive}\\\hline
\end{tabular}}
\caption{Pathways and genes involves in each pathway.}\label{tab:gene}
\end{table}

\begin{figure}[!t]
	\centering
	\includegraphics[trim=1cm 2cm 0 0, scale=0.3]{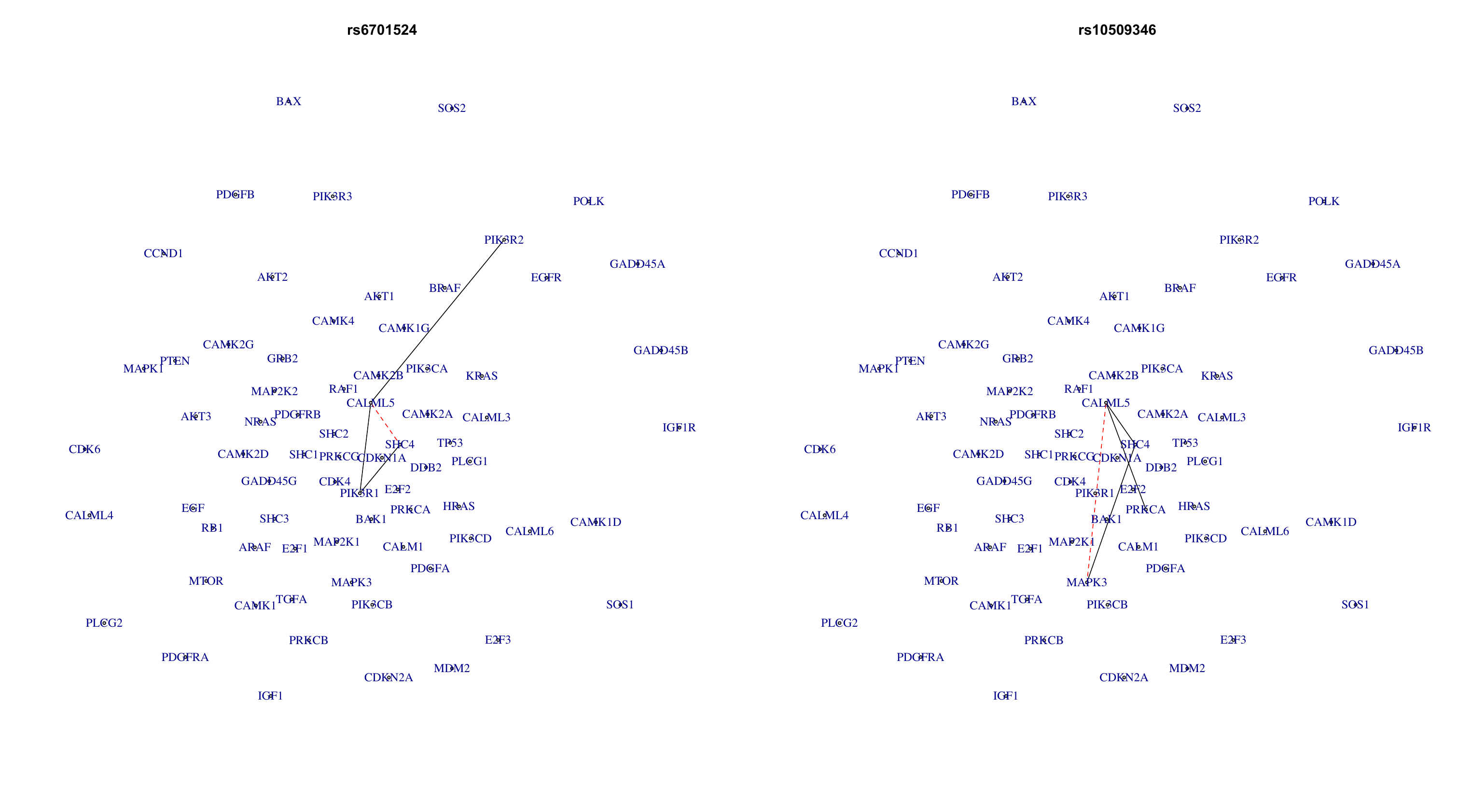}
	\caption{Graphs depending on each covariate (i.e., different SNPs). Edges that have positive (negative) effects on partial correlations are shown in red dashed (black solid) lines.}
	\label{fig:cov}
\end{figure}
 
We next examine the covariate effects on the network. Identified are nine co-expression QTLs, namely, \texttt{rs6701524}, \texttt{rs10509346}, \texttt{rs10492975}, \texttt{rs723211}, \texttt{rs1347069}, \texttt{rs473698}, \texttt{rs4118334}, \texttt{rs882664} and \texttt{rs1267622}.
The network effects of \texttt{rs6701524} are shown in Figure \ref{fig:cov} (left panel). This SNP, residing in MTOR, is found to affect CALML5's co-expression with PIK3R1, and also with PIK3R2 and SHC4. This is an interesting finding as PI3K/MTOR is a key pathway in GBM development and progression, and inhibition of PI3K/MTOR signaling was found effective in increasing survival with GBM tumor \citep{batsios2019pi3k}.
This co-expression QTL can potentially regulate the co-expressions of CALML5, PIK3R1, PIK3R2, MTOR, and play an important role in activating the PI3K/MTOR pathway. 

Shown in Figure \ref{fig:cov} (right panel) are the network effects of \texttt{rs10509346}, a variant of CAMK2G. The figure indicates that this SNP affects the co-expressions of CAMK2G with genes in the Ras-Raf-MEK-ERK pathway. This agrees to the finding that the Ras-Raf-MEK-ERK pathway is modulated by Ca$^{+2}$ and calmodulin \citep{agell2002modulation}.
%\texttt{rs1267622}, a variant of BRAF  commonly activated by somatic mutations in human cancer \citep{davies2002mutations}.  The figure indicates that this SNP notably affects the co-expression of BAX with other genes. This agrees to the finding that RAF suppresses activation and translocation of BAX \citep{koziel2013raf}. 
Moreover, based on our analysis,
\texttt{rs10492975} regulates the co-expressions of CALML5, PIK3R2 and CAMK1; 
\texttt{rs723211} is associated with the co-expressions of CALML5 and other genes; 
\texttt{rs1347069} influences the co-expressions of SHC4 and CDKN2A; 
\texttt{rs473698} may modify the co-expressions of PRKCG and CAMK1; 
\texttt{rs4118334} modulates the co-expressions of SHC2 and CAMK1; 
\texttt{rs882664} influences the co-expressions of PRKCA and CAMK1; 
\texttt{rs1267622}  may alter the co-expressions of SHC3 and RAF1. See details in Table S6.
As co-expression QTL identification has sparked recent interest, these  findings warrant more in-depth investigation.

\section{Discussion}\label{sec:discussion}

As the off-diagonal entries in the precision matrix $\bOmega(\u)$ are covariate dependent, a natural sufficient condition for positive definiteness, {derived from diagonal dominance, is
\\
$\max(1,\Vert\u\Vert_{\infty})\Vert\bbeta_j\Vert_1 < 1$.
With Assumption \ref{ass1} stipulating $|u^{(i)}_h|\le M$, positive definiteness holds when $\Vert\bbeta_j\Vert_1 < 1/\max(1, M)$, $j\in[p]$. 
Assuming $u_h\in[-1,1]$ (if not, rescale first), this sufficient condition can be simplified to $\Vert\bbeta_j\Vert_1 < 1$ for all $j$ (note that $\bbeta_j$ is sparse), suggesting that, to satisfy diagonal dominance, the ``effect sizes"  of $\u$  (i.e., $\|\bbeta_j\|_1$) on partial correlations cannot be too large. }
If the true covariance/precision matrix is positive definite, it then follows from Theorems \ref{thm1}-\ref{thm4} that the estimated precision is asymptotically positive definite. 
However, for finite sample cases, it may be desirable to ensure the positive definiteness of the final estimator. A post-hoc rescaling procedure seems to work well as in Section S1.6.

In our estimation procedure, we could estimate $\bbeta_1,\ldots,\bbeta_p$ jointly by combining $p$ loss functions, and minimizing  $\sum_{j=1}^p\ell_j(\bbeta_j|\mathcal{D})+\lambda\sum_j\Vert\bbeta_j\Vert_1+\lambda_{g}\sum_j\Vert\bbeta_{j,-0}\Vert_{1,2}$. It would have the benefit of preserving symmetry by restricting $[\B_{h}]_{jk}=[\B_{h}]_{kj}$ and {possibly} permitting additional dimension reduction via low-rankness \citep{zhang2018tensor}. However, this would be much more computationally intensive than \eqref{eqn:obj} by optimizing with respect to $\mathcal{O}(p^2q)$ parameters simultaneously.
Moreoever, we can modify our method to accommodate the hierarchy between main effects and interaction terms by re-organizing $\bbeta_j$ as 
\begin{equation}\label{inter}
\bbeta_j=(
\overbrace{\underbrace{\beta_{j10}}_{\text{main effect}},\underbrace{\beta_{j11},\ldots,\beta_{j1q}}_{\text{interactions}}}^{\text{group 1}},\ldots,
\overbrace{\underbrace{\beta_{jp0}}_{\text{main effect}},\underbrace{\beta_{jp1}\ldots,\beta_{jpq}}_{\text{interactions}}}^{\text{group $p$}}),
\end{equation} 
and imposing a modified sparse group lasso penalty $\lambda\Vert\bbeta^{-0}_j\Vert_0+\lambda_g\Vert\bbeta_j\Vert_{1,2}$, where $\bbeta^{-0}_j$ is $\bbeta_j$ after leaving out the main effects $\{\beta_{j10}\ldots,\beta_{jp0}\}$ and groups in $\Vert\bbeta_j\Vert_{1,2}$ are as defined in \eqref{inter}. 
The penalty is designed in such a way that the element-wise sparsity is not imposed on the main effects, and interactions, if selected, will enter the model with non-zero  main effects; a similar regularizer was adopted by \citep{she2018group} for penalized interaction models. With slight modifications, our established theoretical framework can still be used to study the theoretical properties of this modified regularizer.

In model \eqref{eqn:igmm}, we had assumed $\sigma^{jj}$ to be free of $\u$. Our empirical investigations show that our method is not sensitive to this assumption (see simulations in Section S1.1).
{It is possible to further extend our framework to allow the residual variances in \eqref {reg} (or correspondingly, the diagonal elements $\sigma^{jj}$'s) to depend on the covariate $\u$. 
To proceed, we consider $\sigma^{jj}(\u)=g(\bm\nu_j^\top\u)$, where $\bm\nu_j$ is the vector of unknown coefficients;  a viable choice is $g(x)=\exp(x)$.
In this case, the node-wise regression representation, {by scaling the response by $g(\bm\nu_j^\top\u)$,} may be reformulated as
\begin{equation}\label{eqn:du}  
Z_j\times g(\bm\nu_j^\top\u)=\sum_{k\neq j}^p\theta_{jk0}Z_k+\sum_{k\neq j}^p\sum_{h=1}^q\theta_{jkh}u_h  Z_k +\tilde{\epsilon}_j,\quad \text{Var}(\tilde{\epsilon}_j)=g(\bm\nu_j^\top\u)
\end{equation} 
where $\theta_{jkh}=-[\B_{h}]_{jk}$. 
To estimate $\bm\nu_j$ and $\btheta_j=(\theta_{j10},\ldots,\theta_{jp0},\ldots,\theta_{j1q},\ldots,\theta_{jpq})$, we may consider the following loss function
$$
\ell_j(\bm\nu_j,\btheta_j|\mathcal{D})=\frac{1}{2n}\sum_{i=1}^n\Vert\z^{(i)}_j\times g(\bm\nu_j^\top\u^{(i)})-\w_i\btheta_j\Vert_2^2,
$$
where $\mathcal{D}$ denotes the observed data,  $\z_j^{(i)}$ collects the samples of the $j$th variable and $\w_i$ is the $i$th row of the design matrix $\W_{-j}$; see their definitions in Section \ref{sec:estimation}. Through $\ell_j(\bm\nu_j,\btheta_j|\mathcal{D})$, $\bm\nu_j$ and $\btheta_j$ can be estimated with sparse penalties. 
This new iterative estimation procedure is more computationally demanding and requires {a} new theoretical analysis. We leave its full investigation as future research.
}

\section*{Acknowledgements}
We are grateful to the Editor, the AE and three anonymous referees for their
insightful comments that have substantially improved the quality and
the presentation of the manuscript. 
The work is partially supported by grants from NIH and NSF.

\bibliographystyle{asa}
\begingroup
\baselineskip=15pt
\bibliography{ref-sgmm}
\endgroup

\newpage
\renewcommand{\thesubsection}{S\arabic{subsection}}
\renewcommand{\theequation}{S\arabic{equation}}
\renewcommand{\thefigure}{S\arabic{figure}}
\renewcommand{\thetable}{S\arabic{table}}
\setcounter{equation}{0}
\setcounter{figure}{0}
\setcounter{table}{0}
\setcounter{page}{1}

\begin{center}
{\large\bf Supplementary Materials for ``High Dimensional Gaussian Graphical Regression Models with Covariates"} 
\medskip
%{\large\bf Jingfei Zhang and Yi Li}
\end{center}

\subsection{Additional Numerical Results}

%\subsubsection{ROC curves}
%\label{sec:roc}
%{Figure \ref{fig:mean_roc} shows the ROC curves for model selection in the mean estimation.
%\begin{figure}[!htb]
%	\centering
%	\includegraphics[trim=0 1cm 0 0, scale=0.45]{mean.pdf}
%	\caption{{ROC curves for varying sample sizes $n$ and signal strength $t$, where the nonzero entries in $\bGamma$ are set to $t$. The values selected by the five-fold cross validation are marked on the curves.}}
%	\label{fig:mean_roc}
%\end{figure}
%}
% maybe we do not care about selection in mean coefficient

\subsubsection{Sensitivity analysis}
\label{sec:sensitivity}
We conduct a sensitivity analysis to examine the performance of our method when $\sigma^{jj}$ is covariate dependent. Specifically, under the same setting as in Section \ref{sec:simulation}, we set $\sigma^{jj}(\u)=1+\sum_{h=1}^q\beta_{jjh}u_h$. If $u_h$ is an effective covariate with a nonzero effect on the graph, we set $s_{\sigma}$ proportion of $\{\beta_{jjh}\}_{j\in[p]}$ to $\sigma_0$, where $\sigma_0$ reflects the strength of covariate dependence for $\sigma^{jj}(\u)$, and set the rest of $\beta_{jjh}$'s to zero. We let $n=200$, $p=25$ and $s_{\sigma}=0.1$. The tuning parameters are selected using cross validation. Table \ref{tab4} reports the average evaluation criteria over 200 data replicates over $q=25,100$ and $\sigma_0=0,0.1,0.2$. Note that $\sigma_0=0$ corresponds to the case that $\sigma^{jj}$'s are not covariate dependent.  When $\sigma_0=0.2$, we have $\sigma^{jj}(\u)\in[1,2]$. 
Reported criteria are estimation errors of the mean $\bm\mu_{\text{error}}=\sum_{i=1}^n\Vert\hat{\bm\mu}_i-\bm\mu_i\Vert^2/n$ and the precision matrix $\bm\bOmega_{\text{error}}=\sum_{i=1}^n\Vert\hat{\bOmega}_i-\bOmega_i\Vert_{F,\text{off}}^2/n$, where $\Vert\cdot\Vert_{F,\text{off}}$ denotes the off-diagonal Frobenius norm and $(\bm\mu_i,\bOmega_i)$ and $(\hat{\bm\mu_i},\hat\bOmega_i)$ are the true and estimated values, respectively, the true positive rate (TPR) and false positive rate (FPR) in estimating the precision coefficient $(\bbeta_1,\ldots,\bbeta_p)$.
As seen from Table \ref{tab4},  under this misspecified setting with covariate dependent residual variances, our method still gives a reasonable performance.
\begin{table}[!t]
\setlength{\tabcolsep}{5pt}
\centering
{\renewcommand{\arraystretch}{0.8}
\begin{tabular}{|c|c|cccc|} \hline
   &     & $\bm\mu_{\text{error}}$ & $\bm\bOmega_{\text{error}}$ & $\text{TPR}_{\bbeta}$ & $\text{FPR}_{\bbeta}$ \\\hline
\multirow{3}{*}{\begin{tabular}[c]{@{}l@{}} $q=50$\end{tabular}}
& $\sigma_0=0$       
&\begin{tabular}[c]{@{}c@{}}3.041\\ \small{(0.018)}\end{tabular}  
&\begin{tabular}[c]{@{}c@{}}2.011\\ \small{(0.018)}\end{tabular}  
%&\begin{tabular}[c]{@{}c@{}}0.872\\ \small{(0.003)}\end{tabular} 
%&\begin{tabular}[c]{@{}c@{}}0.156\\ \small{(0.002)}\end{tabular} 
&\begin{tabular}[c]{@{}c@{}}0.817\\ \small{(0.004)}\end{tabular} 
&\begin{tabular}[c]{@{}c@{}}0.003\\ \small{(0.000)}\end{tabular} \\\cline{2-6}
& $\sigma_0=0.1$       
&\begin{tabular}[c]{@{}c@{}}2.871\\ \small{(0.022)}\end{tabular}  
&\begin{tabular}[c]{@{}c@{}}2.167\\ \small{(0.015)}\end{tabular}  
%&\begin{tabular}[c]{@{}c@{}}0.888\\ \small{(0.004)}\end{tabular} 
%&\begin{tabular}[c]{@{}c@{}}0.151\\ \small{(0.002)}\end{tabular} 
&\begin{tabular}[c]{@{}c@{}}0.700\\ \small{(0.004)}\end{tabular} 
&\begin{tabular}[c]{@{}c@{}}0.004\\ \small{(0.000)}\end{tabular} \\\cline{2-6}
& $\sigma_0=0.2$      
&\begin{tabular}[c]{@{}c@{}}2.842\\ \small{(0.018)}\end{tabular}  
&\begin{tabular}[c]{@{}c@{}}2.663\\ \small{(0.020)}\end{tabular}  
%&\begin{tabular}[c]{@{}c@{}}0.896\\ \small{(0.003)}\end{tabular} 
%&\begin{tabular}[c]{@{}c@{}}0.155\\ \small{(0.001)}\end{tabular} 
&\begin{tabular}[c]{@{}c@{}}0.645\\ \small{(0.004)}\end{tabular} 
&\begin{tabular}[c]{@{}c@{}}0.003\\ \small{(0.000)}\end{tabular}  \\\hline
 \multirow{3}{*}{\begin{tabular}[c]{@{}l@{}}$q=100$\end{tabular}} 
& $\sigma_0=0$       
&\begin{tabular}[c]{@{}c@{}}3.603\\ \small{(0.019)}\end{tabular}  
&\begin{tabular}[c]{@{}c@{}}2.147\\ \small{(0.016)}\end{tabular}  
%&\begin{tabular}[c]{@{}c@{}}0.804\\ \small{(0.003})\end{tabular} 
%&\begin{tabular}[c]{@{}c@{}}0.095\\ \small{(0.001)}\end{tabular} 
&\begin{tabular}[c]{@{}c@{}}0.777\\ \small{(0.005)}\end{tabular} 
&\begin{tabular}[c]{@{}c@{}}0.002\\ \small{(0.000)}\end{tabular} \\\cline{2-6}
& $\sigma_0=0.1$      
&\begin{tabular}[c]{@{}c@{}}3.465\\ \small{(0.022)}\end{tabular}  
&\begin{tabular}[c]{@{}c@{}}2.382\\ \small{(0.016)}\end{tabular}  
%&\begin{tabular}[c]{@{}c@{}}0.829\\ \small{(0.004)}\end{tabular} 
%&\begin{tabular}[c]{@{}c@{}}0.101\\ \small{(0.001)}\end{tabular} 
&\begin{tabular}[c]{@{}c@{}}0.668\\ \small{(0.004)}\end{tabular} 
&\begin{tabular}[c]{@{}c@{}}0.002\\ \small{(0.000)}\end{tabular} \\\cline{2-6}
& $\sigma_0=0.2$     
&\begin{tabular}[c]{@{}c@{}}3.370\\ \small{(0.023)}\end{tabular}  
&\begin{tabular}[c]{@{}c@{}}2.878\\ \small{(0.018)}\end{tabular}  
%&\begin{tabular}[c]{@{}c@{}}0.845\\ \small{(0.003)}\end{tabular} 
%&\begin{tabular}[c]{@{}c@{}}0.103\\ \small{(0.001)}\end{tabular} 
&\begin{tabular}[c]{@{}c@{}}0.627\\ \small{(0.003)}\end{tabular} 
&\begin{tabular}[c]{@{}c@{}}0.002\\ \small{(0.000)}\end{tabular}  \\\hline
\end{tabular}}
\caption{Average evaluation criteria with varying covariate dimension $q$ and covariate dependence parameter $\sigma_0$ with standard errors in parenthesis.}
\label{tab4}
\end{table}

\subsubsection{Estimation accuracy of $\bGamma$}
\label{sec:mean}
Table \ref{tab1} shows that the first step of our estimation procedure achieves good performance, and the estimation error of $\bGamma$ decreases as $n$ increases, or as $p$ and $q$ decrease. Such observations conform to Theorem \ref{thm3}.
\begin{table}[!t]
\centering
\begin{tabular}{|c|c|ccc|}\hline    
$n$  &  $(p, q)$  & TPR$_{\bGamma}$  & FPR$_{\bGamma}$  & Error of $\bGamma$\\\hline
\multirow{4}{*}{200} 
& (25,50)   & 0.872 (0.003)      & 0.156 (0.002)      &  1.811 (0.006)         \\
& (25,100)  & 0.804 (0.003)      & 0.095 (0.001)      &  1.980 (0.006)        \\\cline{2-5}
& (50,50)   & 0.867 (0.003)      & 0.110 (0.001)      &  1.938 (0.006)        \\
& (50,100)  & 0.789 (0.003)      & 0.072 (0.000)      &  2.162 (0.005)         \\\hline
\multirow{4}{*}{400}
& (25,50)   & 0.996 (0.000)      & 0.172 (0.001)     &  1.273 (0.004)         \\
& (25,100)  & 0.986 (0.001)      & 0.113 (0.001)     &  1.386 (0.005)        \\\cline{2-5}
& (50,50)   & 0.992 (0.001)      & 0.122 (0.000)     &  1.351 (0.004)        \\
& (50,100)  & 0.987 (0.001)      & 0.082 (0.001)     &  1.514 (0.004)         \\\hline
\end{tabular}
\caption{Estimation accuracy of $\bGamma$ in simulations with various sample sizes $n$, network sizes $p$ and covariate dimensions $q$.}\label{tab1}
\end{table}

\subsubsection{Higher dimensional cases}
\label{sec:highd}
{We have furthered increased $(p,q)$ to (25,400) and (300, 300). The simulation results over 200 data replicates are reported in Table \ref{tab-400}. 
For both settings in Table \ref{tab-400}, the mean coefficient $\bGamma$ are set following Section \ref{sec:simulation}. The non-sparse entries in $\{\B_h\}_{h=0}^q$ for $(p,q)=(25,400)$ and $(p,q)=(300, 300)$ are set to be the same as $p=25$ and $p=50$ in Section \ref{sec:simulation}, respectively. These settings involve an extremely large number of parameters; for example,
with $(p,q)=(300,300)$, the total number of precision parameters is almost 27 million (or 26,999,700 precisely).
It appears that our proposed method still achieves a reasonable performance in these high dimensional settings.
\begin{table}[!htb]
\centering
{\begin{tabular}{|c|c|cccc|} \hline
$n$ &  ($p$, $q$)   & Error of $\bGamma$  & Error of $\bbeta$ & TPR$_{\bbeta}$ & FPR$_{\bbeta}$\\\hline
\multirow{2}{*}{200} & (25, 400)
& 2.195 (0.005)  & 1.534 (0.005)  &  0.721 (0.004)       & 0.0003 (0.0000)            \\
%& (200, 200)
%&  2.376 (0.007) & 2.422 (0.005)   & 0.550 (0.004)   & 0.0001 (0.0000)              \\
& (300, 300)
&  2.525 (0.007) & 2.461 (0.005)   & 0.527 (0.004)   & 0.0000 (0.0000)               \\\hline
\end{tabular}}
\caption{{Estimation and selection accuracy of \texttt{RegGMM} with larger network sizes $p$ and covariate dimensions $q$.}}\label{tab-400}
\end{table}
}

\subsubsection{Comparison with benchmark methods}
\label{sec:benchmark}
We compare our proposed method with several benchmark solutions, including the standard i.i.d Gaussian graphical model estimated by the neighborhood selection method  \citep{meinshausen2006high}, referred to as \texttt{MB}, and by the graphical lasso estimation method \citep{friedman2008sparse}, referred to as \texttt{glasso},  the conditional mean Gaussian graphical model \citep{cai2012covariate}, referred to as \texttt{cmGMM},  and the stratified Gaussian graphical model \citep{danaher2014joint}, referred to as \texttt{stGMM}. 
Our evaluation criteria include the average estimation error of the mean defined as $\bm\mu_{\text{error}}=\sum_{i=1}^n\Vert\hat{\bm\mu}_i-\bm\mu_i\Vert^2/n$, where $\bm\mu_i=\bGamma\u^{(i)}$ and $\hat{\bm\mu_i}$ is estimated from a given method, average estimation error of the precision matrix defined as $\bm\bOmega_{\text{error}}=\sum_{i=1}^n\Vert\hat{\bOmega}_i-\bOmega_i\Vert_{F,\text{off}}^2/n$, where $\bOmega_i=\B_0+\sum_{h=1}^q\B_h\u^{(i)}_{h}$ and $\hat\bOmega_i$ is estimated from a given method, and the average selection error of the precision matrix $\text{TPR}_{\bOmega}=\sum_{i=1}^n\text{TPR}(\hat\bOmega_i)/n$ and $\text{FPR}_{\bOmega}=\sum_{i=1}^n\text{FPR}(\hat\bOmega_i)/n$, where $\text{TPR}(\hat\bOmega_i)$ and $\text{TPR}(\hat\bOmega_i)$ are true and false positive errors calculated by comparing the entries in $\hat\bOmega_i$ and $\bOmega_i$, respectively. 

\begin{table}[!t]
\setlength{\tabcolsep}{4pt}
\centering
{\renewcommand{\arraystretch}{0.9}
{\begin{tabular}{|c|c|c|cccc|} \hline
$n$ &  $p$, $q$  &  Method   & $\bm\mu_{\text{error}}$ & $\bm\bOmega_{\text{error}}$ & $\text{TPR}_{\bOmega}$ & $\text{FPR}_{\bOmega}$ \\\hline
\multirow{20}{*}{200} & \multirow{3}{*}{\begin{tabular}[c]{@{}l@{}} $p=25$\\ $q=50$\end{tabular}} 
& \texttt{RegGMM}   &{\bf 3.042} (0.019) & {\bf 2.011} (0.018) & 0.853 (0.004) & 0.122 (0.002)            \\
&& \texttt{MB}     &7.067 (0.010) & 3.371 (0.011) & 0.666 (0.003) & 0.129 (0.002)           \\
&& \texttt{glasso} &7.067 (0.010)& 10.824 (0.027) & 0.706 (0.004) & 0.158 (0.002)           \\
&& \texttt{cmGMM} & 3.919 (0.040) & 29.248 (0.443) & {\bf 0.982} (0.003) & 0.477 (0.002)           \\
&& \texttt{stGMM} & 7.219 (0.015) & 8.807 (0.054) & 0.492 (0.008) & {\bf 0.003} (0.001)           \\\cline{2-7}
& \multirow{3}{*}{\begin{tabular}[c]{@{}l@{}}$p=25$\\ $q=100$\end{tabular}} 
& \texttt{RegGMM} & {\bf 3.603} (0.019) & {\bf 2.147} (0.016) & 0.809 (0.005) & 0.121 (0.002)           \\
&& \texttt{MB}     & 6.916 (0.010) & 3.168 (0.012) & 0.642 (0.004) & 0.107 (0.002)          \\
&& \texttt{glasso} & 6.916 (0.010) & 11.394 (0.040) & 0.698 (0.004) & 0.149 (0.002)            \\
&& \texttt{cmGMM}  & 5.724 (0.023)& 15.236 (0.225) & {\bf 0.837} (0.005) & 0.340 (0.004)           \\
&& \texttt{stGMM} & 7.226 (0.011) & 9.131 (0.047) & 0.502 (0.007) & {\bf 0.028} (0.001)   \\\cline{2-7}        
& \multirow{3}{*}{\begin{tabular}[c]{@{}l@{}} $p=50$\\ $q=50$\end{tabular}} 
& \texttt{RegGMM}   & {\bf 3.753} (0.020) & {\bf 5.036} (0.024) & 0.657 (0.004) & 0.104 (0.002)              \\
&& \texttt{MB}     & 7.735 (0.010) & 6.650 (0.014) & 0.448 (0.003) & 0.060 (0.001)    \\
&& \texttt{glasso} & 7.735 (0.010) & 15.661 (0.031) & 0.511 (0.002) & 0.090 (0.001)            \\
&& \texttt{cmGMM}  & 5.491 (0.027) & 34.644 (0.232) & {\bf 0.890} (0.003) & 0.430 (0.002)            \\
&& \texttt{stGMM} & 8.414 (0.013) & 11.345 (0.071) & 0.182 (0.008) & {\bf 0.003} (0.000)            \\\cline{2-7}
& \multirow{3}{*}{\begin{tabular}[c]{@{}l@{}}  $p=50$\\ $q=100$\end{tabular}} 
& \texttt{RegGMM}   & {\bf 4.440} (0.019) & {\bf 5.332} (0.020) & 0.629 (0.003) & 0.094 (0.001)              \\
&& \texttt{MB}     & 7.791 (0.010) & 6.656 (0.015) & 0.447 (0.002) & 0.061 (0.001)            \\
&& \texttt{glasso} & 7.791 (0.010) & 15.697 (0.039) & 0.524 (0.003) & 0.095 (0.001)            \\
&& \texttt{cmGMM}  & 6.199 (0.028) & 29.146 (0.080) & {\bf 0.829} (0.003) & 0.396 (0.000)           \\
&& \texttt{stGMM} & 8.667 (0.013) & 11.322 (0.077) & 0.182 (0.008) & {\bf 0.004} (0.000)            \\\hline
\end{tabular}}}
\caption{Evaluation criteria with varying network size $p$ and covariate dimension $q$. Numbers achieving the best evaluation criteria in each setting are marked in boldface.}
\label{tab3}
\end{table}

With both \texttt{MB} and \texttt{glasso} stemming from an i.i.d. model,  we have  $\hat{\bm\mu}_i=\sum_{j=1}^n\x^{(j)}/n$ and $\hat\bOmega_i=\hat\bOmega$, where $\hat\bOmega$ is the output from \texttt{MB} or \texttt{glasso}. 
As \texttt{cmGMM} assumes heterogeneous means but a common covariance, we let 
$\hat{\bm\mu}_i=\hat\bGamma\u^{(i)}$ and $\hat\bOmega_i=\hat\bOmega$, where $\hat\bGamma$ and $\hat\bOmega$ are the output from \texttt{cmGMM}. For \texttt{stGMM} and given $q$ binary covariates, there are $2^q$ groups (or conditions) {to be considered, far exceeding the number of subjects $n$}. 
%comment: The concern is more about the number of clusters being larger than the number of samples to be clustered.
To implement \texttt{stGMM}, we cluster the subjects into $K<n$ groups by applying $K$-means clustering to the covariates and denote the cluster label of subject $i$ as $c_i\in\{1,\ldots,K\}$. We then implement \texttt{stGMM} and let $\hat{\bm\mu}_i=\sum_{j=1}^n\x^{(j)}\1_{\{c_j=c_i\}}/(\sum_{j=1}^n\1_{\{c_j=c_i\}})$ and $\hat\bOmega_i=\hat\bOmega_{c_i}$, where $\bOmega_1,\ldots,\bOmega_K$ are the output from \texttt{stGMM}. All benchmark methods are tuned using the tuning methods suggested in the respective papers. 

Under the same setting as in Section \ref{sec:simulation} and let $p=25,50$, $q=50,100$, $n=200$ and $K=5$,  Table \ref{tab3} reports the average evaluation criteria over 200 data replicates, revealing that \texttt{RegGMM}  improves upon the benchmark methods notably in  estimation accuracy and selection accuracy. 
Specifically, because \texttt{MB}, \texttt{glasso} and \texttt{stGMM} are not designed to estimate the covariate dependent means or precision matrices well, they tend to incur large estimation errors and low selection accuracy.
While \texttt{cmGMM} estimates well the covariate dependent means, it is not designed to accommodate covariate dependent precision matrices and, therefore, tends to overselect nonzero coefficients in $\hat\bOmega$ in the settings examined.

\subsubsection{Role of standardization}
\label{sec:stand}
We comment on the model interpretability after {standardizing  covariates} \citep{tibshirani1997lasso} to have mean zero and variance one.
As such, each graphical regression \eqref{eqn:greg} is in effect
\begin{equation*}
\hat\z_j=\sum_{k\neq j}^p\beta_{jk0}\frac{\hat\z_k}{\text{sd}(z_k)}+\sum_{k\neq j}^p\sum_{h=1}^q\beta_{jkh}\frac{(\u_h-\bar \u_h)}{\text{sd}(\u_h)}\odot\frac{\hat\z_k}{\text{sd}(\hat\z_k)} +\epsilon_j,
\end{equation*} 
where $\bar\u_h$ denotes the mean of $\u_h$, $\text{sd}(\cdot)$ denotes the standard deviation, and $\hat\z_j$, residual from the first estimation step, has mean zero. The above equation can be re-written as
\begin{equation}\label{eqn:final}
\hat\z_j=\sum_{k\neq j}^p\tilde\beta_{jk0}\hat\z_k+\sum_{k\neq j}^p\sum_{h=1}^q\tilde\beta_{jkh}\u_h\odot\hat\z_k +\epsilon_j,
\end{equation}
where $\tilde\beta_{jk0}=\frac{1}{\text{sd}(\hat\z_k)}\left\{\beta_{jk0}-\sum_{h=1}^q\frac{\bar\u_h}{\text{sd}(\u_h)}\beta_{jkh}\right\}$ and $\tilde\beta_{jkh}=\frac{\beta_{jkh}}{\text{sd}(\u_h)\text{sd}(\z_k)}$. Notice $\tilde\beta_{jkh}$ and $\beta_{jkh}$ only differs in scale by a positive scalar.
Therefore, parameter estimates can be interpreted with non-standardized covariates after a re-calculation as in \eqref{eqn:final}.

\subsubsection{Positive definiteness}
\label{sec:pd}
As our goal is to identify edges that are modulated by external covariates, we choose to employ the computationally efficient node-wise regressions and bypass the need to work with the entire covariance/precision matrix. If the true covariance/precision matrix is positive definite, it then follows from Theorems \ref{thm1}-\ref{thm4} that the estimated precision is asymptotically positive definite. 
However, for finite sample cases, it may be desirable to ensure the positive definiteness of the final estimator.
We find that, assuming the true precision matrix is diagonal dominant and the covariates are in known and bounded intervals, we can adopt a post-hoc re-scaling step that gives the same estimator asymptotically as guaranteed by Theorems \ref{thm1}-\ref{thm4}. Specifically, without loss of generality, assume $u_h\in[0,1]$ (if not, rescale $u_h$ first).
For any $j$ such that $\Vert\hat\bbeta_j\Vert_1>1$, we set the final estimate to $[\hat\B_h]_{j\cdot}/\hat\sigma{}^{jj}$ and $[\hat\B_h]_{\cdot j}/\hat\sigma{}^{jj}$.

We have evaluated the performance of the above procedure under the same setting as in Section \ref{sec:simulation}. Recall in our simulations, the true precision matrix is diagonal dominant and $u_h\in[0,1]$. The results over 200 data replicates are reported in Table \ref{tab-psd}, where the rescaled $\hat\bbeta_j$ is denoted as $\hat\bbeta_j^r$. Table \ref{tab-psd} suggests that the original and rescaled estimators have similar estimation errors. 
\begin{table}[!t]
\centering
\begin{tabular}{|c|c|c|c|c|} \hline
 & \begin{tabular}[c]{@{}l@{}} $p=25$\\ $q=50$\end{tabular} 
 & \begin{tabular}[c]{@{}l@{}} $p=25$\\ $q=100$\end{tabular} 
 & \begin{tabular}[c]{@{}l@{}} $p=50$\\ $q=50$\end{tabular} 
 & \begin{tabular}[c]{@{}l@{}} $p=50$\\ $q=100$\end{tabular} \\\hline
$\bm\bbeta_{\text{error}}$   &  1.378 (0.006)       & 1.417 (0.006)   & 2.228 (0.005)      & 2.291 (0.005)           \\
$\bm\bbeta^r_{\text{error}}$  & 1.372 (0.005)  & 1.446 (0.006)   & 2.226 (0.005)   & 2.292 (0.004) \\\hline
\end{tabular}
\caption{Estimation accuracy of $\bbeta_j$'s and $\bbeta^r_j$'s in simulations with $n=200$, varying network size $p$ and covariate dimension $q$.}\label{tab-psd}
\end{table}

\subsection{Technical Lemmas}
We state the technical lemmas that will be used in our proofs. 
\begin{lemma}[Lemma 1 in \cite{bellec2018prediction}]\label{lemma1}
Let $\text{pen}: \mathbb{R}^d\rightarrow\mathbb{R}$ be any convex function and $\hat\bbeta$ be defined by 
$$
\hat\bbeta\in\arg\min_{\bbeta\in\mathbb{R}^d}\left\{\Vert\y-\W\bbeta\Vert_2^2+\text{pen}(\bbeta)\right\},
$$
where $\W\in\mathbb{R}^{n\times d}$, $\y\in\mathbb{R}^n$. Then for $\bbeta\in\mathbb{R}^d$, 
$$
\Vert\y-\W\hat\bbeta\Vert_2^2+\text{pen}(\hat\bbeta)+\Vert\W(\hat\bbeta-\bbeta)\Vert_2^2\le\Vert\y-\W\bbeta\Vert_2^2+\text{pen}(\bbeta).
$$
\end{lemma}

\begin{lemma}[Theorem F in \cite{graybill1957idempotent}]\label{lemma2}
Let $\bepsilon_j\sim\mathcal{N}_p(\0,\sigma^2\bm I)$ and $\A$ be an $p\times p$ idempotent matrix with rank equals to $r\le p$. Then, $\bepsilon_j^\top\A\bepsilon_j/\sigma^2$ follows a $\chi^2$ distribution with $r$ degrees of freedom.
\end{lemma}

\begin{lemma}[Lemma 1 in \cite{laurent2000adaptive}]\label{lemma3}
Suppose that $U$ follows a $\chi^2$ distribution with $r$ degrees of freedom. For any $x>0$, it holds that 
$$
P(U-r\ge 2\sqrt{rx}+2x)\le\exp(-x).
$$
\end{lemma}

\begin{lemma}[Proposition 5.16 in \cite{vershynin2010introduction}]\label{lemma3.2}
Let $X_1,\ldots,X_n$ be independent mean zero sub-exponential random variables. Let $v_1=\max_{i}\|X_i\|_{\psi_1}$, where $\|X_i\|_{\psi_1}=\sup_{d\ge1}d^{-1}(E|X_i|^d)^{1/d}$ denotes the sub-exponential norm. There exists a constant $c$ such that, for any $t>0$, 
$$
P\left(\left|\sum_{i=1}^nX_i\right|\ge t\right)\le 2\exp\left\{ -c\min\left(\frac{t^2}{v_1^2n},\frac{t}{z_1}\right) \right\}. 
$$
\end{lemma}

%\begin{lemma}[Lemma 3 in \cite{bala2018}]\label{lemma3.3}
%Let $X_1,\ldots,X_n$ be independent centered sub-Gaussian random variables and $Y_1,\ldots,Y_n$ be independent centered sub-exponential random variables. Let $v_1=\max_{i}\|X_i\|_{\psi_1}$, where $\|X_i\|_{\psi_1}=\sup_{d\ge1}d^{-1/2}(E|X_i|^d)^{1/d}$ denotes the sub-Gaussian norm and $v_2=\max_{i}\|Y_i\|_{\psi_2}$, where $\|Y_i\|_{\psi_2}=\sup_{d\ge1}d^{-1}(E|Y_i|^d)^{1/d}$ denotes the sub-exponential norm. There exists a constant $c$ such that, for any $t>0$, 
%$$
%P\left(\left|\sum_{i=1}^n\{X_iY_i-\mathbb{E}(X_iY_i)\}\right|\ge t\right)\le 2\exp\left\{ %-c\min\left(\frac{t^2}{v_1^2v_2^2n},\frac{t^{2/3}}{v_1^{2/3}v_2^{2/3}}\right) \right\}. 
%$$

\begin{lemma}\label{lemma5}
Consider independent vectors $(y_1,\x_1),\ldots,(y_n,\x_n)$ in $\mathbb{R}\times\mathbb{R}^p$ such that $y_i=\x_i^\top\bbeta+\epsilon_i$ and $\x_i$ is elementwise sub-Gaussian, $i\in[n]$. Let $\epsilon_i$'s be {independent} Gaussian errors with non-constant variances and assume that $\sup_{i\in[n]}\text{Var}(\epsilon_i)$ is bounded by a constant $K_1>0$. Suppose $\Vert\bbeta\Vert_0=k$ and $\lambda_{\min}(\text{Cov}(\x_1))\ge1/\phi_0$ for some $\phi_0>0$. Let 
$$
\hat\bbeta_{\lambda}=\arg\min_{\btheta\in\mathbb{R}^p}\frac{1}{2n}\sum_{i=1}^n(y_i-\x_i^\top\btheta)^2+\lambda\Vert\btheta\Vert_1.
$$
When $\lambda=14\sigma_n\sqrt{\tau_1\log\,p/n}+\tau_1K(\log\,p\,\log\,n)^{1/2}/n$ for any $\tau_1>0$, the lasso estimate satisfies with probability at least $1-3p^{-\tau_1}$ that
$$
\Vert\hat\bbeta_{\lambda}-\bbeta\Vert_2\precsim \sigma_n\sqrt{\frac{k\,\log\, p}{n}}+\frac{k^{1/2}\log\,p}{n},
$$
where $\sigma_n=\frac{1}{n}\sum_{i=1}^n\text{Var}(\epsilon_i)$.
\end{lemma}
\noindent
The above result is adapted from Theorem 4.5 in \cite{kuchibhotla2018moving} by considering Gaussian errors.

\begin{lemma}[Theorem 4.1 in \cite{kuchibhotla2018moving}]\label{lemma6.1}
Let $\X_1,\ldots,\X_n$ be independent random vectors in $\mathbb{R}^p$. Assume each element of $\X_i$ is sub-exponential with $\Vert X_{i,j}\Vert_{\psi_1}<K_2$, $i\in[n],j\in[p]$.
where {$\|X_{i,j}\|_{\psi_1}=\sup_{d\ge1}d^{-1}(E|X_{i,j}|^d)^{1/d}$} denotes the sub-exponential norm.
Let $\hat\bSigma_{\X}=\X^\top\X/n$ and $\bSigma_{\X}=\mathbb{E}(\X^\top\X/n)$. Define 
$$
\Upsilon_{n,k}=\max_{j,k}\frac{1}{n}\sum_{i=1}^n\text{Var}\{X_{i,j}X_{i,k}\}. 
$$
Then for any $t>0$, with probability at least $1-\mathcal{O}(p^{-1})$, 
$$
\sup_{\Vert\v\Vert_0\le k,\,\Vert\v\Vert_2\le 1}\left\vert\v^\top(\hat\bSigma_{\X}-\bSigma_{\X})\v\right\vert\precsim k\sqrt{\frac{\Upsilon_{n,k}\log\,p}{n}}+K_2^2\frac{k(\log\,n\log\,p)^2}{n}.
$$
\end{lemma}

\begin{lemma}[Lemma 12 in \cite{loh2011high}]\label{lemma7}
For any symmetric matrix $\bSigma\in\mathbb{R}^{p\times p}$ and if $\vert\v^\top\bSigma\v\vert\le\delta_1$ for any $\v\in\{\v:\Vert\v\Vert_0\le 2s\,\text{and}\,\Vert\v\Vert_2=1\}$, then 
$$
\vert\v^\top\bSigma\v\vert\le27\delta_1(\Vert\v\Vert_2^2+\frac{1}{s}\Vert\v\Vert_1^2),\,\text{for any}\quad\v\in\mathbb{R}^p.
$$ 
\end{lemma}

\subsection{Proofs of Main Results}
{We begin by recalling the notation used throughout the paper. We denote the true parameters by $\bbeta_j$, $j\in[p]$, though, in some places and without ambiguities, we  use them to denote the corresponding parameters or the arguments in functions. 
The index set $\{1,\ldots,(p-1)(q+1)\}$ is partitioned into $q+1$ groups, indexed by $(0),(1),\ldots,(q)\subset\{1,\ldots,(p-1)(q+1)\}$. For a group index subset $\mathcal{G}\in\{1,\ldots,q\}$, we let $(\mathcal{G})=\cup_{h\in\mathcal{G}}(h)$, $(\mathcal{G}^c)=\cup_{h\notin\mathcal{G}}(h)$, and $(\bbeta_j)_{(\mathcal{G})}$ represent a sub-vector of $\bbeta_j$ 
%in the union of groups $h\in\mathcal{G}$.
indexed by $(\mathcal{G})$.}
We define $\Vert\bbeta_{j,-0}\Vert_{1,2}=\sum_{h=1}^q\Vert(\bbeta_j)_{(h)}\Vert_2$. 
We use $\mS_j$ to denote the element-wise support set of $\bbeta_j$, i.e, $\mS_j=\{l:(\bbeta_j)_l\neq 0, l\in[(p-1)(q+1)]\}$, and $\mG_{j}$ to denote the group-wise support set of $\bbeta_j$, i.e, $\mG_j=\{h:(\bbeta_j)_{(h)}\neq \0, h\in[q]\}$. Moreover, we let $s_j=|\mS_j|$ and $s_{j,g}=|\mG_{j}|$.

\subsubsection{Proof of Theorem \ref{thm1}} 
\label{sec:thm1}
As $\hat\bbeta_j$ is a minimizer of the objective function \eqref{eqn:obj} and the {sparse group penalty} function in \eqref{eqn:obj} is convex,  Lemma  \ref{lemma1} implies that
\begin{eqnarray*}
&&\frac{1}{2n}\Vert \z_j-\W\hat\bbeta_j\Vert_2^2+\lambda\Vert\hat\bbeta_j\Vert_1+\lambda_{g}\Vert\hat\bbeta_{j,-0}\Vert_{1,2}+\frac{1}{2n}\Vert\W({\hat\bbeta_j-\bbeta_j})\Vert_2^2\\
&\le&\frac{1}{2n}\Vert \z_j-\W\bbeta_j\Vert_2^2+\lambda\Vert\bbeta_j\Vert_1+\lambda_{g}\Vert\bbeta_{j,-0}\Vert_{1,2}.
\end{eqnarray*}
Writing $\Delta=\hat\bbeta_j-\bbeta_j$ and reorganizing terms in the above inequality gives   
$$
\frac{1}{n}\Vert\W\Delta\Vert_2^2+\lambda\Vert\hat\bbeta_j\Vert_1+\lambda_{g}\Vert\hat\bbeta_{j,-0}\Vert_{1,2}
\le \frac{1}{n}\langle\bepsilon_j,\W\Delta\rangle+\lambda\Vert\bbeta_j\Vert_1+\lambda_{g}\Vert\bbeta_{j,-0}\Vert_{1,2}.
$$
Using the fact that $\Vert\hat\bbeta_j\Vert_1=\Vert(\hat\bbeta_j)_{\mS_j}\Vert_1+\Vert(\hat\bbeta_j)_{\mS^c_j}\Vert_1$, $\Vert\bbeta_j\Vert_1=\Vert(\bbeta_j)_{\mS_j}\Vert_1$, $\Vert\hat\bbeta_j\Vert_{1,2}=\Vert(\hat\bbeta_j)_{(\mG_j)}\Vert_{1,2}+\Vert(\hat\bbeta_j)_{(\mG^c_j)}\Vert_{1,2}$, $\Vert\bbeta_j\Vert_{1,2}=\Vert(\bbeta_j)_{(\mG_j)}\Vert_{1,2}$ and applying the triangle inequalities of $\Vert\cdot\Vert_1$ and $\Vert\cdot\Vert_{1,2}$, we arrive at 
\begin{equation}\label{eqn:eq1}
\frac{1}{n}\Vert\W\Delta\Vert_2^2+\lambda\Vert\Delta_{\mS_j^c}\Vert_1+\lambda_{g}\Vert\Delta_{(\mG_j^c)}\Vert_{1,2}
\le \frac{1}{n}\langle\bepsilon_j,\W\Delta\rangle+\lambda\Vert\Delta_{\mS_j}\Vert_1+\lambda_{g}\Vert\Delta_{(\mG_j)}\Vert_{1,2}.
\end{equation}
Defining $\hat\mS_j=\{l:(\hat\bbeta_j)_l\neq 0, l\in[(p-1)(q+1)]\}$ and letting $\tilde\mS_j=\mS_j\cup\hat\mS_j$,  we obtain 
\begin{eqnarray}\label{eqn:eq2}
&&\langle\bepsilon_j,\W\Delta\rangle=\langle\bepsilon_j,\mP_{\tilde\mS_j}\W_{\tilde\mS_j}\Delta_{\tilde\mS_j}\rangle\\\nonumber
&=&\langle\mP_{\tilde\mS_j}\bepsilon_j,\W\Delta\rangle {\le} \frac{1}{2a_1}\Vert\W\Delta\Vert_2^2+\frac{a_1}{2}\Vert\mP_{\tilde\mS_j}\bepsilon_j\Vert_2^2,
\end{eqnarray}
where {the second equality holds with $\tilde\mS_j=\mS_j\cup\hat\mS_j$} and the last inequality comes from  that $2ab\le ta^2+b^2/t$ {for any $t>0$}. Here, $\mP_{\tilde\mS_j}$ is the orthogonal projection matrix onto the column space of $\W_{\tilde\mS_j}$. 
{As opposed to the classic techniques that bound $\langle\bepsilon,\W\Delta\rangle$ with $\Vert\W^\top\bepsilon\Vert_{\infty}\Vert\Delta\Vert_1$ or $\Vert\W^\top\bepsilon\Vert_{\infty,2}\Vert\Delta\Vert_{1,2}$ \citep{bickel2009simultaneous,lounici2011oracle,negahban2012unified}, we bound this term in \eqref{eqn:eq2} with $\Vert\W\Delta\Vert_2^2/(2a_1)+a_1\Vert\mP_{\tilde\mS_j}\bepsilon_j\Vert_2^2/2$, {which is useful in our proof to more sharply bound the term $\Vert\mP_{\tilde\mS_j}\bepsilon_j\Vert_2^2$. This  indeed is a challenging step }as the group lasso penalty term in \eqref{eqn:obj} is not decomposable with respect to $\mS_j$ and, hence, the existing techniques based on decomposable regularizers and null space properties are non-applicable. We provide a new proof, which is divided into three steps.}

\medskip
\noindent
\textbf{Step 1:} We first bound $\Vert\mP_{\mS_j}\bepsilon_j\Vert_2^2$ with some cardinality measures. Given any $\mJ\subset[(p-1)(q+1)]$ and $\bgamma\in\{0,1\}^{(p-1)(q+1)}$ satisfying $\bgamma_{\mJ}=\1$ and $\bgamma_{\mJ^c}=\0$, we write $\mG(\mJ)=\{h:(\bgamma)_{(h)}\neq \0, h\in[q]\}$. In this step, we aim to show that, given $0\le s_{j,g}\le q+1$ and $0\le s_j\le (p-1)(q+1)$, {the following  holds} 
\begin{eqnarray*}
&&\mathbb{P}\left[\sup_{|\mJ|=s_{j},|\mG(\mJ)|=s_{j,g}}\Vert\mP_{\mJ}\bepsilon_j\Vert_2^2\ge 6\sigma_{\epsilon_j}^2\left\{s_j\log(ep)+s_{j,g}\log(eq/s_{j,g})\right\}+t\sigma_{\epsilon_j}^2\right]\le c_1\exp(-c_2t),
\end{eqnarray*}
where $c_1$, $c_2$ are {positive constants}.  

As the projection matrix $\mP_{\mJ}$ is idempotent, Lemma \ref{lemma2} implies that 
$$
\Vert\mP_{\mJ}\bepsilon_j\Vert_2^2/\sigma_{\epsilon_j}^2\sim\chi^2_d, \quad d<s_j,
$$
{where $d$ is the rank of $\mP_{\mJ}$}.
Next, we find the size of $\{\mJ\subset[(p-1)(q+1)], |\mJ|=s_{j},|\mG(\mJ)|=s_{j,g}\}$ by  considering two  cases (i) $s_{j,g}=s_j$ and (ii) $s_{j,g}<s_j$. These are the only two possible cases since $\mS_j$ includes all nonzero elements in $(\bbeta_j)_{(0)}$ and $(\bbeta_j)_{(1)}$,\ldots,$(\bbeta_j)_{(q)}$.

\noindent
\textbf{case (i):} $s_{j,g}=s_j$. Here, $\{\mJ\subset[(p-1)(q+1)], |\mJ|=s_{j},|\mG(\mJ)|=s_{j,g}\}$ contains ${q\choose s_{j,g}}(p-1)^{s_j}$ elements. It follows from Stirling's approximation that $\log{q\choose s_{j,g}}\le s_{j,g}\log(eq/s_{j,g})$.
{Therefore, $\log\left\{{q\choose s_{j,g}}(p-1)^{s_j}\right\}\le s_j\log p+s_{j,g}\log(eq/s_{j,g})$.}

\noindent
\textbf{case (ii):} $s_{j,g}<s_j$. The number of elements in $\{\mJ\subset[(p-1)(q+1)], |\mJ|=s_{j},|\mG(\mJ)|=s_{j,g}\}$ is  bounded above by ${q\choose s_{j,g}}{(p-1)+(p-1)s_{j,g}\choose s_j}$. By Stirling's approximation, we have\\ $\log{(p-1)+(p-1)s_{j,g}\choose s_j}\le s_j\log(e(p-1)(s_{j,g}+1)/s_j)\le s_j\log(ep)$. 
{Therefore, we have \\$\log\left\{{q\choose s_{j,g}}{(p-1)+(p-1)s_{j,g}\choose s_j}\right\}\le s_j\log (ep)+s_{j,g}\log(eq/s_{j,g})$.}

Combining these two cases, we  conclude that $|\{\mJ\subset[(p-1)(q+1)], |\mJ|=s_{j},|\mG(\mJ)|=s_{j,g}\}|$ is bounded above by $s_j\log(ep)+s_{j,g}\log(eq/s_{j,g})$.
Applying Lemma \ref{lemma3} and applying the union bound leads to the desired result in Step 1. 

\medskip
\noindent
\textbf{Step 2:} Using the result from Step 1, we find an upper bound for $\Vert\mP_{\tilde\mS_j}\bepsilon_j\Vert_2^2$. First, we define 
$$
r_{s,s_g}=\left(\sup_{|\mJ|=s,|\mG(\mJ)|=s_g}\Vert\mP_{\mJ}\bepsilon_j\Vert_2^2-M\sigma_{\epsilon_j}^2\left\{s\log(ep)+s_g\log(eq/s_g)\right\}\right)_+,
$$
and $r=\sup_{1\le s\le (p-1)(q+1),\,0\le s_g\le q}r_{s,s_g}$, {where $M>0$ is a constant to be specified later}. Then, 
\begin{eqnarray}\label{eqn:sbound1}
\Vert\mP_{\tilde\mS_j}\bepsilon_j\Vert_2^2 &\le& M\sigma_{\epsilon_j}^2\left\{\tilde s_j\log(ep)+\tilde s_{j,g}\log(eq/s_{j,g})\right\}+r\\\nonumber
&\le&M\sigma_{\epsilon_j}^2\left\{(s_j+\hat s_j)\log(ep)+(s_{j,g}+\hat s_{j,g})\log(eq/s_{j,g})\right\}+r.
\end{eqnarray}
With  $M=9$, {the result from Step 1 gives}
\begin{eqnarray*}
\mathbb{P}\{r\ge t\sigma_{\epsilon_j}^2\}&\le&\sum_{s=1}^{(p-1)(q+1)}\sum_{s_g=0}^q\mathbb{P}\{r_{s,s_g}\ge t\sigma_{\epsilon_j}^2\}\\
&\le&\sum_{s=1}^{(p-1)(q+1)}\sum_{s_g=0}^qc_1\exp[-c_2t-3c_2\left\{s\log(ep)+s_g\log(eq/s_g)\right\}].
\end{eqnarray*}

\medskip
\noindent
\textbf{Step 3:} This step  derives an inequality for $\Vert\mP_{\tilde\mS_j}\bepsilon_j\Vert_2^2$ by utilizing the computational optimality of $\hat\bbeta_j$. Since the objective function is convex, $\hat\bbeta_j$ is a stationary point of
$$
\frac{1}{2n}\Vert\z_j-\W\bbeta_j\vert_2^2+\lambda\Vert\bbeta_j\Vert_1+\lambda_{g}\Vert\bbeta_{j,-0}\Vert_{1,2}.
$$
 By the KKT conditions, for any $l\in\hat\mS_j\cap(0)$, $h\in[q]$,  $\hat\bbeta_j$ must satisfy that 
\begin{equation}\label{kkt1}
\lambda\,\text{sign}\{(\hat\bbeta_j)_l\}=\frac{1}{n}\langle\w_l,\z_j-\W\hat\bbeta_j\rangle.
\end{equation}
Similarly, for any $l\in\hat\mS_j\cap(h)$, it must satisfy that 
\begin{equation}\label{kkt2}
\lambda\,\text{sign}\{(\hat\bbeta_j)_l\}+\lambda_g\frac{(\hat\bbeta_j)_l}{\|(\hat\bbeta_j)_{(h)}\|_2^2}=\frac{1}{n}\langle\w_l,\z_j-\W\hat\bbeta_j\rangle.
\end{equation}
Squaring both sides of \eqref{kkt1} and \eqref{kkt2} and summing over all $l\in\hat\mS_j$ gives
$$
\lambda^2\hat s_j+\lambda_g^2\hat s_{j,g}\le\frac{1}{n^2}\Vert\W^\top_{\hat\mS}(\z_j-\W\hat\bbeta_j)\Vert^2_2,
$$
due to  $\text{sign}\{(\hat\bbeta_j)_l\}(\hat\bbeta_j)_l \ge 0.$

Consider $\W_{i.}^\top\v$, where $\v=(\v_1,\ldots,\v_p)$, $\v_l\in\mathbb{R}^{q+1}$ and $\Vert\v\Vert=1$. We have \\$\W_{i\cdot}=(z^{(i)}_1,z^{(i)}_1u^{(i)}_1,\ldots,z^{(i)}_1u^{(i)}_q,\ldots,z^{(i)}_p,z^{(i)}_pu^{(i)}_1,\ldots,z^{(i)}_pu^{(i)}_q)$. With a slight abuse of notation, we include the intercept term into $\u^{(i)}$ in the subsequent development. Letting $\V=[\v^\top_1,\ldots,\v^\top_p]\in\mathbb{R}^{(q+1)\times p}$, we can reexpress $\W_{i\cdot}\v$ as $\W_{i\cdot}\v=\u^{(i)}{}^\top\V\z^{(i)}$, $i\in[n]$.
Consequently, by the law of total expectation and Assumption \ref{ass1}, we have 
\begin{eqnarray}\label{eqn:mineg}
\mathbb{E}\left(\v^\top\frac{\W^\top\W}{n}\v\right)&=&\frac{1}{n}\sum_{i=1}^n\mathbb{E}\left(\u^{(i)}{}^\top\V\z^{(i)}\right)^2\\\nonumber
&=&\mathbb{E}\left[\mathbb{E}\left\{\frac{1}{n}\sum_{i=1}^n\mathbb{E}\left(\u^{(i)}{}^\top\V\z^{(i)}\right)^2\Big\vert \{\u^{(i)}\}_{i\in[n]}\right\}\right]\\\nonumber
&=&\mathbb{E}\left\{\frac{1}{n}\sum_{i=1}^n\u^{(i)}{}^\top\V\bSigma(\u^{(i)})\V^\top{\u^{(i)}}\right\}\\\nonumber
&\ge&\phi_1\times\mathbb{E}\left\{\text{tr}\left(\u^{(1)}{}^\top\V\V^\top\u^{(1)}\right)\right\}\\\nonumber%=\phi_0\times\text{tr}\left(\frac{1}{n}\sum_{i=1}^n{\u^{(i)}}\u^{(i)}{}^\top\V\V^\top\right)\\\nonumber
&\ge&\phi_1\times\lambda_{\min}\left(\text{Cov}(\u^{(1)})\right)\text{tr}(\V\V^\top)=\phi_1/\phi_0,
\end{eqnarray}
where we have used the fact $\text{tr}(\A\B)\ge\lambda_{\min}(\A)\text{tr}(\B)$ for positive semi-definite matrices $\A$ and $\B$ and $\text{tr}(\V\V^\top)=1$. 
By the Cauchy–Schwarz inequality, we deduce
%\begin{eqnarray}\label{eqn:fourth}
%\frac{1}{n}\sum_{i=1}^n\mathbb{E}\{(\W_{i.}^\top\v)^4\}&=& %\mathbb{E}\left\{\frac{1}{n}\sum_{i=1}^n\mathbb{E}(\W_{i.}^\top\v)^4\Big\vert \{\u^{(i)}\}_{i\in[n]}\right\}\\\nonumber
%&\le&3\mathbb{E}\left\{\frac{1}{n}\sum_{i=1}^n\left(\u^{(i)}{}^\top\V\bSigma(\u^{(i)})\V^\top{\u^{(i)}}\right)^2\right\}\\\nonumber
%&\le&3\phi^2_2\mathbb{E}\left(\u^{(1)}{}^\top\V\V^\top\u^{(1)}\right)^2\le3\phi^2_2\mathbb{E}\left\{\lambda_{\min}(\u^{(1)}{}\u^{(1)^\top})\right\}^2=\mathcal{O}(1),
%\end{eqnarray}
\begin{eqnarray}\label{eqn:fourth}
\max_{j,k}\frac{1}{n}\sum_{i=1}^n\mathbb{E}\{(\W_{ij}\W_{ik})^2\}&=&\max_{l_1,l_2,l_3,l_4}\mathbb{E}\left({z_{l_1}^{(1)}}^2{z_{l_2}^{(1)}}^2{u_{l_3}^{(1)}}^2{u_{l_4}^{(1)}}^2\right)=\mathcal{O}(1),
\end{eqnarray}
where we have used the fact that $z^{(i)}_{l_1}$ and $u^{(i)}_{l_2}$ have bounded eighth moments, as they are both sub-Gaussian with a bounded sub-Gaussian norm.
It then follows from Lemma \ref{lemma6.1} and with \eqref{eqn:mineg} and Assumption 3 that, with probability  at least $1-C_0\exp\{-(\log\,p+\log\,q)\}$, we have $\Vert\W_{\hat\mS}\Vert^2/n\le M_1$ for some $M_1>0$.
Since $\hat\mS_j\in\tilde\mS$, it holds that
\begin{equation}\label{eqn:kkt30}
{\lambda}^2\hat s_j+{\lambda_g}^2\hat s_{j,g}\le\frac{2M_1}{n}\Vert\W\Delta\Vert_2^2+\frac{2M_1}{n}\Vert \mP_{\tilde\mS_j}\bepsilon_j\Vert_2^2.
\end{equation}
Combining \eqref{eqn:sbound1} and \eqref{eqn:kkt30} and letting 
$$
\lambda=C\sigma_{\epsilon_j}\sqrt{\log(ep)/n+s_{j,g}\log(eq/s_{j,g})/(ns_j)},\quad \lambda_{g}=\sqrt{s_j/s_{j,g}}\lambda,
$$
where $C=3(M_1a_2)^{1/2}$ for some $a_2>0$, we {arrive at}
\begin{equation}\label{eqn:proj}
(1-\frac{2}{a_2})\Vert\mP_{\tilde\mS_j}\bepsilon_j\Vert_2^2\le 9\sigma_{\epsilon_j}^2\left\{s_j\log(ep)+s_{j,g}\log(eq/s_{j,g})\right\}+\frac{2}{a_2}\Vert\W\Delta\Vert_2^2+r.
\end{equation}
{This, together with \eqref{eqn:eq1} and \eqref{eqn:eq2},  implies}
\begin{eqnarray}\label{eqn:verify}
&&\frac{\Vert\W\Delta\Vert_2^2}{n}+\lambda\Vert\Delta_{\mS_j^c}\Vert_1+\lambda_{g}\Vert\Delta_{(\mG_j^c)}\Vert_{1,2}\\\nonumber
&\le&\frac{1}{2a_1}\frac{\Vert\W\Delta\Vert_2^2}{n}+\frac{9a_1a_2}{2(a_2-2)}\frac{\sigma_{\epsilon_j}^2\{s_j\log(ep)+s_{j,g}\log(eq/s_{j,g})\}}{n}\\\nonumber
&&+\frac{a_1}{a_2-2}\frac{\Vert\W\Delta\Vert_2^2}{n}+\frac{a_1a_2}{2(a_2-2)n}r+\lambda\Vert\Delta_{\mS_j}\Vert_1+\lambda_{g}\Vert\Delta_{(\mG_j)}\Vert_{1,2}.
\end{eqnarray}
Next, we have that
$$
\frac{\Vert\Delta_{\mS_j}\Vert_1}{\sqrt{s_j}}+\frac{\Vert\Delta_{(\mG_{j})}\Vert_{1,2}}{\sqrt{s_{j,g}}}\le\Vert\Delta_{\mS_j}\Vert_2+\Vert\Delta_{(\mG_j)}\Vert_2\le2\frac{\phi_1}{\phi_0}\Vert\bSigma^{1/2}_{\W}\Delta\Vert_2,
$$
where the first inequality is due to that $\Vert\Delta_{\mS_j}\Vert_1\le\sqrt{s_j}\Vert\Delta_{\mS_j}\Vert_2$, $\Vert\Delta_{(\mG_{j})}\Vert_{1,2}\le\sqrt{s_{j,g}}\Vert\Delta_{(\mG_j)}\Vert_2$ and the second inequality holds because  $\Vert\Delta_{\mS_j}\Vert_2+\Vert\Delta_{(\mG_j)}\Vert_2 <  2\Vert\Delta \Vert_2 $ trivially with $\Delta_{\mS_j}$ and $\Delta_{(\mG_j)}$ being the sub-vectors of $\Delta$, and $\lambda_{min}(\bSigma_{\W})\ge\phi_1/\phi_0>0$ in \eqref{eqn:mineg}.
Consequently, 
\begin{eqnarray*}
\lambda\Vert\Delta_{\mS_j}\Vert_1+\lambda_g\Vert\Delta_{(\mG_{j})}\Vert_{1,2}&\le&2C\frac{\phi_1}{{\phi_0}}\sqrt{e_j}\Vert\bSigma_{\W}^{1/2}\Delta\Vert_2\le a_3C\frac{\phi_1}{{\phi_0}}e_j+{\frac{1}{a_3}\Vert\bSigma_{\W}^{1/2}\Delta\Vert^2_2},
\end{eqnarray*}
where {$e_j=\sigma_{\epsilon_j}^2\{s_j\log(ep)+s_{j,g}\log(eq/s_{j,g})\}/n$} and the last inequality comes from  that $2ab\le ta^2+b^2/t$ for any $t>0$.
Plugging this into \eqref{eqn:verify}, we obtain
\begin{eqnarray}\label{eqn:verify20}
&&\left\{1-\frac{1}{2a_1}-\frac{a_1}{a_2-2}\right\}\frac{\Vert\W\Delta\Vert_2^2}{n}\\\nonumber
&\le&\left\{\frac{9a_1a_2}{2(a_2-2)}+Ca_3\frac{\phi_1}{{\phi_0}}\right\}\frac{\sigma_{\epsilon_j}^2\{s_j\log(ep)+s_{j,g}\log(eq/s_{j,g})\}}{n}+\frac{1}{a_3}\Vert\bSigma_{\W}^{1/2}\Delta\Vert^2_2+\frac{a_1a_2}{2(a_2-2)n}r.
\end{eqnarray}

It remains to bound the distance between $\Vert\W\Delta\Vert_2^2/n$ and $\Vert\bSigma_{\W}^{1/2}\Delta\Vert_2^2$. To proceed, we first show {with probability  at least $1-C'\exp\{-(\log\,p+\log\,q)\}$}, 
\begin{equation}\label{eqn:step10}
\sup_{\v\in\mathbb{K}_0(2C_{\bbeta_j}s_j)}\left\vert\v^\top\left(\frac{\W^\top\W}{n}-\bSigma_{\W}\right)\v\right\vert\le1/L,
\end{equation}
where $L$ is an arbitrarily large constant and $\mathbb{K}_0(2C_{\bbeta_j}s_j)=\{\v:\Vert\v\Vert_0\le 2C_{\bbeta_j}s_j\,\text{and}\,\Vert\v\Vert_2=1\}$ for some positive constant $C_{\bbeta_j}$. 

Given \eqref{eqn:fourth}, it follows from Lemma \ref{lemma6.1} and Assumption 3 that with probability  at least $1-C'\exp\{-(\log\,p+\log\,q)\}$
$$
\left\vert\v^\top\left(\frac{\W^\top\W}{n}-\bSigma_{\W}\right)\v\right\vert=o(1).
$$
Combing this with the result in Lemma \ref{lemma7}, we have, {with probability at least $1-C'\exp\{-(\log\,p+\log\,q)\}$,} 
\begin{equation}\label{eqn:340}
\left\vert\Delta^\top\left(\frac{\W^\top\W}{n}-\bSigma_{\W}\right)\Delta\right\vert\le\frac{1}{L'}\left(\Vert\Delta\Vert_2^2+\frac{1}{s_j}\Vert\Delta\Vert^2_1\right),
\end{equation}
where $L'$ is an arbitrarily large positive constant.
Plugging \eqref{eqn:340} into \eqref{eqn:verify20} and choosing proper constants $a_1$, $a_2$ and $a_3$ (e.g., $a_1=2$, $a_2=6$, $a_3=6$), we have 
\begin{equation}\label{eqn:350}
\frac{1}{2}\Vert\bSigma^{1/2}_{\W}\Delta\Vert_2^2\precsim\frac{\sigma_{\epsilon_j}^2\{s_j\log(ep)+s_{j,g}\log(eq/s_{j,g})\}}{n}+\frac{1}{L'}\left(\Vert\Delta\Vert_2^2+\frac{1}{s_j}\Vert\Delta\Vert^2_1\right)+\frac{\sigma_{\epsilon_j}^2}{n},
\end{equation}
with probability  at least $1-c_1\exp[-c'_2\{s_j\log(ep)+s_{j,g}\log(eq/s_{j,g})\}]$,
due to that 
\begin{eqnarray*}
&&\mathbb{P}\left[r\ge M_0\sigma_{\epsilon_j}^2\{s_j\log(ep)+s_{j,g}\log(eq/s_{j,g})\}\right]\le c_1\exp[-c'_2\{s_j\log(ep)+s_{j,g}\log(eq/s_{j,g})\}],
\end{eqnarray*}
for a large positive constant $M_0$.
Next, taking $a_1=2-\sqrt{2}$ and $a_2=6$ in \eqref{eqn:verify} and using the expressions for $\lambda$, $\lambda_g$, we have with probability at least $1-c_1\exp[-c'_2\{s_j\log(ep)+s_{j,g}\log(eq/s_{j,g})\}]$ that
\begin{equation}\label{eqn:360}
\frac{\Vert\Delta_{\mS_j^c}\Vert_1}{\sqrt{s_j}}+\frac{\Vert\Delta_{(\mG_j^c)}\Vert_{1,2}}{\sqrt{s_{j,g}}}\le \sigma_{\epsilon_j}\sqrt{\frac{s_j\log(ep)+s_{j,g}\log(eq/s_{j,g})}{n}}+\frac{\Vert\Delta_{\mS_j}\Vert_1}{\sqrt{s_j}}+\frac{\Vert\Delta_{(\mG_j)}\Vert_{1,2}}{\sqrt{s_{j,g}}}.
\end{equation}
Adding $\Vert\Delta_{\mS_j}\Vert_1/\sqrt{s_j}$ to  both sides of \eqref{eqn:360}, we get
\begin{equation}\label{eqn:370}
\frac{\Vert\Delta\Vert_1}{\sqrt{s_j}}\le\sqrt{e_j}+3\Vert\Delta\Vert_2.
\end{equation}
Plugging \eqref{eqn:370} into \eqref{eqn:350} and with $\lambda_{min}(\bSigma_{\W})\ge\phi_1/\phi_0>0$ in \eqref{eqn:mineg}, we have
$$
\|\Delta\|_2^2\precsim \frac{\sigma_{\epsilon_j}^2}{n}\left\{s_j\log(ep)+s_{j,g}\log(eq/s_{j,g})\right\}+\frac{\sigma_{\epsilon_j}^2}{n},
$$
with probability at least $1-C_1\exp[-C_2\{s_j\log(ep)+s_{j,g}\log(eq/s_{j,g})\}$, for some positive constants $C_1$, $C_2$.
 
\eop

\subsubsection{Proof of Theorem \ref{thm2}}  
We first establish the element-wise error bound of $\hat \bbeta_j$ in \eqref{eqn:infybound}, which is to be  accomplished in three steps.

\noindent
\textbf{Step 1:} With $\bPsi=\W^\top\W/n$, this step shows that with probability at least $1-2\exp\{-c_3(\log\,p+\log\,q)\}$,
\begin{equation}\label{eqn:step1t}
\left\Vert\bPsi\Delta\right\Vert_{\infty}\le\frac{3\eta_j\lambda}{2}
\end{equation}
for a constant $c_3>0$.
We also show with probability at least  
$1-2\exp\{-c'_3\log\,p\}$ that
\begin{equation}\label{eqn:step1n}
\Vert\Delta_{\mS_j^c}\Vert_1\le 4\eta_j\Vert\Delta_{\mS_j}\Vert_1.
\end{equation}

The KKT conditions state that  $\btheta$, an optimizer of \eqref{eqn:obj}, satisfies
$$
\begin{cases}
\left(\W^\top(\z_j-\W\btheta)/n\right)_l=\text{sign}(\theta_l)\lambda      & \quad \text{if } \theta_l\neq 0,l\in(0)\\
\left(\W^\top(\z_j-\W\btheta)/n\right)_l=\text{sign}(\theta_l)\lambda+\lambda_g\frac{\theta_l}{\|\btheta_{(h)}\|_2}      & \quad \text{if } \theta_l\neq 0,l\in(h)\\
\left\vert\left(\W^\top(\z_j-\W\btheta)/n\right)_l\right\vert<\eta_j\lambda      & \quad \text{if } \theta_l= 0.
\end{cases}
$$
Thus,  $\hat \bbeta_j$ must satisfy that
$$
\left\Vert\frac{1}{n}\W^\top(\z_j-\W \hat \bbeta_j)\right\Vert_{\infty}\le\eta_j\lambda.
$$
If we can show that with high probability
\begin{equation}\label{eqn:41t}
\frac{1}{n}\left\Vert\W^\top\bepsilon_j\right\Vert_{\infty}\le\frac{\eta_j\lambda}{2},
\end{equation}
we conclude that  (\ref{eqn:step1t}) holds with at least the same probability.
%$$\left\Vert\bPsi\Delta\right\Vert_{\infty}\le\frac{3\eta_j\lambda}{2}.$$

To prove \eqref{eqn:41t}, we first define $V_l=\w_l^\top\bepsilon_j/n$, $l\in[(p-1)(q+1)]$, where $\w_l$ is the $l$th column of $\W$. As $\u^{(i)}_h$ is bounded as assumed in Assumption 1, $w^{(i)}_l$ is sub-Gaussian. Hence, $V_l$ is a sum of independent sub-exponential random variables and 
$\text{Var}(w^{(i)}_{l})\le M^2\phi_2$ for some $M>0$. Applying Lemma \ref{lemma3.2} gives that 
\begin{equation}\label{eqn:41s}
\mathbb{P}\left(\vert V_l\vert>\frac{\eta_j\lambda}{2}\right)\le 2\exp\left(-\frac{c\eta^2_j\lambda^2n}{4M^2\phi_2\sigma_{\epsilon_j}^2}\right)\le2\exp\{-C'_0(\log\,p+\log\,q)\},
\end{equation} 
where $C'_0=cC^2/(4M^2\phi_2)$, and {$C$ is as defined in \eqref{eqn:lambda},}  and the last inequality is due to
$$
\eta_j\lambda\ge C\sigma_{\epsilon_j}\sqrt{\frac{\log\,p}{n}}+C\sigma_{\epsilon_j}\sqrt{\frac{\log\,q}{n}}.
$$
With constant $C$ in \eqref{eqn:lambda}  chosen sufficiently large, we have $C'_0>1$.
{Applying the union bound inequality  gives}
\begin{eqnarray*}
\mathbb{P}\left(\frac{1}{n}\left\Vert\W^\top\bepsilon_j\right\Vert_{\infty}\ge\frac{\eta_j\lambda}{2}\right)&\le&\mathbb{P}\left(\max_{l}\vert V_l\vert\ge\frac{\eta_j\lambda}{2}\right)\\
&\le&2\exp\{-(C'_0-1)(\log\,p+\log\,q)\}.
\end{eqnarray*}
We have finished showing \eqref{eqn:step1t} of Step 1 by
{taking $c_3=C'_0-1$}.

Next, we  show that $\Vert\Delta_{\mS_j^c}\Vert_1\le3\eta_j\Vert\Delta_{\mS_j}\Vert_1$ with  probability {greater than $1-\exp\{-c'_3(\log\,p)\}$.} {The definition of $\hat\bbeta_j$ implies that} 
$$
\frac{1}{2n}\Vert \z_j-\W\hat\bbeta_j\Vert_2^2+\lambda\Vert\hat\bbeta_j\Vert_1+\lambda_g\Vert\hat\bbeta_{j,-0}\Vert_{1,2}\le\frac{1}{2n}\Vert\bepsilon_j\Vert_2^2+\lambda\Vert\bbeta_j\Vert_1+\lambda_g\Vert\bbeta_{j,-0}\Vert_{1,2}.
$$
Developing the left hand side of the above inequality {gives}
\begin{equation}\label{eg:max0}
\lambda\Vert\hat\bbeta_j\Vert_1+\lambda_g\Vert\hat\bbeta_{j,-0}\Vert_{1,2}
\le\lambda\Vert\bbeta_j\Vert_1+\lambda_g\Vert\bbeta_{j,-0}\Vert_{1,2}+\frac{1}{n}\Delta^\top\W^\top\bepsilon_j.
\end{equation}
Recall that $V_l=\w_l^\top\bepsilon_j/n$, $l\in[(p-1)(q+1)]$, where $\w_l$ is the $l$th column of $\W$.
Using similar arguments as in \eqref{eqn:41t} and again applying Lemma \ref{lemma3.2} gives that 
\begin{equation}\label{eq:inf2}
\mathbb{P}\left(\vert V_l\vert>{\frac{\lambda}{2}}\right)\le 2\exp\left(-\frac{c\lambda^2n}{{4}M^2\phi_2\sigma_{\epsilon_j}^2}\right)\le2\exp\{-C''_0(\log\,p)\},
\end{equation} 
where $C''_0=cC^2/({4}M^2\phi_2)$,  {$C$ is as defined in \eqref{eqn:lambda},} and the last inequality is due to $\lambda^2n\ge C^2\sigma^2_{\epsilon_j}\log\,p$.
{Applying the union bound inequality then gives} 
$$
\mathbb{P}\left(\frac{1}{n}\left\Vert\W^\top\bepsilon_j\right\Vert_{\infty}\ge{\frac{\lambda}{2}}\right)\le\mathbb{P}\left(\max_{l}\vert V_l\vert\ge{\frac{\lambda}{2}}\right)\le2\exp\{\log\,q-(C''_0-1)(\log\,p)\}.
$$
As $\log\,p\asymp\log\,q$, when $C$ is chosen sufficiently large, we have, for some $c_3'>0$, $\exp\{\log\,q-(C''_0-1)(\log\,p)\}\le \exp\{-c_3'\log\,p\}$. 
Defining event $\mathcal{A}$ as
\begin{equation}\label{eg:A}
\mathcal{A}=\left\{\frac{1}{n}\left\Vert\W^\top\bepsilon_j\right\Vert_{\infty}\le{\frac{\lambda}{2}}\right\},
\end{equation}
we have $\mathbb{P}(\mathcal{A}^c)\le 2\exp\{-c_3'\log\,p\}$.
Given $\mathcal{A}$, \eqref{eg:max0} leads to  
$$
2\Vert\hat\bbeta_j\Vert_1+2\sqrt{\frac{s_j}{s_{j,g}}}\Vert\hat\bbeta_{j,-0}\Vert_{1,2}
\le2\Vert\bbeta_j\Vert_1+2\sqrt{\frac{s_j}{s_{j,g}}}\Vert\bbeta_{j,-0}\Vert_{1,2}+\Vert\Delta\Vert_1,
$$
as $\lambda_g=\sqrt{s_j/s_{j,g}}\lambda$  specified in \eqref{eqn:lambda}.

Adding $\Vert\hat\bbeta_j-\bbeta_j\Vert_1$ and $\sqrt{\frac{s_j}{s_{j,g}}}\Vert\hat\bbeta_{j,-0}-\bbeta_{j,-0}\Vert_{1,2}$ to both sides and noting $\Vert\Delta\Vert_1=\Vert\hat\bbeta_j-\bbeta_j\Vert_1$, we obtain
\begin{eqnarray*}
&&\Vert\hat\bbeta_j-\bbeta_j\Vert_1+2\Vert\hat\bbeta_j\Vert_1+2\sqrt{\frac{s_j}{s_{j,g}}}\Vert\hat\bbeta_{j,-0}\Vert_{1,2}+\sqrt{\frac{s_j}{s_{j,g}}}\Vert\hat\bbeta_{j,-0}-\bbeta_{j,-0}\Vert_{1,2}\\
&\le&2\Vert\hat\bbeta_j-\bbeta_j\Vert_1+2\Vert\bbeta_j\Vert_1+2\sqrt{\frac{s_j}{s_{j,g}}}\Vert\bbeta_{j,-0}\Vert_{1,2}+2\sqrt{\frac{s_j}{s_{j,g}}}\Vert\hat\bbeta_{j,-0}-\bbeta_{j,-0}\Vert_{1,2}, 
\end{eqnarray*}
which leads to
\begin{eqnarray*}
&&\Vert\hat\bbeta_j-\bbeta_j\Vert_1+\sqrt{\frac{s_j}{s_{j,g}}}\Vert\hat\bbeta_{j,-0}-\bbeta_{j,-0}\Vert_{1,2}\\
&&\le2(\Vert\hat\bbeta_j-\bbeta_j\Vert_1+\Vert\bbeta_j\Vert_1-\Vert\hat\bbeta_j\Vert_1)+2\sqrt{\frac{s_j}{s_{j,g}}}(\Vert\bbeta_{j,-0}-\bbeta_{j,-0}\Vert_{1,2}+\Vert\bbeta_{j,-0}\Vert_{1,2}-\Vert\hat\bbeta_{j,-0}\Vert_{1,2}).
\end{eqnarray*}
Since  $\Vert(\hat\bbeta_j)_l-(\bbeta_j)_l\Vert_1+\Vert(\bbeta_j)_l\Vert_1-\Vert(\hat\bbeta_j)_l\Vert_1=0$ for $l\in\mS_j^c$ and 
$\Vert(\hat\bbeta_j)_{(h)}-(\bbeta_j)_{(h)}\Vert_{1,2}+\Vert(\bbeta_j)_{(h)}\Vert_{1,2}-\Vert(\hat\bbeta_j)_{(h)}\Vert_{1,2}=0$ for $h\in\mG_j^c$, using  the triangular inequality yields that
$$
\Vert\Delta_{\mS_j^c}\Vert_1\le\Vert\Delta\Vert_1\le 4\Vert\Delta_{\mS_j}\Vert_1+4\sqrt{\frac{s_j}{s_{j,g}}}\Vert\Delta_{\mG_j}\Vert_{1,2}.
$$
Therefore, conditional on event $\mathcal{A}$ {in \eqref{eg:A}}, we have that
$$
\Vert\Delta_{\mS_j^c}\Vert_1+\sqrt{\frac{s_j}{s_{j,g}}}\Vert\Delta_{\mG_j^c}\Vert_{1,2}\le 4\Vert\Delta_{\mS_j}\Vert_1+4\sqrt{\frac{s_j}{s_{j,g}}}\Vert\Delta_{\mG_j}\Vert_{1,2},
$$
which further implies  $\Vert\Delta_{\mS_j^c}\Vert_1\le 4\eta_j\Vert\Delta_{\mS_j}\Vert_1$ because $\Vert\Delta_{\mG_j}\Vert_{1,2}\le\Vert\Delta_{\mS_j}\Vert_{1}$.

\medskip
\noindent
\textbf{Step 2:} The step bounds the diagonal and off-diagonal elements of $\bPsi$. 
We first bound the diagonal elements, i.e., $\bPsi_{ll}=\Vert\w_l\Vert_2^2/n$, where $w^{(i)}_l$ is sub-Gaussian as $u^{(i)}_l$ is bounded. Under Assumptions \ref{ass1} and \ref{ass2}, we have $\text{Var}(w^{(i)}_l)\le M^2\phi_2$ and by \eqref{eqn:mineg}, we have $\bSigma_{\W}(l,l)\ge\phi_1/\phi_0$. 
Using the concentration inequality for sub-exponential random variables in Lemma \ref{lemma3.2}, we have
$$
\mathbb{P}\left(\left\vert\bPsi_{ll}-\bSigma_{\W}(l,l)\right\vert>\bSigma_{\W}(l,l)/2\right)\le 2\exp(-c_4n),
$$
for some positive constant $c_4$. Immediately, 
\begin{equation}\label{eq:dibound}
\mathbb{P}(\bPsi_{ll}\notin[{\phi_0}/(2\phi_1),2M^2\phi_2])\le2\exp(-c_4n)
\end{equation}
 because \[\mathbb{P}(\bPsi_{ll}\notin[{\phi_0}/(2\phi_1),2M^2\phi_2])\le\mathbb{P}\left(\left\vert\bPsi_{ll}-\bSigma_{\W}(l,l)\right\vert>\bSigma_{\W}(l,l)/2\right).
\]
Similarly, for the off diagonal elements, i.e., $\bPsi_{kl}=\w_k^\top\w_l/n$,  by noting that $\Vert\bSigma(\u^{(i)})\Vert_{\max}\le \phi_2$, we have
\begin{eqnarray}\label{eq:offbound}
&&\mathbb{P}\left\{\bPsi_{kl}\notin[-\frac{1}{c_0(1+8\eta_j)s_j},\frac{3}{c_0(1+8\eta_j)s_j}]\right\}\\\nonumber
&\le&\mathbb{P}(\vert\bPsi_{kl}-\bSigma_{\W}(k,l)\vert\ge2\bSigma_{\W}(k,l))\le2\exp(-c_5n),
\end{eqnarray}
for a positive constant $c_5$.

\medskip
\noindent
\textbf{Step 3:} This step establishes that, {conditional on event $\mathcal{A}$ and that
\begin{equation}\label{eq:psibound}
\bPsi_{ll}\in[{\phi_0}/(2\phi_1),2M^2\phi_2],\quad \bPsi_{kl}\in[-\frac{1}{c_0(1+8\eta_j)s_j},\frac{3}{c_0(1+8\eta_j)s_j}],
\end{equation}
we have} 
$$
\min_{\Vert\v_{\mS^c_j}\Vert_1\le3\eta_j\Vert\v_{\mS_j}\Vert_1}\frac{\Vert\W\v\Vert_2}{\sqrt{n}\Vert\v_{\mS_j}\Vert_2}\ge\sqrt{\frac{{\phi_0}}{2\phi_1}-\frac{1}{c_0}}>0.
$$
First, we have that 
\begin{eqnarray*}
\frac{\Vert\W\v_{\mS_j}\Vert^2_2}{n\Vert\v_{\mS_j}\Vert^2_2}&=&\frac{\v_{\mS_j}^\top\text{diag}(\bPsi)\v_{\mS_j}}{\Vert\v_{\mS_j}\Vert^2_2}+\frac{\v_{\mS_j}^\top(\bPsi-\text{diag}(\bPsi))\v_{\mS_j}}{\Vert\v_{\mS_j}\Vert^2_2}\\
&\ge&\frac{{\phi_0}}{2\phi_1}-\frac{1}{c_0(1+8\eta_j)s_j}\frac{\Vert\v_{\mS_j}\Vert_1^2}{\Vert\v_{\mS_j}\Vert_2^2},
\end{eqnarray*}
where $\text{diag}(\bPsi)$ is a diagonal matrix with the diagonal elements being identical to those of $\bPsi$, and the last inequality follows from \eqref{eq:psibound}. Furthermore,
\begin{eqnarray*}
\frac{\Vert\W\v\Vert^2_2}{n\Vert\v_{\mS_j}\Vert^2_2}&\ge&\frac{\Vert\W\v_{\mS_j}\Vert^2_2}{n\Vert\v_{\mS_j}\Vert^2_2}+2\frac{\v_{\mS_j}^\top\bPsi\v_{\mS^c_j}}{n\Vert\v_{\mS_j}\Vert_2}\\
&\ge&\frac{{\phi_0}}{2\phi_1}-\frac{1}{c_0(1+8\eta_j)s_j}\frac{\Vert\v_{\mS_j}\Vert_1^2}{\Vert\v_{\mS_j}\Vert_2^2}-\frac{2}{c_0(1+8\eta_j)s_j}\frac{\Vert\v_{\mS_j}\Vert_1\Vert\v_{\mS^c_j}\Vert_1}{\Vert\v_{\mS_j}\Vert_2^2}\\
&\ge&\frac{{\phi_0}}{2\phi_1}-\frac{1+8\eta_j}{c_0(1+8\eta_j)s_j}\frac{\Vert\v_{\mS_j}\Vert_1^2}{\Vert\v_{\mS_j}\Vert_2^2}\ge\frac{{\phi_0}}{2\phi_1}-\frac{1}{c_0}>0,
\end{eqnarray*}
where we have used the results that $\Vert\Delta_{\mS_j^c}\Vert_1\le4\eta_j\Vert\Delta_{\mS_j}\Vert_1$ on event $\mathcal{A}$ and the fact that $\Vert\v_{\mS_j}\Vert_1\le \sqrt{s_j}\Vert\v_{\mS_j}\Vert_2$. Thus, we have finished Step 3.

\medskip
Lastly, based on the results from Steps 1-3, we find the $\ell_{\infty}$ bound {of the error of $\hat \bbeta_j$.}
For $l\in[(p-1)(q+1)]$, it is true that  
$$
\left(\bPsi(\hat\bbeta_j-\bbeta_j)\right)_l=\bPsi_{ll}(\hat\bbeta_j-\bbeta_j)_l+\sum_{k\neq l} \bPsi_{kl}(\hat\bbeta_j-\bbeta_j)_k.
$$
Given \eqref{eq:psibound},  we have
$$
\left\vert\left(\bPsi(\hat\bbeta_j-\bbeta_j)\right)_l-\bPsi_{ll}(\hat\bbeta_j-\bbeta_j)_l\right\vert\le\frac{3}{c_0(1+8\eta_j)s_j}\sum_{k\neq l}\vert(\hat\bbeta_j-\bbeta_j)_k\vert,
$$
and also
\begin{equation}\label{eqn:infy1t}
\Vert\hat\bbeta_j-\bbeta_j\Vert_{\infty}\le2\phi_1\left\Vert\bPsi\Delta\right\Vert_{\infty}+\frac{6\phi_1}{c_0{\phi_0}(1+8\eta_j)s_j}\Vert\Delta\Vert_1.
\end{equation}
With $\Delta=\hat\bbeta_j-\bbeta_j$,
and conditioning on $\left\Vert\bPsi\Delta\right\Vert_{\infty}\le\frac{3\eta_j\lambda}{2}$ and $\Vert\Delta_{\mS_j^c}\Vert_1\le4\eta_j\Vert\Delta_{\mS_j}\Vert_1$ from Step 1, we have that  
\begin{equation}\label{eqn:u1}
\frac{\Vert\W {\Delta} \Vert^2_2}{n}\le\Vert\bPsi\Delta\Vert_{\infty}\Vert\Delta\Vert_1\le\frac{3\eta_j\lambda}{2}(1+4\eta_j)\sqrt{s_j}\Vert\Delta_{\mS_j}\Vert_2,
\end{equation}
while Step 3 also gives that $\Vert\W\Delta\Vert^2_2/n\ge \{{\phi_0}/(2\phi_1)-1/c_0\}\Vert\Delta_{\mS_j}\Vert^2_2$. Combining the above two inequalities yields that 
{$\{{\phi_0}/(2\phi_1)-1/c_0\}\Vert\Delta_{\mS_j}\Vert^2_2\le\frac{3\eta_j\lambda}{2}(1+4\eta_j)\sqrt{s_j}\Vert\Delta_{\mS_j}\Vert_2$, and, therefore,
$$
\Vert\Delta_{\mS_j}\Vert_2\le 3\eta_j\lambda(1+4\eta_j)\frac{c_0\phi_1}{c_0{\phi_0}-2\phi_1}\sqrt{s_j}.
$$
With $\Vert\Delta_{\mS_j^c}\Vert_1\le4\eta_j\Vert\Delta_{\mS_j}\Vert_1$, it follows that $\Vert\Delta\Vert_1\le(1+4\eta_j)\Vert\Delta_{\mS_j}\Vert_1\le(1+4\eta_j)\sqrt{s_j}\Vert\Delta_{\mS_j}\Vert_2$, and, therefore,}   
$$
\Vert\Delta\Vert_1\le3\eta_j\lambda(1+4\eta_j)^2\frac{c_0\phi_1}{c_0{\phi_0}-2\phi_1}s_j.
$$
Plugging this into \eqref{eqn:infy1t}, we get the desired result that with probability at least $1-C'_1\exp\{-C'_2\log p\}$, 
$$
\Vert\hat\bbeta_j-\bbeta_j\Vert_{\infty}\le\left(3\phi_1\eta_j+\frac{18\phi^2_1(1+4\eta_j)^2\eta_j}{{\phi_0}(c_0{\phi_0}-2\phi_1)(1+8\eta_j)}\right)\lambda,
$$
for some positive constants $C'_1$, $C'_2$.

\subsubsection{Proof of Theorem \ref{thm3}} 
{We start the proof by noting that the two events in \eqref{eqn:bound2} hold with the specified probability by applying Lemma \ref{lemma5}, which is applicable because }
$$
\sup_{i\in[n]}\text{Var}(z_j^{(i)})=\mathcal{O}(1),\quad\frac{1}{n}\sum_{i=1}^n\text{Var}(z_j^{(i)})=\mathcal{O}(1), 
$$
due to $\{\bSigma(\u^{(i)})\}_{jj}\le\phi_2$, $j\in[p]$ as assumed in Assumption \ref{ass2}.
In what follows, we show \eqref{eqn:bound3}, conditional on that  the  two events, as specified in \eqref{eqn:bound2}, hold.

When $\bGamma$ is unknown, compared to the oracle regression equation $\z_j=\W\bbeta_j+\bepsilon_j$,  we only have access to the noisy equation 
$$
\hat\z_j=\hat\W\bbeta_j+\E_j,
$$
where $\E_j=\bepsilon_j+(\hat\z_j-\z_j)+(\W-\hat\W)\bbeta_j$.  As $\hat\bbeta_j$ is a minimizer of {the convex} objective function \eqref{eqn:obj3},  Lemma \ref{lemma1} implies
\begin{eqnarray*}
&&\frac{1}{2n}\Vert \hat\z_j-\hat\W\hat\bbeta_j\Vert_2^2+\lambda\Vert\hat\bbeta_j\Vert_1+\lambda_g\Vert\hat\bbeta_{j,-0}\Vert_{1,2}+\frac{1}{2n}\Vert\hat\W(\hat\bbeta_j-\bbeta_j)\Vert_2^2\\
&\le&\frac{1}{2n}\Vert \hat\z_j-\hat\W\bbeta_j\Vert_2^2+\lambda\Vert\bbeta_j\Vert_1+\lambda_g\Vert\bbeta_{j,-0}\Vert_{1,2}.
\end{eqnarray*}
With $\Delta=\hat\bbeta_j-\bbeta_j$, reorganizing terms in the above inequality leads to 
$$
\frac{1}{n}\Vert\hat\W\Delta\Vert_2^2+\lambda\Vert\hat\bbeta_j\Vert_1+\lambda_g\Vert\hat\bbeta_{j,-0}\Vert_{1,2}
\le \frac{1}{n}\langle\E_j,\hat\W\Delta\rangle+\lambda\Vert\bbeta_j\Vert_1+\lambda_g\Vert\bbeta_{j,-0}\Vert_{1,2}.
$$
Next,  {with $\Delta_{\E_j}=(\hat\z_j-\z_j)+(\W-\hat\W)\bbeta_j$,} we have that 
\begin{eqnarray*}
\frac{1}{n}\langle\E_j,\hat\W\Delta\rangle&=&\frac{1}{n}\langle\bepsilon_j,\hat\W\Delta\rangle+\frac{1}{n}\langle\Delta_{\E_j},\hat\W\Delta\rangle\\
&\le&\frac{1}{n}\langle\bepsilon_j,\hat\W\Delta\rangle+\frac{1}{2n}\Vert\Delta_{\E_j}\Vert_2^2+\frac{1}{2n}\Vert\hat\W\Delta\Vert_2^2.
\end{eqnarray*}
Using similar arguments as in \eqref{eqn:eq1}, we obtain
\begin{eqnarray}\label{eqn:eq31}
&&\frac{1}{2n}\Vert\hat\W\Delta\Vert_2^2+\lambda\Vert\Delta_{\mS_j^c}\Vert_1+\lambda_g\Vert\Delta_{(\mG_j^c)}\Vert_{1,2}\\\nonumber
&&\le\frac{1}{n}\langle\bepsilon_j,\hat\W\Delta\rangle+\frac{1}{2n}\Vert\Delta_{\E_j}\Vert_2^2+\lambda\Vert\Delta_{\mS_j}\Vert_1+\lambda_g\Vert\Delta_{(\mG_j)}\Vert_{1,2}.
\end{eqnarray}
Consider the two stochastic terms $\langle\bepsilon_j,\hat\W\Delta\rangle$ and $\Vert\Delta_{\E_j}\Vert_2^2$. For the latter, recall that our proof is conditional on the events of \eqref{eqn:bound2}. Then,
 with probability  at least 
 $1-\exp({\log p} +\log q-\tau_1\log q)$, 
\begin{equation}\label{eqn:de}
\frac{1}{n}\Vert\Delta_{\E_j}\Vert_2^2\le\frac{1}{n}\Vert\hat\z_j-\z_j\Vert_2^2+\frac{\max_j\Vert\W_{\cdot j}-\hat\W_{\cdot j}\Vert_2^2}{n}\cdot\Vert\bbeta_j\Vert_1^2\precsim \sigma^2_{\epsilon_j}\frac{t\,\log\,q}{n},
\end{equation}
where the last inequality is true due to Assumption \ref{ass4} and \eqref{eqn:bound2}.

Next, we  bound  $\langle\bepsilon_j,\hat\W\Delta\rangle$. Defining $\hat\mS_j=\left\{l:(\hat\bbeta_j)_l\neq 0, l\in[(p-1)(q+1)]\right\}$ and letting $\tilde\mS_j=\mS_j\cup\hat\mS_j$, we have that 
\begin{eqnarray}\label{eqn:eq32}
&&\langle\bepsilon_j,\hat\W\Delta\rangle=\langle\bepsilon_j,\hat\mP_{\tilde\mS_j}\hat\W_{\tilde\mS_j}\Delta_{\tilde\mS_j}\rangle\\\nonumber
&=&\langle\hat\mP_{\tilde\mS_j}\bepsilon_j,\hat\W\Delta\rangle\le\frac{1}{2a_1}\Vert\hat\W\Delta\Vert_2^2+\frac{a_1}{2}\Vert\hat\mP_{\tilde\mS_j}\bepsilon_j\Vert_2^2,
\end{eqnarray}
where $\hat\mP_{\tilde\mS_j}$ is the orthogonal projection matrix onto the column space of $\hat\W_{\tilde\mS_j}$. Using the same argument as in \eqref{eqn:sbound1}, we have 
\begin{equation}\label{eqn:stat3}
\Vert\hat\mP_{\tilde\mS_j}\bepsilon_j\Vert_2^2<M\sigma_{\epsilon_j}^2\left\{(s_j+\hat s_j)\log(ep)+(s_{j,g}+\hat s_{j,g})\log(eq/s_{j,g})\right\}+\hat r, 
\end{equation}
where 
$$
\hat r=\sup_{\substack{{1\le s\le (p-1)(q+1)}\\{0\le s_g\le q}}}\left(\sup_{|\mJ|=s,|\mG(\mJ)|=s_g}\Vert\hat\mP_{\mJ}\bepsilon_j\Vert_2^2-M\sigma_{\epsilon_j}^2\left\{s\log(ep)+s_g\log(eq/s_g)\right\}\right)_+.
$$
Setting $M=9$, by Step 1 in Section \ref{sec:thm1}, we have
$$
\mathbb{P}\{\hat r\ge t\sigma_{\epsilon_j}^2\}<\sum_{s=1}^{(p-1)(q+1)}\sum_{s_g=0}^qc_1\exp(-c_2t)\exp[-3c_2\left\{s\log(ep)+s_g\log(eq/s_g)\right\}].
$$

We move to bound  $\Vert\hat\mP_{\tilde\mS_j}\bepsilon_j\Vert_2^2$ by using the computational optimality of $\hat\bbeta_j$.
As in \eqref{kkt1} and \eqref{kkt2}, it follows that
\begin{equation}\label{eqn:kkt3}
{\lambda}^2\hat s_j+{\lambda_g}^2\hat s_{j,g}\le\frac{1}{n^2}\Vert\hat\W_{\hat\mS_j}(\hat\z_j-\hat\W\hat\bbeta_j)\Vert_2^2.
\end{equation}

We also have that $\Vert\hat\W_{\hat\mS_j}-\W_{\hat\mS_j}\Vert/\sqrt{n}\precsim\sqrt{\hat s_jt\log\,q/n}$ {conditional on the events of \eqref{eqn:bound2}}. 
As $\hat s_j<s_{\lambda}=\mathcal{O}(n^{1/2})$, $t=o(n^{1/3})$and $\log\,q=\mathcal{O}(n^{1/6})$, we have $\Vert\hat\W_{\hat\mS_j}-\W_{\hat\mS_j}\Vert/\sqrt{n}=o(1)$. Together with the result $\Vert\W_{\hat\mS_j}\Vert^2/n\le M_1$ from Step 3 in Section \ref{sec:thm1}, we have that $\Vert\hat\W_{\hat\mS_j}\Vert^2\le M_3$ for some $M_3>0$.
It then follows from the Cauchy-Schwarz inequality that
\begin{eqnarray}\label{eqn:comp3}
{\lambda}^2\hat s_j+{\lambda_{g}}^2\hat s_{j,g}&\le&\frac{3M_3}{n}\Vert\hat\W\Delta\Vert_2^2\\\nonumber
&+&\frac{3M_3}{n}\Vert\hat\mP_{\tilde\mS_j}\bepsilon_j\Vert_2^2+\frac{3M_3}{n}\Vert\Delta_{\E_j}\Vert_2^2.
\end{eqnarray}
Set $\lambda=C\sigma_{\epsilon_j}\sqrt{\{\log(ep)/n+s_{j,g}\log(eq/s_{j,g})/(ns_j)\}}$ and $\lambda_{g}=\sqrt{s_j/s_{j,g}}\lambda$, where $C=3(a_2M_3)^{1/2}$ for some $a_2>0$. Combining \eqref{eqn:stat3} and \eqref{eqn:comp3}, we have that
\begin{eqnarray}\label{eqn:proj2}
(1-\frac{3}{a_2})\Vert\hat\mP_{\tilde\mS_j}\bepsilon_j\Vert_2^2&\le& 9\sigma_{\epsilon_j}^2\left\{s_j\log(ep)+s_{j,g}\log(eq/s_{j,g})\right\}\\\nonumber
&+&\frac{3}{a_2}\Vert\hat\W\Delta\Vert_2^2+\frac{3}{a_2}\Vert\Delta_{\E_j}\Vert_2^2+\hat r.
\end{eqnarray}
Plugging the above inequality and \eqref{eqn:eq32} into \eqref{eqn:eq31}, we have
\begin{eqnarray}\label{eqn:verify2}
&&\frac{\Vert\hat \W\Delta\Vert_2^2}{2n}+\lambda\Vert\Delta_{\mS_j^c}\Vert_1+\lambda_g\Vert\Delta_{(\mG_j^c)}\Vert_{1,2}\\\nonumber
&\le&\frac{1}{2a_1}\frac{\Vert\hat\W\Delta\Vert_2^2}{n}+\frac{9a_1a_2}{2(a_2-3)}\frac{\sigma_{\epsilon_j}^2\{s_j\log(ep)+s_{j,g}\log(eq/s_{j,g})\}}{n}\\\nonumber
&&+\frac{3a_1}{2(a_2-3)}\frac{\Vert\hat\W\Delta\Vert_2^2}{n}+\frac{a_1a_2}{2(a_2-3)n}\hat r+\frac{1}{2n}\Vert\Delta_{\E_j}\Vert_2^2\\\nonumber
&&+\frac{a_1a_2}{2(a_2-3)n}\Vert\Delta_{\E_j}\Vert_2^2+\lambda\Vert\Delta_{\mS_j}\Vert_1+\lambda_g\Vert\Delta_{(\mG_j)}\Vert_{1,2}.
\end{eqnarray}
%Using Lemma \ref{lemma4}, we have
As in Section \ref{sec:thm1},
{we have that
$$
\frac{\Vert\Delta_{\mS_j}\Vert_1}{\sqrt{s_j}}+\frac{\Vert\Delta_{(\mG_{j})}\Vert_{1,2}}{\sqrt{s_{j,g}}}\le\Vert\Delta_{\mS_j}\Vert_2+\Vert\Delta_{(\mG_j)}\Vert_2\le2\frac{\phi_1}{{\phi_0}}\Vert\bSigma^{1/2}_{\W}\Delta\Vert_2.
$$
Consequently, 
\begin{equation}\label{eqn:eq33}
\lambda\Vert\Delta_{\mS_j}\Vert_1+\lambda_g\Vert\Delta_{(\mG_j)}\Vert_{1,2}\le2C\frac{\phi_1}{{\phi_0}}\sqrt{e_j'}\Vert\bSigma^{1/2}_{\W}\Delta\Vert_2.
\end{equation}}
where $e_j'=\sigma_{\epsilon_j}^2\{s_j\log(ep)+s_{j,g}\log(eq/s_{j,g})\}/n$.
Combining \eqref{eqn:verify2} and \eqref{eqn:eq33}, we have
\begin{eqnarray*}
&&\left\{\frac{1}{2}-\frac{1}{2a_1}-\frac{3a_1}{2(a_2-3)}\right\}\frac{\Vert\hat\W\Delta\Vert_2^2}{n}\\
&\le&\left\{\frac{9a_1a_2}{2(a_2-3)}+Ca_3\frac{\phi_1}{{\phi_0}}\right\}\frac{\sigma_{\epsilon_j}^2\{s_j\log(ep)+s_{j,g}\log(eq/s_{j,g})\}}{n}\\
&&+\frac{1}{a_3}\frac{\Vert\bSigma^{1/2}_{\W}\Delta\Vert_2^2}{n}+\frac{a_1a_2}{2(a_2-2)n}\hat r+C''\left\{\frac{1}{2n}+\frac{a_1a_2}{2(a_2-3)n}\right\}.
\end{eqnarray*}
Therefore, by choosing proper constants $a_1$, $a_2$ and $a_3$ (e.g., $a_1=4$, $a_2=51$, $a_3=4$), we have, with probability at least $1-c_1\exp[-c'_2\{s_j\log(ep)+s_{j,g}\log(eq/s_{j,g})\}]$,
\begin{equation}\label{eqn:combine}
\frac{\Vert\hat\W\Delta\Vert_2^2}{n}-\frac{\Vert\bSigma^{1/2}_{\W}\Delta\Vert_2^2}{2n}\precsim\frac{\sigma_{\epsilon_j}^2\{s_j\log(ep)+s_{j,g}\log(eq/s_{j,g})\}}{n}, 
\end{equation}
where we have used the fact that
\begin{eqnarray*}
&&\mathbb{P}\left[\hat r\ge M_0\sigma_{\epsilon_j}^2\{s_j\log(ep)+s_{j,g}\log(eq/s_{j,g})\}\right]\\
&&\le c_1\exp[-c'_2\{s_j\log(ep)+s_{j,g}\log(eq/s_{j,g})\}],
\end{eqnarray*}
for a large constant $M_0$.

We then bound the difference between $\Vert\hat \W\Delta\Vert_2^2/n$ and $\Vert\bSigma^{1/2}_{\W}\Delta\Vert_2^2/n$. To do this, we first show that, with probability  at least $1-c_6\exp[c_7\{\log\,p-(\tau_1-1)\log\,q\}]$,
\begin{equation}\label{eqn:what}
\sup_{\v\in\mathbb{K}_0(2C_{\bbeta_j}s_j)}\left\vert\v^\top\left(\frac{\hat\W^\top\hat\W}{n}-\bSigma_{\W}\right)\v\right\vert\le1/L,
\end{equation}
where $L$ is a large constant and $\mathbb{K}_0(2C_{\bbeta_j}s_j)=\{\v:\Vert\v\Vert_0\le 2C_{\bbeta_j}s_j\,\text{and}\,\Vert\v\Vert_2=1\}$ for some positive constant $C_{\bbeta_j}$. 
Notice that 
$$
\left\vert\v^\top\left(\frac{\hat\W^\top\hat\W}{n}-\frac{\W^\top\W}{n}\right)\v\right\vert\le\frac{2\vert\v^\top\W^\top(\hat\W-\W)\v\vert}{n}
+\frac{\Vert(\hat\W-\W)\v\Vert_2^2}{n},
$$
where we have used that $\Vert\v\Vert_1\le\sqrt{2C_{\bbeta_j}s_j}\Vert\v\Vert_2$.
For the second term on the right hand-side, we have with probability  at least $1-3\exp\{\log\,p-(\tau_1-1)\log\,q\}$,
$$
\frac{\Vert(\hat\W-\W)\v\Vert_2}{\sqrt{n}}\le\frac{\max_j\Vert\W_{\cdot j}-\hat\W_{\cdot j}\Vert_2}{\sqrt{n}}\cdot\Vert\v\Vert_1\precsim\sqrt{\frac{2C_{\bbeta_j}s_jt\log\,q}{n}}=o(1),
$$
and, moreover,  
\begin{equation}\label{eqn:w2}
\frac{\vert\v^\top\W^\top(\hat\W-\W)\v\vert}{n}\le\frac{\Vert(\hat\W-\W)\v\Vert_2}{\sqrt{n}}\times\frac{\Vert\W\v\Vert_2}{\sqrt{n}}.
\end{equation}
Next, using a similar argument as in Section \ref{sec:thm1}, it follows from \eqref{eqn:fourth} and Lemma \ref{lemma6.1} that with probability  at least $1-\mathcal{O}\{(pq)^{-1}\}$ 
$$
\left\vert\v^\top\left(\frac{\W^\top\W}{n}-\bSigma_{\W}\right)\v\right\vert=o(1).
$$
Using the same argument as in \eqref{eqn:mineg} and by Assumption \ref{ass4}, we can show $\Vert\bSigma_{\W}\Vert$ is upper bounded by $\phi'_0\phi_1$. Thus, we have that $\Vert\W\v\Vert_2/\sqrt{n}=\mathcal{O}(1)$ with probability  at least $1-\mathcal{O}\{(pq)^{-1}\}$. Putting this together with \eqref{eqn:w2}, we have shown \eqref{eqn:what}.

Next, conditioning on \eqref{eqn:what} and using the result in Lemma \ref{lemma7}, we have 
\begin{equation}\label{eqn:34}
\left\vert\Delta^\top\left(\frac{\hat\W^\top\hat\W}{n}-\bSigma_{\W}\right)\Delta\right\vert\le\frac{1}{L'}\left(\Vert\Delta\Vert_2^2+\frac{1}{s_j}\Vert\Delta\Vert^2_1\right).
\end{equation}
Plugging this into \eqref{eqn:combine}, we have 
\begin{eqnarray}\label{eqn:35}
\frac{\Vert\W\Delta\Vert_2^2}{2n}&\precsim&\frac{\sigma_{\epsilon_j}^2\{s_j\log(ep)+s_{j,g}\log(eq/s_{j,g})\}}{n}\\\nonumber
&+&\frac{1}{L'}\left(\Vert\Delta\Vert_2^2+\frac{1}{s_j}\Vert\Delta\Vert^2_1\right)+\frac{\sigma_{\epsilon_j}^2}{n}.
\end{eqnarray}
By choosing an appropriate $a_1$ given $a_2$ in \eqref{eqn:verify2}, we have with probability  at least $1-c_1\exp[-c'_2\{s_j\log(ep)+s_{j,g}\log(eq/s_{j,g})\}]$,
\begin{equation}\label{eqn:36}
\frac{\Vert\Delta_{\mS_j^c}\Vert_1}{\sqrt{s_j}}+\frac{\Vert\Delta_{(\mG_j^c)}\Vert_{1,2}}{\sqrt{s_{j,g}}}\le\sqrt{e'_j}+\frac{\Vert\Delta_{\mS_j}\Vert_1}{\sqrt{s_j}}+\frac{\Vert\Delta_{(\mG_j)}\Vert_{1,2}}{\sqrt{s_{j,g}}}.
\end{equation}
Adding $\Vert\Delta_{\mS_j}\Vert_1/\sqrt{s_j}$ to both sides of \eqref{eqn:36}, we get
\begin{equation}\label{eqn:37}
\frac{\Vert\Delta\Vert_1}{\sqrt{s_j}}\le\sqrt{e'_j}+3\Vert\Delta\Vert_2
\end{equation}
Plugging \eqref{eqn:37} into \eqref{eqn:35} and by \eqref{eqn:mineg}, we have
$$
\|\Delta\|_2^2\precsim \frac{\sigma_{\epsilon_j}^2}{n}\left\{s_j\log(ep)+s_{j,g}\log(eq/s_{j,g})\right\}+\frac{\sigma_{\epsilon_j}^2}{n},
$$
with probability at least $1-C_3\exp[C_4\{\log\,p-(\tau_1-1)\log\,q\}]$, for some positive constants $C_3$ and $C_4$.

\eop

\subsubsection{Proof of Theorem \ref{thm4}} 
We first establish  {the $\ell_\infty$ norm bound of $\hat \bbeta_j$ in \eqref{eqn:infybound2} with three steps.}

\noindent
\textbf{Step 1:} In this step, we show that, with probability  at least $1-{c_8}\exp[{c_9}\{\log\,p-(\tau_1-1)\log\,q)\}]$ for some {$c_8, c_9>0$}, 
\begin{equation}\label{eqn:step1}
\left\Vert\hat\bPsi\Delta\right\Vert_{\infty}\le\frac{3\eta_j\lambda}{2}.
\end{equation}
And it also holds with probability  at least $1-{c'_8}\exp[{c'_9}\{\log\,p-(\tau_1-1)\log\,q)\}]$ that, for some {$c'_8, c'_9>0$}, 
\begin{equation}
\Vert\Delta_{\mS_j^c}\Vert_1\le4\eta_j\Vert\Delta_{\mS_j}\Vert_1.
\end{equation}

Under the KKT conditions, if $\btheta$ is an optimum of \eqref{eqn:obj3}, then
$$
\begin{cases}
\left(\hat\W^\top(\hat\z_j-\hat\W\btheta)/n\right)_l=\text{sign}(\theta_l)\lambda      & \quad \text{if } \theta_l\neq 0,l\in(0)\\
\left(\hat\W^\top(\hat\z_j-\hat\W\btheta)/n\right)_l=\text{sign}(\theta_l)\lambda+\lambda_g\frac{\theta_l}{\|\btheta_{(h)}\|_2}      & \quad \text{if } \theta_l\neq 0,\,l\in(h)\\
\left\vert\left(\hat\W^\top(\hat\z_j-\hat\W\btheta)/n\right)_l\right\vert<\eta_j\lambda      & \quad \text{if } \theta_l= 0.
\end{cases}
$$
As such, any solution $\hat\bbeta_j$ satisfies that
$$
\left\Vert\frac{1}{n}\hat\W^\top(\hat\z_j-\hat\W\hat\bbeta_j)\right\Vert_{\infty}\le\eta_j\lambda.
$$
If we can show that with high probability
\begin{equation}\label{eqn:41}
\frac{1}{n}\left\Vert\hat\W^\top\E_j\right\Vert_{\infty}\le\frac{\eta_j\lambda}{2},
\end{equation}
then we can reach the desired conclusion that with high probability
$$
\left\Vert\hat\bPsi\Delta\right\Vert_{\infty}\le\frac{3\eta_j\lambda}{2}.
$$
Now we consider the inequality in \eqref{eqn:41}, and  note that 
$$
\frac{1}{n}\left\Vert\hat\W^\top\E_j\right\Vert_{\infty}\le\underbrace{\frac{1}{n}\Vert\W^\top\bepsilon_j\Vert_{\infty}}_{\text{I}}+\underbrace{\frac{1}{n}\Vert(\hat\W-\W)^\top\bepsilon_j\Vert_{\infty}}_{\text{II}}+\underbrace{\frac{1}{n}\Vert\hat\W^\top\Delta_{\E_j}\Vert_{\infty}}_{\text{III}}.
$$
Consider term (I). Define $V_l=\w_l^\top\bepsilon_j/n$, $j\in[(p-1)(q+1)]$. 
Using a similar argument as in \eqref{eqn:41s}, we have
$$
\mathbb{P}\left(\vert V_l\vert>\frac{\eta_j\lambda}{2}\right)\le 2\exp\left(-\frac{c\eta^2_j\lambda^2n}{4M^2\phi_2\sigma_{\epsilon_j}^2}\right)\le2\exp\{-C'_0(\log\,p+\log\,q)\},
$$ 
where $C'_0=cC/(4M^2\phi_2)>1$. 
Using the union bound inequality, we have 
\begin{eqnarray*}
\mathbb{P}\left(\frac{1}{n}\left\Vert\W^\top\bepsilon_j\right\Vert_{\infty}\ge\frac{\eta_j\lambda}{2}\right)&\le&\mathbb{P}\left(\max_{l}\vert V_l\vert\ge\frac{\eta_j\lambda}{2}\right)\\
&\le&2\exp\{-(C'_0-1)(\log\,p+\log\,q)\}.
\end{eqnarray*}
For term (II), it is true that
$$
\frac{1}{n}\Vert(\hat\W-\W)^\top\bepsilon_j\Vert_{\infty}=\frac{1}{n}\max_{l}\vert\langle\hat\w_l-\w_l,\bepsilon_j\rangle\vert\le\frac{\max_l\Vert\hat\w_l-\w_l\Vert_2}{\sqrt{n}}\cdot\frac{\Vert\bepsilon_j\Vert_2}{\sqrt{n}}.
$$
Using Lemma \ref{lemma3.2}, we have that for any large constant $M_4>0$,
$$
\mathbb{P}\left(\frac{\Vert\bepsilon_j\Vert_2}{\sqrt{n}}>M_4^{1/2}\sigma_{\epsilon_j}\right)\le\mathbb{P}\left(\left\vert\frac{\Vert\bepsilon_j\Vert^2_2}{n}-\sigma^2_{\epsilon_j}\right\vert>(M_4-1)\sigma^2_{\epsilon_j}\right)\le2\exp(-b_0\log\,q).
$$
We further have $\mathbb{P}(\Vert\hat\w_l-\w_l\Vert_2/\sqrt{n})\precsim t^{1/2}\lambda_1$ with probability  at least $1-3\exp(-\tau_1\log\,q)$ and $t^{1/2}\lambda_1=o(\eta_j\lambda)$, which is true as $t=o(\sqrt{n/\log\,q})$. Applying the union bound, we have
\begin{equation}\label{eqn:term21}
\mathbb{P}\left(\frac{1}{n}\Vert(\hat\W-\W)^\top\bepsilon_j\Vert_{\infty}\ge\frac{\eta_j\lambda}{2}\right)\le b'_0\exp\{\log p+\log q-\tau_1\log q)\}.
\end{equation} 
Moving to  term (III), notice that there exists a large constant $M'_4>0$ such that 
\begin{eqnarray*}
\frac{1}{n}\Vert\hat\W^\top\Delta_{\E_j}\Vert_{\infty}&=&\Vert(\W+\hat\W-\W)^\top\Delta_{\E_j}/n\Vert_{\infty}\\
&\le&\frac{1}{n}\Vert\W^\top\Delta_{\E_j}\Vert_{\infty}+\frac{\max_{l}\Vert\hat\w_l-\w_l\Vert_2}{\sqrt{n}}\cdot\frac{\Vert\Delta_{\E_j}\Vert_2}{\sqrt{n}}\\
&\le&\frac{1}{n}\Vert\W^\top\Delta_{\E_j}\Vert_{\infty}+M'_4t\lambda_1^2,
\end{eqnarray*}
with probability  at least $b''_0\exp\{\log p+\log q-\tau_1\log q)\}$, where the last inequality follows from \eqref{eqn:bound2} and \eqref{eqn:de}.
Since $t^{1/2}\lambda_1=o(\eta_j\lambda)$, it suffices to bound the term of $\Vert\W^\top\Delta_{\E_j}\Vert/n$. To this end, we have 
$$
\frac{1}{n}\Vert\W^\top\Delta_{\E_j}\Vert_{\infty}<\frac{\max_l\Vert\w_l\Vert_2}{\sqrt{n}}\cdot\frac{\Vert\Delta_{\E_j}\Vert_2}{\sqrt{n}}.
$$
As $\w_l$ has independent sub-Gaussian entries with a bounded sub-Gaussian norm, using a similar argument as for term (II) we have
\begin{equation}\label{eqn:term31}
\mathbb{P}\left(\frac{1}{n}\Vert\W^\top\Delta_{\E_j}\Vert_{\infty}\ge\frac{\eta_j\lambda}{2}\right)\le b'''_0\exp\{\log p+\log q-\tau_1\log q)\}.
\end{equation} 
Combining  terms (I)-(III), we have finished showing \eqref{eqn:step1} of Step 1.

Next, we prove that $\Vert\Delta_{\mS_j^c}\Vert_1\le 4\eta_j\Vert\Delta_{\mS_j}\Vert_1$. By the definition {of $\hat\bbeta_j$}, we have 
$$
\frac{1}{2n}\Vert \hat\z_j-\hat\W\hat\bbeta_j\Vert_2^2+\lambda\Vert\hat\bbeta_j\Vert_1+\lambda_g\Vert\hat\bbeta_{j,-0}\Vert_{1,2}\le\frac{1}{2n}\Vert\E_j\Vert_2^2+\lambda\Vert\bbeta_j\Vert_1+\lambda_g\Vert\bbeta_{j,-0}\Vert_{1,2}.
$$
Developing the left hand side of the above inequality, we have
$$
\lambda\Vert\hat\bbeta_j\Vert_1+\lambda_g\Vert\hat\bbeta_{j,-0}\Vert_{1,2}
\le\lambda\Vert\bbeta_j\Vert_1+\lambda_g\Vert\bbeta_{j,-0}\Vert_{1,2}+\frac{1}{n}\Delta^\top\hat\W^\top\E_j.
$$
Define an event 
\begin{equation}\label{eqn:A1}
\mathcal{A}_1=\left\{\frac{1}{n}\Vert\W^\top\Delta_{\E_j}\Vert_{\infty}\le{\frac{\lambda}{2}}\right\}.
\end{equation}
We again write 
$$
\frac{1}{n}\left\Vert\hat\W^\top\E_j\right\Vert_{\infty}\le\underbrace{\frac{1}{n}\Vert\W^\top\bepsilon_j\Vert_{\infty}}_{\text{I}}+\underbrace{\frac{1}{n}\Vert(\hat\W-\W)^\top\bepsilon_j\Vert_{\infty}}_{\text{II}}+\underbrace{\frac{1}{n}\Vert\hat\W^\top\Delta_{\E_j}\Vert_{\infty}}_{\text{III}}.
$$
For term (I),  first define $V_l=\w_l^\top\bepsilon_j/n$, $j\in[(p-1)(q+1)]$. 
Using a similar argument as in \eqref{eq:inf2}, we have
$$
\mathbb{P}\left(\frac{1}{n}\left\Vert\W^\top\bepsilon_j\right\Vert_{\infty}\ge {\frac{\lambda}{2}} \right)\le\mathbb{P}\left(\max_{l}\vert V_l\vert\ge {\frac{\lambda}{2}} \right)\le \exp\{-c_3'\log\,p\};
$$
for term (II), using the same argument as in \eqref{eqn:term21} and by noting $t^{1/2}\lambda_1=o(\lambda)$, which is true as $t=o(\sqrt{n/\log\,q})$, we have
$$
\mathbb{P}\left(\frac{1}{n}\Vert(\hat\W-\W)^\top\bepsilon_j\Vert_{\infty}\ge{\frac{\lambda}{2}}\right)\le b'_0\exp\{\log p+\log q-\tau_1\log q)\};
$$ 
for term (III), using the same argument as in \eqref{eqn:term31} and again noting $t^{1/2}\lambda_1=o(\lambda)$, we have
$$
\mathbb{P}\left(\frac{1}{n}\Vert\W^\top\Delta_{\E_j}\Vert_{\infty}\ge{\frac{\lambda}{2}}\right)\le b'''_0\exp\{\log p+\log q-\tau_1\log q)\}.
$$ 
Combining terms (I)-(III), we have $\mathbb{P}(\mathcal{A}_1^c)\le {c'_8}\exp[{c'_9}\{\log\,p-(\tau_1-1)\log\,q)\}]$.
Conditioning on $\mathcal{A}_1$ {in \eqref{eqn:A1}}, we have that 
$$
2\Vert\hat\bbeta_j\Vert_1+2\sqrt{\frac{s_j}{s_{j,g}}}\Vert\hat\bbeta_{j,-0}\Vert_{1,2}
\le2\Vert\bbeta_j\Vert_1+2\sqrt{\frac{s_j}{s_{j,g}}}\Vert\bbeta_{j,-0}\Vert_{1,2}+\Vert\Delta\Vert_1.
$$
The rest of the arguments is similar to Step 1 in the proof of Theorem \ref{thm2}, which leads us to the desired result of $\Vert\Delta_{\mS_j^c}\Vert_1\le4\eta_j\Vert\Delta_{\mS_j}\Vert_1$ given $\mathcal{A}_1$.

\medskip
\noindent
\textbf{Step 2:} In this step, we bound the diagonal and off-diagonal elements of $\hat\bPsi$. First, we consider the diagonal elements of $\hat\bPsi$, i.e.,  $\hat\bPsi_{ll}$'s.
Note that 
$$
\frac{\Vert\w_l\Vert_2}{\sqrt{n}}-\frac{\Vert\hat\w_l-\w_l\Vert_2}{\sqrt{n}}\le\frac{\Vert\hat\w_l\Vert_2}{\sqrt{n}}\le\frac{\Vert\w_l\Vert_2}{\sqrt{n}}+\frac{\Vert\hat\w_l-\w_l\Vert_2}{\sqrt{n}}.
$$
Since $\Vert\hat\w_l-\w_l\Vert_2/\sqrt{n}=\mathcal{O}(t^{1/2}\lambda_1)=o(1)$ and $\mathbb{P}(\bPsi_{ll}\notin[{\phi_0}/(2\phi_1),2M^2\phi_2])\le2\exp(-c_4n)$ from \eqref{eq:dibound}, we have that ${\phi_0}/(3\phi_1)\le\hat\bPsi_{ll}\le 3M^2\phi_2$ with probability  at least $1-2\exp(-c'_4n)$. Next, we consider the off-diagonal elements. It holds for $k\neq l$ that
$$
\Vert\w_k^\top\w_l-\hat\w_k^\top\hat\w_l\Vert_2\le\Vert\w_k^\top(\w_l-\hat\w_l)\Vert_2+\Vert\hat\w_l^\top(\hat\w_k-\w_k)\Vert_2.
$$
Since $\Vert\hat\w_l-\w_l\Vert_2/\sqrt{n}=\mathcal{O}(t^{1/2}\lambda_1)=o(1)$ and \eqref{eq:offbound} shows
$$
\mathbb{P}\left\{\bPsi_{kl}\notin[-\frac{1}{c_0(1+8\eta_j)s_j},\frac{3}{c_0(1+8\eta_j)s_j}]\right\}\le2\exp(-c_5n),
$$
the following holds with probability  at least $1-\exp(-c'_5n)$,
$$
\max_{k\neq l}\hat\bPsi_{kl}\in[-\frac{2}{c_0(1+8\eta_j)s_j},\frac{4}{c_0(1+8\eta_j)s_j}].
$$

\medskip
\noindent
\textbf{Step 3:} We show that conditional on $\mathcal{A}_1$ and that
\begin{equation}\label{eqn:c20}
\max_l\hat\bPsi_{ll}\in[{\phi_0}/(3\phi_1),3M^2\phi_2],\quad \max_{k\neq l}\hat\bPsi_{kl}\in[-\frac{2}{c_0(1+8\eta_j)s_j},\frac{4}{c_0(1+8\eta_j)s_j}],
\end{equation}
it holds that
$$
\min_{\Vert\v_{\mS^c_j}\Vert_1\le3\eta_j\Vert\v_{\mS_j}\Vert_1}\frac{\Vert\hat\W\v\Vert_2}{\sqrt{n}\Vert\v_{\mS_j}\Vert_2}\ge\sqrt{\frac{{\phi_0}}{3\phi_1}-\frac{2}{c_0}}>0.
$$
First, given \eqref{eqn:c20}, we have that 
\begin{eqnarray*}
\frac{\Vert\hat\W\v_{\mS_j}\Vert^2_2}{n\Vert\v_{\mS_j}\Vert^2_2}&=&\frac{\v_{\mS_j}^\top\text{diag}(\hat\bPsi)\v_{\mS_j}}{\Vert\v_{\mS_j}\Vert^2_2}+\frac{\v_{\mS_j}^\top(\hat\bPsi-\text{diag}(\hat\bPsi))\v_{\mS_j}}{\Vert\v_{\mS_j}\Vert^2_2}\\
&\ge&\frac{{\phi_0}}{3\phi_1}-\frac{2}{c_0(1+8\eta_j)s_j}\frac{\Vert\v_{\mS_j}\Vert_1^2}{\Vert\v_{\mS_j}\Vert_2^2}.
\end{eqnarray*}
Furthermore, given \eqref{eqn:c20} and $\mathcal{A}_1$, we have that 
\begin{eqnarray*}
\frac{\Vert\hat\W\v\Vert^2_2}{n\Vert\v_{\mS_j}\Vert^2_2}&\ge&\frac{\Vert\hat\W\v_{\mS_j}\Vert^2_2}{n\Vert\v_{\mS_j}\Vert^2_2}+2\frac{v_{\mS_j}^\top\hat\bPsi\v_{\mS^c_j}}{n\Vert\v_{\mS_j}\Vert_2}\\
&\ge&\frac{{\phi_0}}{3\phi_1}-\frac{2}{c_0(1+8\eta_j)s_j}\frac{\Vert\v_{\mS_j}\Vert_1^2}{\Vert\v_{\mS_j}\Vert_2^2}-\frac{4}{c_0(1+8\eta_j)s_j}\frac{\Vert\v_{\mS_j}\Vert_1\Vert\v_{\mS^c_j}\Vert_1}{\Vert\v_{\mS_j}\Vert_2^2}\\
&\ge&\frac{{\phi_0}}{3\phi_1}-\frac{2(1+8\eta_j)}{c_0(1+8\eta_j)s_j}\frac{\Vert\v_{\mS_j}\Vert_1^2}{\Vert\v_{\mS_j}\Vert_2^2}\ge\frac{{\phi_0}}{3\phi_1}-\frac{2}{c_0}>0,
\end{eqnarray*}
where we have used the results that $\Vert\Delta_{\mS_j^c}\Vert_1\le 4\eta_j\Vert\Delta_{\mS_j}\Vert_1$ and the fact that $\Vert\v_{\mS_j}\Vert_1\le \sqrt{s_j}\Vert\v_{\mS_j}\Vert_2$. 

Lastly, with results from Steps 1-3, we  find the $\ell_{\infty}$ bound of $\bbeta_j$.
For $l\in[(p-1)(q+1)]$, it is true that
$$
\left(\hat\bPsi({\hat\bbeta_j-\bbeta_j})\right)_l=\hat\bPsi_{ll}({\hat\bbeta_j-\bbeta_j})_l+\sum_{k\neq l}\hat \bPsi_{kl}({\hat\bbeta_j-\bbeta_j})_k
$$
Given \eqref{eqn:c20} from Step 2, we have 
$$
\left\vert\left(\hat\bPsi({\hat\bbeta_j-\bbeta_j})\right)_l-\hat\bPsi_{ll}({\hat\bbeta_j-\bbeta_j})_l\right\vert\le\frac{4}{c_0(1+8\eta_j)s_j}\sum_{k\neq l}\vert({\hat\bbeta_j-\bbeta_j})_k\vert,
$$
and also
\begin{equation}\label{eqn:infy1}
\Vert{\hat\bbeta_j-\bbeta_j}\Vert_{\infty}\le3\phi_1\left\Vert\hat\bPsi\Delta\right\Vert_{\infty}+\frac{12\phi_1}{c_0{\phi_0}(1+8\eta_j)s_j}\Vert\Delta\Vert_1.
\end{equation}
With $\Delta=\hat\bbeta_j-\bbeta_j$ and
given $\left\Vert\hat\bPsi\Delta\right\Vert_{\infty}\le\frac{3\eta_j\lambda}{2}$ and $\Vert\Delta_{\mS_j^c}\Vert_1\le4\eta_j\Vert\Delta_{\mS_j}\Vert_1$ from Step 1, we have
$$
\frac{\Vert\hat\W\Delta\Vert^2_2}{n}\le\Vert\hat\bPsi\Delta\Vert_{\infty}\Vert\Delta\Vert_1\le\frac{3\eta_j\lambda}{2}(4\eta_j+1)\sqrt{s_j}\Vert\Delta_{\mS_j}\Vert_1.
$$
We also have from Step 3 that $\Vert\hat\W\Delta\Vert^2_2/n\ge \{{\phi_0}/(3\phi_1)-2/c_0\}\Vert\Delta_{\mS_j}\Vert^2_2$ given \eqref{eqn:c20} and $\mathcal{A}_1$. Combining these two inequalities and by noting $\Vert\Delta\Vert_1\le(1+4\eta_j)\sqrt{s_j}\Vert\Delta_{\mS_j}\Vert_2$, we have that 
$$
\Vert\Delta\Vert_1\le\frac{3\eta_j\lambda}{2}(1+4\eta_j)^2\frac{3c_0\phi_1}{c_0{\phi_0}-6\phi_1}s_j.
$$
Plugging this into \eqref{eqn:infy1}, we obtain that 
$$
\Vert{\hat\bbeta_j-\bbeta_j}\Vert_{\infty}\le\frac{9}{2}\left\{\phi_1\eta_j+\frac{12\phi^2_1(1+4\eta_j)^2}{{\phi_0}(c_0{\phi_0}-6\phi_1)(1+8\eta_j)}\right\}\lambda.
$$
with probability  at least $1-C_5\exp[C_6\{\log\,p-(\tau_1-1)\log\,q\}]$, for some positive constants $C_5$ and $C_6$.

\subsection{Additional results of data analysis}
\begin{figure}[!hbt]
	\centering
	\includegraphics[scale=0.375, angle =-90]{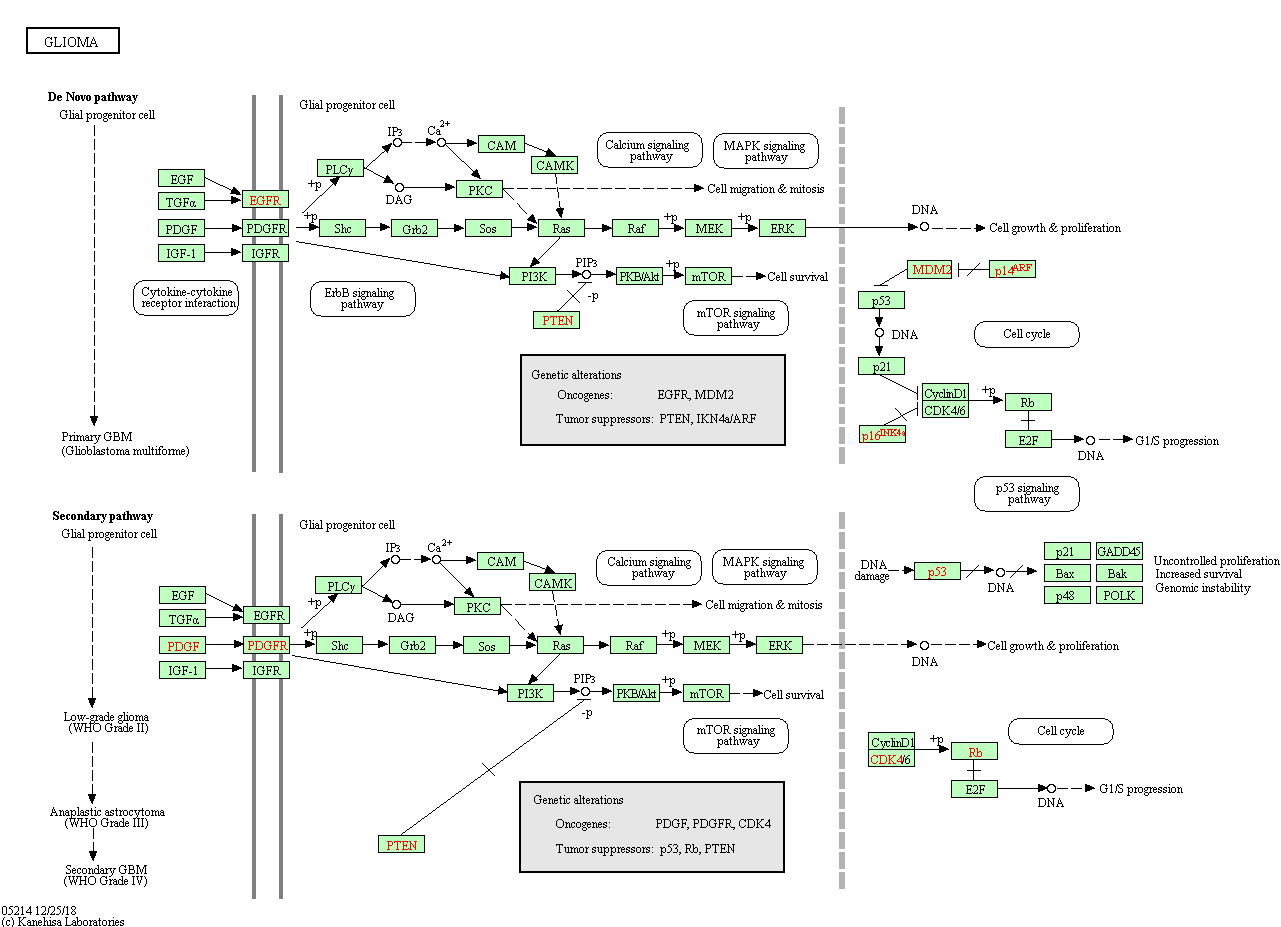}
	\caption{The KEGG human glioma pathway. This figure is downloaded from \url{https://www.genome.jp/kegg/} \citep{kanehisa2000kegg}.}
	\label{fig:pathway}
\end{figure}

%\begin{figure}[!t]
%	\centering
%	\includegraphics[trim=40mm 0 0 0, scale=0.28]{cov2.pdf}
%	\includegraphics[trim=40mm 0 0 0, scale=0.28]{cov3.pdf}
%	\caption{Graphs depending on each covariate (i.e., different SNPs). Edges that have positive (negative) effects on partial correlations are shown in red dashed (black solid) lines.}
%	\label{fig:cov2}
%\end{figure}

\begin{table}[!t]
\centering
\begin{tabular}{l|l}\hline   
 SNP  &  co-expressed genes \\\hline
 rs10492975   & (CALML5,PIK3R2), (CALML5,CAMK1)      \\
 rs723211   & (HRAS,CALML5), (CALML5,CAMK1G)      \\
 rs1347069  & (SHC4,CDKN2A)      \\
 rs473698   & (PRKCG,CAMK1)  \\
 rs4118334  & (SHC2,CAMK1)  \\
 rs882664   & (PRKCA,CAMK1)  \\
 rs1267622  & (SHC3,RAF1)  \\\hline
\end{tabular}
\caption{Identified co-expression QTLs and the corresponding co-expressed genes.}\label{tab:cov}
\end{table}

\clearpage
\newpage
\subsection*{Additional references}

\begin{description}\setlength{\itemsep}{-0.5ex}

\bibitem[{Balasubramanian et~al.(2018)}]{bala2018}
Balasubramanian, K., Fan, J., and Yang, Z. (2018), {Tensor methods for additive index models under discordance and heterogeneity,} \textit{arXiv preprint arXiv:1807.06693}.

\bibitem[{Bellec et~al.(2018)Bellec, Dalalyan, Grappin, and Paris}]{bellec2018prediction}
Bellec, P.~C., Dalalyan, A.~S., Grappin, E., and Paris, Q. (2018), {On
  the prediction loss of the lasso in the partially labeled setting,}
  \textit{Electronic Journal of Statistics}, 12, 3443--3472.
  
\bibitem[{Graybill and Marsaglia(1957)}]{graybill1957idempotent}
Graybill, F.~A. and Marsaglia, G. (1957), {Idempotent matrices and
  quadratic forms in the general linear hypothesis,} \textit{The Annals of
  Mathematical Statistics}, 28, 678--686.
  
 \bibitem[{Kuchibhotla and Chakrabortty(2018)}]{kuchibhotla2018moving}
Kuchibhotla, A.~K. and Chakrabortty, A. (2018), {Moving beyond
  sub-gaussianity in high-dimensional statistics: Applications in covariance
  estimation and linear regression,} \textit{arXiv preprint arXiv:1804.02605}.
   
\bibitem[{Laurent and Massart(2000)}]{laurent2000adaptive}
Laurent, B. and Massart, P. (2000), {Adaptive estimation of a quadratic
  functional by model selection,} \textit{Annals of Statistics}, 1302--1338.
 
\bibitem[{Loh and Wainwright(2011)}]{loh2011high}
Loh, P.-L. and Wainwright, M.~J. (2011), {High-dimensional regression
  with noisy and missing data: Provable guarantees with non-convexity,} in
  \textit{Advances in Neural Information Processing Systems}, pp. 2726--2734.
   
\bibitem[{Vershynin(2010)}]{vershynin2010introduction}
Vershynin, R. (2010), {Introduction to the non-asymptotic analysis of
  random matrices,} \textit{arXiv preprint arXiv:1011.3027}.

\end{description}

\end{document}